\documentclass[useAMS,usenatbib]{mn2e}

\usepackage{lscape}
\usepackage{graphicx}
\usepackage{longtable}
\usepackage{natbib}
\usepackage{multirow}
\usepackage{amsmath}
\usepackage{deluxetable}

\newcommand\ion[2]{\mbox{#1$\;${\sc #2}}}%

\title[]{A large spectroscopic sample of L and T dwarfs from UKIDSS LAS: peculiar objects, binaries, and space density}
\author[F. Marocco et al.]{F. Marocco$^{1}$\thanks{E-mail: f.marocco@herts.ac.uk; Based on observations made with ESO telescopes at the La Silla Paranal Observatory under programs 086.C-0450, 087.C-0639, 088.C-0048, 091.C-0452.}, H. R. A. Jones$^{1}$, A. C. Day-Jones$^{1}$, D. J. Pinfield$^{1}$, P. W. Lucas$^{1}$, \newauthor B. Burningham$^{1,2}$, Z. H. Zhang$^{1,3}$, R. L. Smart$^{4}$, J. I. Gomes$^{1}$, L. Smith$^{1}$\\
$^{1}$Centre for Astrophysics Research, Science and Technology Research Institute, University of Hertfordshire, Hatfield AL10 9AB \\
$^{2}$NASA Ames Research Center, Mail Stop 245-3, Moffett Field, CA 94035, USA \\
$^{3}$Instituto de Astrofisica de Canaria (IAC), C/V\'ia L\'actea s/n, E-38200 La Laguna, Tenerife, Spain\\
$^{4}$INAF/Osservatorio Astrofisico di Torino, Strada Osservatorio 20, 10025 Pino Torinese, Italy
}
\begin{document}

\pagerange{\pageref{firstpage}--\pageref{lastpage}} \pubyear{2015}

\maketitle

\label{firstpage}

\begin{abstract}
We present the spectroscopic analysis of a large sample of late-M, L, and T dwarfs from the United Kingdom Deep Infrared Sky Survey. Using the YJHK photometry from the Large Area Survey and the red-optical photometry from the Sloan Digital Sky Survey we selected a sample of 262 brown dwarf candidates and we have followed-up 196 of them using the echelle spectrograph X-shooter on the Very Large Telescope. The large wavelength coverage ($0.30 - 2.48 \mu$m) and moderate resolution (R$\sim 5000 - 9000$) of X-shooter allowed us to identify peculiar objects including 22 blue L dwarfs, 2 blue T dwarfs, and 2 low gravity M dwarfs. Using a spectral indices-based technique we identified 27 unresolved binary candidates, for which we have determined the spectral type of the potential components via spectral deconvolution. The spectra allowed us to measure the equivalent width of the prominent absorption features and to compare them to atmospheric models. Cross-correlating the spectra with a radial velocity standard, we measured the radial velocity for our targets, and we determined the distribution of the sample, which is centred at -1.7$\pm$1.2 km s$^{-1}$ with a dispersion of 31.5 km s$^{-1}$. Using our results we estimated the space density of field brown dwarfs and compared it with the results of numerical simulations. Depending on the binary fraction, we found that there are $(0.85 \pm 0.55) \times 10^{-3}$ to $(1.00 \pm 0.64) \times 10^{-3}$ objects per cubic parsec in the L4-L6.5 range, $(0.73 \pm 0.47) \times 10^{-3}$ to $(0.85 \pm 0.55) \times 10^{-3}$ objects per cubic parsec in the L7-T0.5 range, and $(0.74 \pm 0.48)  \times 10^{-3}$ to $(0.88 \pm 0.56) \times 10^{-3}$ objects per cubic parsec in the T1-T4.5 range. We notice that there seem to be an excess of objects in the L to T transition with respect to the late T dwarfs, a discrepancy that could be explained assuming a higher binary fraction than expected for the L to T transition, or that objects in the high-mass end and low-mass end of this regime form in different environments, i.e. following different Initial Mass Functions.
\end{abstract}


\section{Introduction}
The study of sub-stellar objects still presents a number of open questions. A very intriguing one is the understanding of the physical and chemical processes taking place at the transition between the spectral types L and T.

The sharp near-infrared colour turnaround that characterizes the transition between the spectral types L7 to T5 \citep{2005ARA&A..43..195K} is particularly challenging to model. The dust settling and the onset of the methane and molecular hydrogen absorption are now believed to be the main causes of the turnaround, but the details of these processes, in particular of the dust settling, are still not well understood. A number of different scenarios have been proposed \citep[e.g. ][]{2003ApJ...585L.151T, 2004AJ....127.3553K, 2002ApJ...568..335M}, but none of them could successfully reproduce the quickness and the sharpness of the turnaround. An important role is also played by atmospheric parameters like metallicity and surface gravity, which influence the nature and the settling of the dust clouds and can lead to the formation of very peculiar spectra \citep[see for instance ][and references therein]{2010ApJS..190..100K}. Understanding in details the effects of these parameters is another open question.

A significant contribution comes from the modern deep wide-field surveys, like DENIS \citep{1999A&A...349..236E}, SDSS \citep{2000AJ....120.1579Y}, 2MASS \citep{2006AJ....131.1163S}, UKIDSS \citep{2007MNRAS.379.1599L}, VHS \citep{2013Msngr.154...35M}, and WISE \citep{2010AJ....140.1868W}. Mapping thousands of squared degrees to significant depths in both optical and infrared bands, these surveys provide huge datasets, and mining them is the best way of finding large samples of brown dwarfs. The increase in numbers of known objects will give us the statistic significance necessary to better constrain current models of the structure and evolution of L and T dwarfs.

In this contribution we present a detailed spectroscopic analysis of a sample of 196 late-M, L and T dwarfs selected from the United Kingdom Infra-red Deep Sky Survey (UKIDSS) Large Area Survey (LAS). The spectra of the targets have been obtained with X-shooter \citep{2011A&A...536A.105V} on the Very Large Telescope. Spectroscopy is a powerful tool to provide insights to the theory, as the formation of the observed spectra is heavily influenced by the physics and the chemistry of the atmosphere. In particular the wide wavelength coverage delivered by X-shooter ($0.30 - 2.48 \mu$m) coupled with its good resolution makes it an ideal instrument for this kind of analysis, as it allows us to obtain both the optical and the near-infrared spectra of our targets at the same time. As these portions of the spectrum are sensitive to different parameters, their comparison can provide extremely useful insights in understanding the physics of the atmospheres of brown dwarfs.

In Section 2 we summarize the candidate selection process, the observation strategy adopted, and the data reduction procedures. In Section 3 we present the results obtained, focusing in particular on the determination of the spectral types, the identification and analysis of the unresolved binaries, and the identification and analysis of the peculiar objects found. In Section 4 we study the evolution of the main spectral features via the analysis of spectral indices and equivalent widths. In Section 5 we present the radial velocities obtained for the targets. In Section 6 we use the sample to place constraints on the Initial Mass Function (IMF) and formation history (also known as Birth Rate, BR) of the local sub-stellar population. Finally in Section 7 we summarize the results obtained.

\section{Candidate selection, observations and data reduction}
\label{cand_selection}
The objects presented here have been selected from the UKIDSS LAS 7th Data Release. The details of the selection criteria can be found in \citet[hereafter ADJ13]{2013MNRAS.430.1171D} and here we briefly summarize them. We selected objects with declination below 20 degrees and brighter than 18.1 in J band. We applied a colour cut of Y $-$ J $<$ 0.8 to remove field M dwarfs \citep{2006MNRAS.367..454H}, and we selected both K band detections and non-detections. Additional quality flags were considered, and their complete list can be found in ADJ13. 

We then cross-matched the preliminary list of candidates against the Sloan Digital Sky Survey (SDSS) 7th Data Release using a matching radius of 4 arcsec. We applied a number of colour-colour cuts, the basic one being \textit{z}$-$J $\geqslant$ 2.4 and J$-$K $\geqslant$ 1.0 or \textit{z}$-$J $\geqslant$ 2.9 and J$-$K $<$ 1.0 \citep{2010AJ....139.1808S}. Given that mid-T dwarfs have very red \textit{z}$-$J colours  \citep[typically $>$ 3.0, e.g.][]{2008MNRAS.390..304P} some of our objects would be too faint for detection in SDSS, and therefore we also include SDSS non-detections. All the remaining candidates were visually inspected to remove mismatches and cross talk, and we finally removed any previously identified L or T dwarfs. The final list of candidates consisted of 262 objects.

We obtained the spectra of 196 of our targets using X-shooter on the Very Large Telescope under the European Southern Observatory (ESO) programs 086.C-0450(A/B), 087.C-0639(A/B), 088.C-0048(A/B), and 091.C-0452A. Sixty-eight spectra were presented in ADJ13, one in \citet{2014MNRAS.439..372M}, and here we present the remaining 127, spanning the RA range 8-16 hours. 

The targets were observed in echelle slit mode, following an A-B-B-A pattern to allow sky subtraction. Individual integration times were set equal to 800, 1200, 1600 and 2000s for J $\leqslant$ 17, 17.5, 18, 18.1 respectively in the VIS arm (covering the 550-1000nm range), decreased by 70s in the UVB arm (300-550nm) and increased by 90s in the NIR arm (1000-2500nm). The data were reduced using the ESO X-shooter pipeline (version 2.0.0 or later). The pipeline performs all the basic steps, such as non-linear pixels cleaning, bias and dark subtraction, flat fielding, sky subtraction, extraction of the individual orders, merging, wavelength calibration and flexure compensation, and flux calibration. The final products are one dimensional, wavelength and flux calibrated spectra, one for each arm. We corrected the spectra for telluric absorption, and merged the three arms using our own IDL code. Telluric standards were observed following a target-telluric-target strategy, trying to minimize the airmass difference between the targets and the telluric stars. Telluric stars were selected preferentially in the late-B $-$ early-A spectral range, as these types of stars are essentially free of absorption features, except for the \ion{H}{i} lines that are not present in brown dwarfs  and whose influence can be interpolated over. Their spectra were also reduced using the X-shooter pipeline. Further details about the observation strategy and the data reduction can be found in ADJ13 and in Appendix \ref{app_log}.

\section{Results}
\begin{figure*}
\centering
\includegraphics[width=\textwidth]{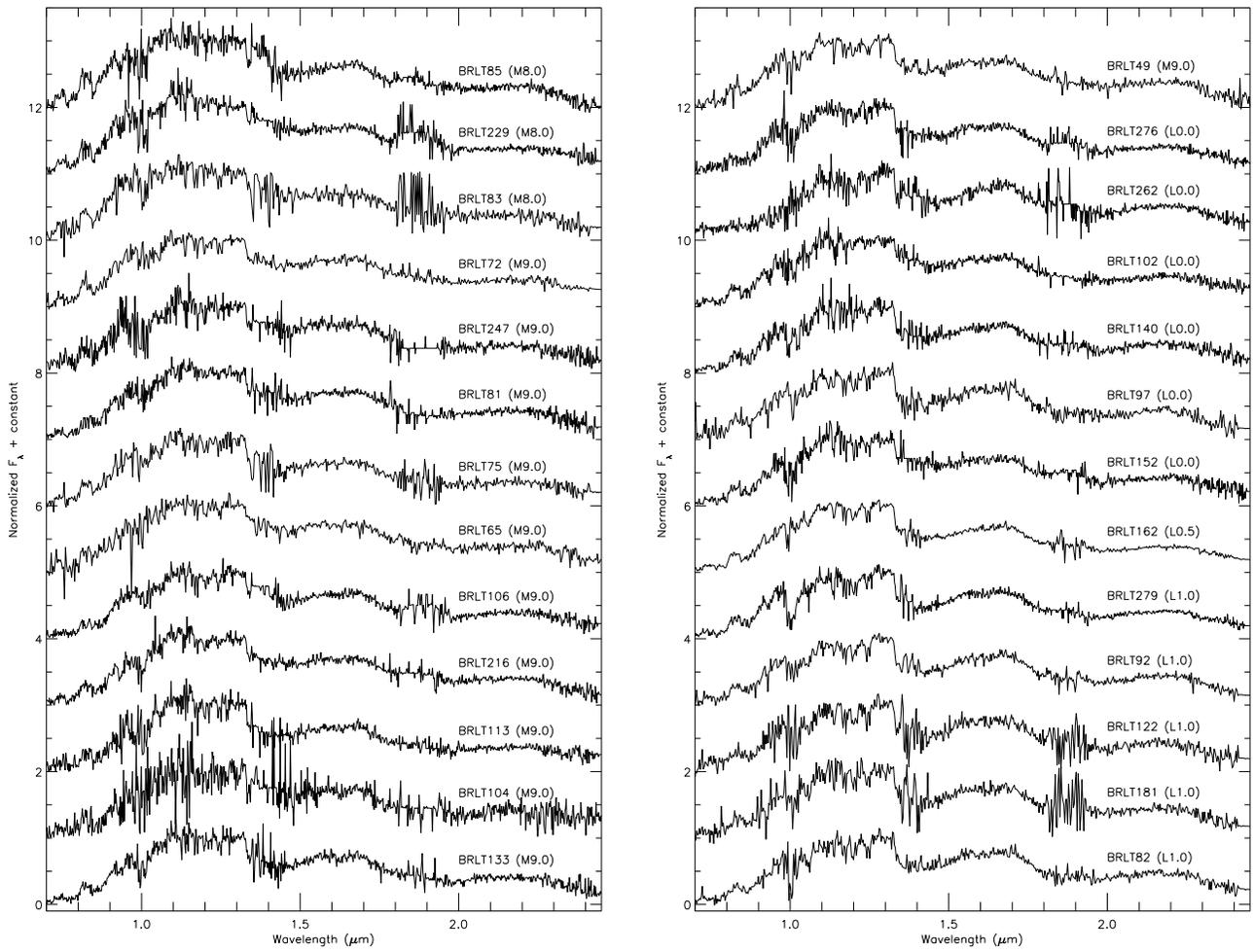}
\caption{The Xshooter spectra of the objects presented here, sorted in ascending order of spectral type (M8.0$-$L1.0). All spectra are normallized at 1.28$\mu$m and offset vertically by increments of one flux unit. \label{spectra0}}
\end{figure*}

\begin{figure*}
\centering
\includegraphics[width=\textwidth]{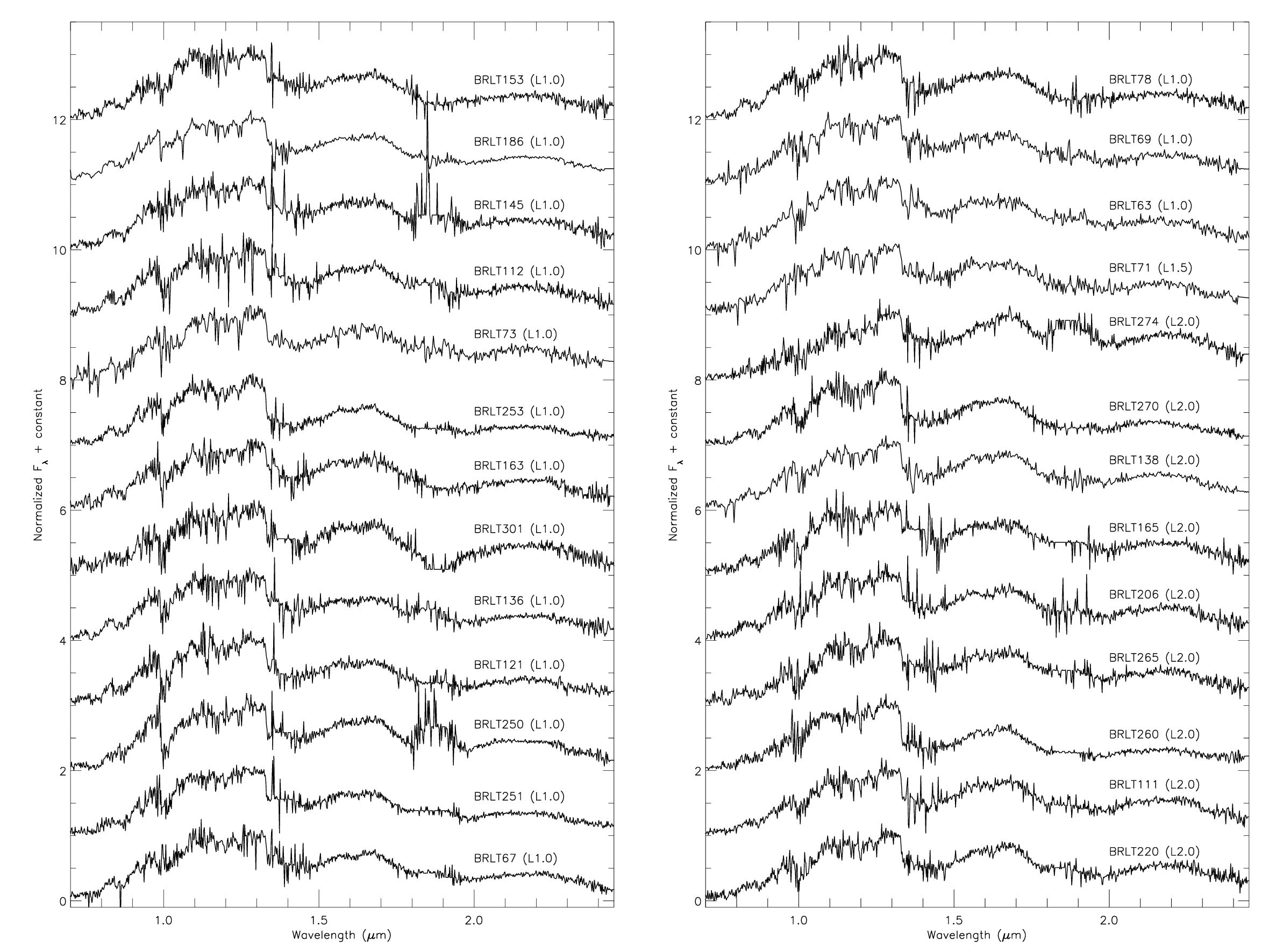}
\caption{The Xshooter spectra of the objects presented here, sorted in ascending order of spectral type (L1.0$-$L2.0). All spectra are normallized at 1.28$\mu$m and offset vertically by increments of one flux unit.  \label{spectra1}}
\end{figure*}

\begin{figure*}
\centering
\includegraphics[width=\textwidth]{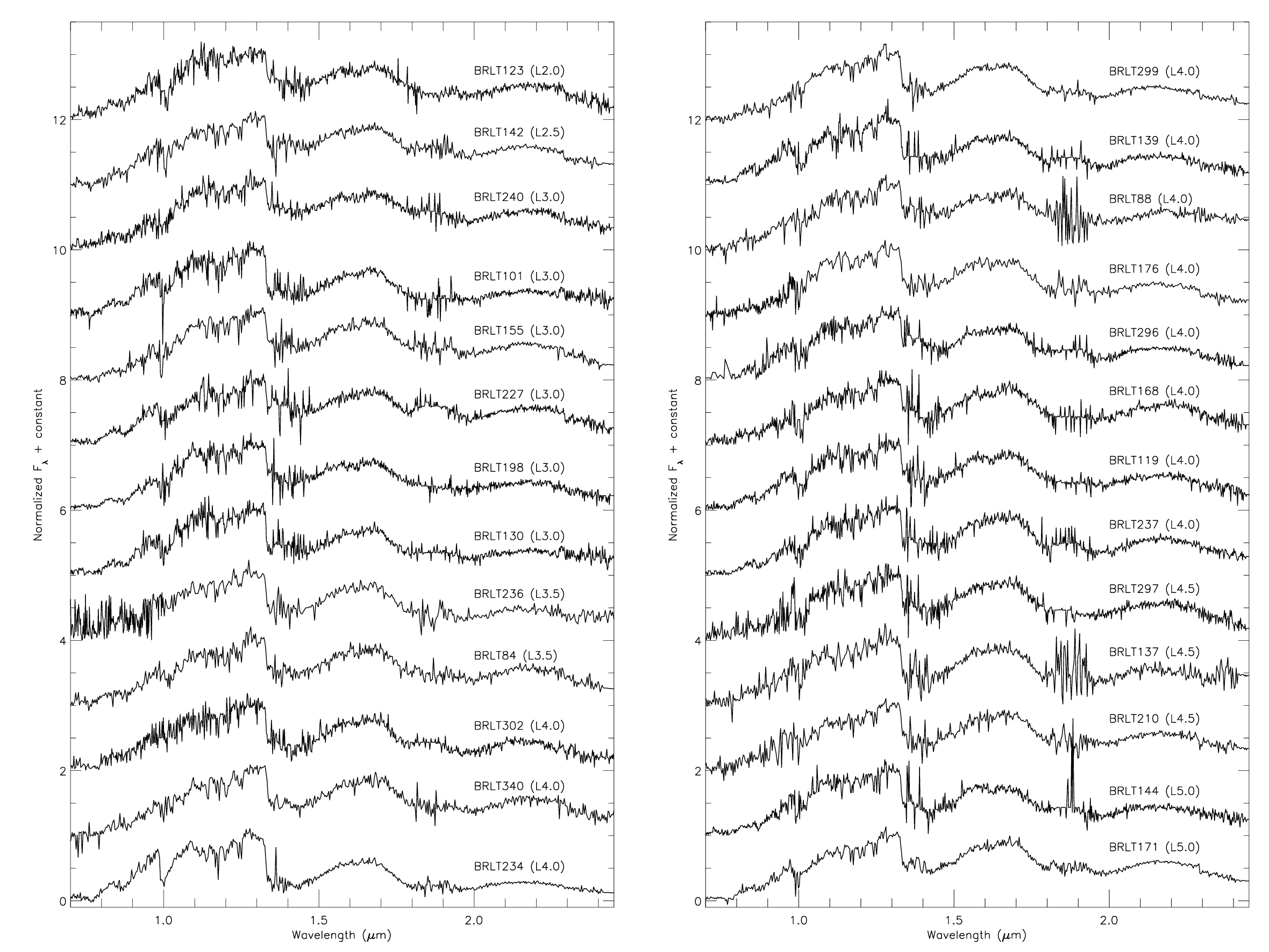}
\caption{The Xshooter spectra of the objects presented here, sorted in ascending order of spectral type (L2.0$-$L5.0). All spectra are normallized at 1.28$\mu$m and offset vertically by increments of one flux unit. \label{spectra2}}
\end{figure*}

\begin{figure*}
\centering
\includegraphics[width=\textwidth]{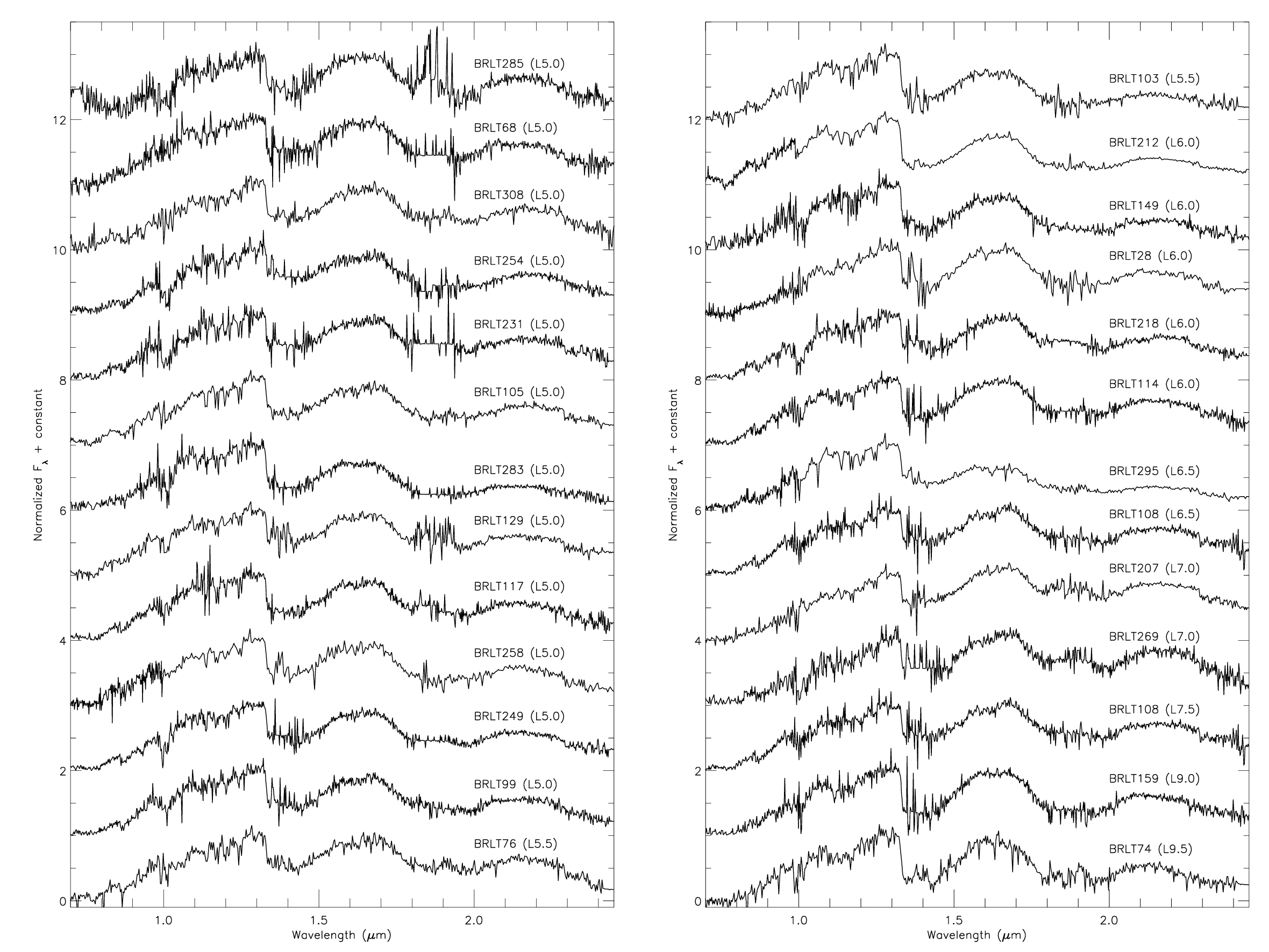}
\caption{The Xshooter spectra of the objects presented here, sorted in ascending order of spectral type (L5.0$-$L9.5). All spectra are normallized at 1.28$\mu$m and offset vertically by increments of one flux unit. \label{spectra3}}
\end{figure*}

\begin{figure*}
\centering
\includegraphics[width=\textwidth]{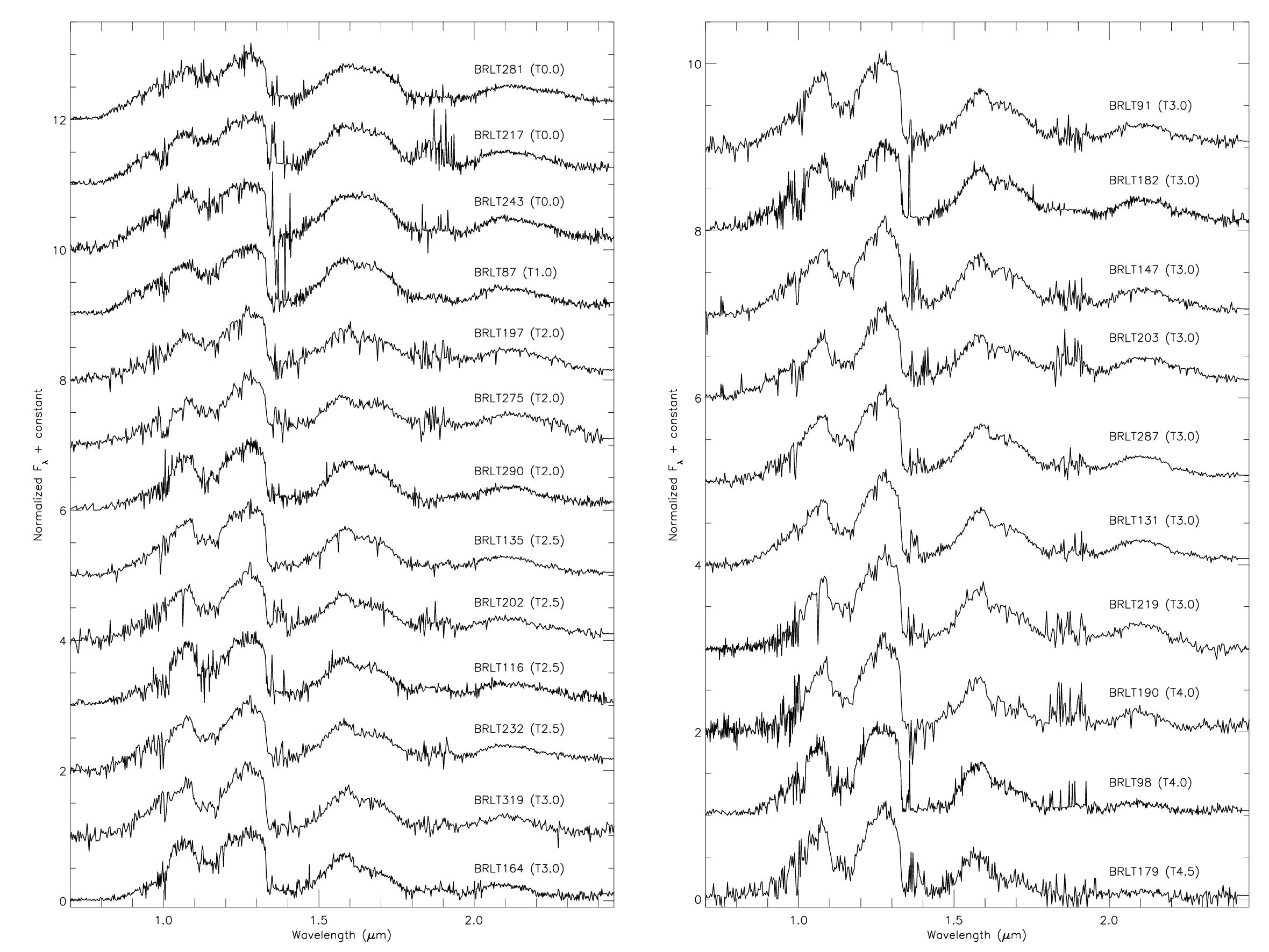}
\caption{The Xshooter spectra of the objects presented here, sorted in ascending order of spectral type (T0.0$-$T4.5). All spectra are normallized at 1.28$\mu$m and offset vertically by increments of one flux unit. \label{spectra4}}
\end{figure*}

Results of the observations are presented in Table \ref{types}. For each object we present the full name, the short ID that will be used in the rest of the paper (see ADJ13 for details), the UKIDSS and SDSS photometry used for candidate selection, and the spectral type derived (see Section \ref{spec_types}). The spectra of our sample are presented in Fig. \ref{spectra0} - \ref{spectra4}, sorted in descending order of spectral type (from early to late). Additional SDSS and WISE photometry can be found in Appendix \ref{app_phot}. 

We note that the red-optical portion of the spectra (i.e. at wavelengths shorter than 1 $\mu$m) tend to be noisier than the infrared portion. In some objects in particular (e.g. BRLT236 and BRLT285) there appear to be strange narrow and broad features, that are due to imperfect background subtraction and/or bad pixels filtering.

\subsection{Spectral classification \label{spec_types}}
Spectral types for the targets were determined via $\chi^2$ fitting with standard templates. The template spectra were taken from the SpeX-Prism online library\footnote{http://pono.ucsd.edu/$\sim$adam/browndwarfs/spexprism}. Each of the targets was smoothed down to the resolution of the templates (R=120), and we excluded the noisy telluric bands when computing the statistic. We visually inspected the three best fit templates to check the accuracy of the fit and to identify possible peculiar objects (see Section \ref{pec}). The spectral types obtained are listed in the second from last column of Table \ref{types}. The uncertainty on the spectral types was determined from the width of the $\chi^2$ distribution.

Unsurprisingly however, a number of objects in the sample did not provide good fits when compared to the standard templates. We discuss in the following sections how we identified the peculiar objects and how we assigned their spectral types.

\onecolumn
\begin{landscape}
\begin{longtable}{l l c c c c c c c c}
\hline
Name & ID & Y & J & H & K & $i$ & $z$ & Spectral type & Ref.\\
\hline
\endfirsthead
\multicolumn{10}{c}{Continued from the previous page.} \\
\hline
Name & ID & Y & J & H & K & $i$ & $z$ & Spectral type & Ref.\\
\hline
\endhead
\hline
\multicolumn{10}{c}{Continued on the next page.} \\
\endfoot
\hline
\caption{The objects observed. YJHK magnitudes are from the UKIDSS LAS DR7, while $iz$ magnitudes are from the SDSS DR7. The spectral types are derived in Section \ref{spec_types}. Discovery reference: (1)~\citet{2013MNRAS.430.1171D}; (2)~\citet{2002ApJ...564..466G}; (3)~\citet{2002AJ....123.3409H}; (4)~\citet{2004AJ....127.3553K}; (5)~\citet{2013MNRAS.433..457B}; (6)~\citet{2010MNRAS.406.1885B}; (7)~\citet{2000ApJ...536L..35L}; (8)~\citet{2006AJ....131.2722C}; (9)~\citet{2008MNRAS.390..304P}; (10)~\citet{2000AJ....119..928F}; (11)~\citet{2014MNRAS.439..372M}. If no discovery reference is listed, the object was previously unpublished. \label{types}}
\endlastfoot
ULAS~J000613.24+154020.7   & BRLT1   & 18.954$\pm$0.079 & 17.876$\pm$0.052 & 16.713$\pm$0.037 & 16.143$\pm$0.036 & 24.161$\pm$0.694 & 20.609$\pm$0.177 & L9.0$\pm$0.5 & 1  \\			    
ULAS~J001040.57+010013.1   & BRLT2   & 19.361$\pm$0.084 & 18.089$\pm$0.038 & 17.457$\pm$0.067 & 16.710$\pm$0.059 & 22.215$\pm$0.198 & 20.797$\pm$0.228 & L1.0$\pm$1.0 & 1  \\			    
ULAS~J001836.51$-$002559.1 & BRLT3   & 18.731$\pm$0.080 & 17.668$\pm$0.052 & 16.608$\pm$0.038 & $\ldots$         & 23.108$\pm$0.287 & 20.274$\pm$0.103 & L9.0$\pm$1.0 & 1  \\			    
ULAS~J002406.37+134705.3   & BRLT6   & 19.281$\pm$0.088 & 18.023$\pm$0.044 & 17.337$\pm$0.060 & 16.528$\pm$0.044 & 24.677$\pm$0.656 & 24.061$\pm$0.386 & L3.0$\pm$1.0 & 1  \\			    
ULAS~J002707.24+142349.0   & BRLT7   & 18.956$\pm$0.059 & 17.981$\pm$0.041 & 17.373$\pm$0.058 & 16.862$\pm$0.063 & 22.638$\pm$0.264 & 20.410$\pm$0.145 & M8.0$\pm$1.0 & 1  \\			    
ULAS~J002827.56+142349.1   & BRLT8   & 18.947$\pm$0.071 & 17.563$\pm$0.034 & 16.616$\pm$0.026 & 15.846$\pm$0.026 & 22.788$\pm$0.263 & 20.287$\pm$0.122 & L8.5$\pm$0.5 & 1  \\			    
ULAS~J002912.25+145604.9   & BRLT9   & 18.812$\pm$0.077 & 17.559$\pm$0.046 & 16.920$\pm$0.053 & 16.330$\pm$0.049 & 22.103$\pm$0.149 & 19.966$\pm$0.098 & L1.0$\pm$1.0 & 1  \\			    
ULAS~J003259.51+141037.1   & BRLT10  & 17.830$\pm$0.027 & 16.645$\pm$0.016 & 15.689$\pm$0.011 & 15.002$\pm$0.013 & 23.083$\pm$0.413 & 19.421$\pm$0.069 & L9.0$\pm$0.5 & 2  \\
\end{longtable}
\end{landscape}
\twocolumn

\subsection{Identification of unresolved binaries \label{unres_bins}}
One possible source of peculiarity in the spectra of brown dwarfs is binarity. Unresolved binaries are in fact characterized by odd spectra, which are the result of the combination of the two components of the system. This is particularly true in L/T transition pairs, where the two components have comparable brightness but significantly different spectra \citep[e.g.][]{2010ApJ...710.1142B}.

In order to select binary candidates within the sample, we followed the method described by \citet{2010ApJ...710.1142B}, who used a combination of index-index and index-spectral type diagrams to define a number of criteria based on the distribution of known unresolved binaries, designed to minimize the number of false positives. The selection is therefore \emph{not} complete. Objects that match two of the six criteria are called ``weak candidates'' while objects that match three or more criteria are called ``strong candidates''. The indices used are summarized in Table \ref{bin_indices}, while the criteria applied are listed in Table \ref{bin_criteria}.

\begin{figure*}
\includegraphics[width=\textwidth]{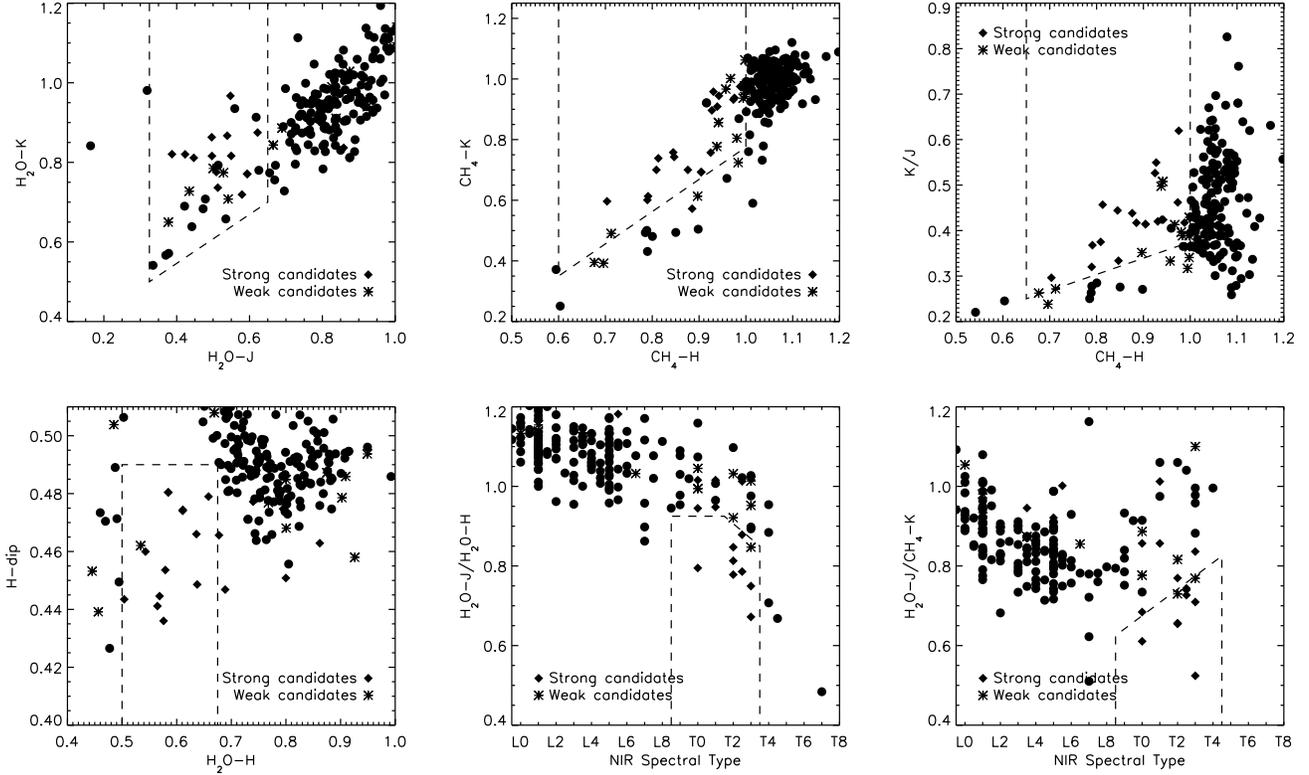}
\caption{The index-index and index-spectral type plots used for binary candidate selection. The dashed lines enclose the selection areas, as defined in Table \ref{bin_criteria}. Weak candidates are marked with stars, while strong candidates are marked with diamonds. \label{bin_selection}}
\end{figure*}

With this technique we were able to identify 27 binary candidates, consisting of 17 weak candidates and 10 strong candidates, which are listed in Table \ref{bin_fit}. The index-index and index-spectral type diagram used are presented in Figure \ref{bin_selection}, where strong candidates are marked with a diamond and weak candidates are marked with an asterisk.

\begin{table}
\centering
\begin{tabular}{l c c c}
\hline
Index & Numerator & Denominator & Feature \\
 & Range & Range & \\
\hline
H$_2$O-\textit{J} & 1.14-1.165  & 1.26-1.285 & 1.15 $\mu$m H$_2$O \\
H$_2$O-\textit{H} & 1.48-1.52   & 1.56-1.60  & 1.4 $\mu$m H$_2$O \\
H$_2$O-\textit{K} & 1.975-1.995 & 2.08-2.10  & 1.9 $\mu$m H$_2$O \\
CH$_4$-\textit{J} & 1.315-1.34  & 1.26-1.285 & 1.32 $\mu$m CH$_4$ \\
CH$_4$-\textit{H} & 1.635-1.675 & 1.56-1.60  & 1.65 $\mu$m CH$_4$ \\
CH$_4$-\textit{K} & 2.215-2.255 & 2.08-2.12  & 2.2 $\mu$m CH$_4$ \\
\textit{K}/\textit{J} & 2.060-2.10  & 1.25-1.29  & \textit{J-K} colour \\
\textit{H}-dip    & 1.61-1.64 & 1.56-1.59 + 1.66-1.69 & 1.65 $\mu$m CH$_4$ \\
\hline
\end{tabular}
\caption{The spectral indices used to identify unresolved binary candidates. All the indices are defined in \citet{2006ApJ...637.1067B} except for \textit{H}-dip which is defined in \citet{2010ApJ...710.1142B}. \label{bin_indices}}
\end{table}

\begin{table}
\centering
\begin{tabular}{l c c}
\hline
Abscissa & Ordinate & Inflection Points \\
 & & (x,y) \\
\hline
H$_2$O-\textit{J} & H$_2$O-\textit{K} & (0.325,0.5),(0.65,0.7) \\
CH$_4$-\textit{H} & CH$_4$O-\textit{K} & (0.6,0.35),(1,0.775) \\
CH$_4$-\textit{H} & \textit{K}/\textit{J} & (0.65,0.25),(1,0.375) \\
H$_2$O-\textit{H} & \textit{H}-dip & (0.5,0.49),(0.875,0.49) \\
SpT & H$_2$O-\textit{J}/H$_2$O-\textit{H} & (L8.5,0.925),(T1.5,0.925),(T3.5,0.85) \\
SpT & H$_2$O-\textit{J}/CH$_4$-\textit{K} & (L8.5,0.625),(T4.5,0.825) \\
\hline
\end{tabular}
\caption{The selection criteria used to identify unresolved binary candidates. Inflection points are defined in \citet{2010ApJ...710.1142B}. \label{bin_criteria}}
\end{table}

\begin{figure*}
\centerline{
\includegraphics[width=0.5\textwidth]{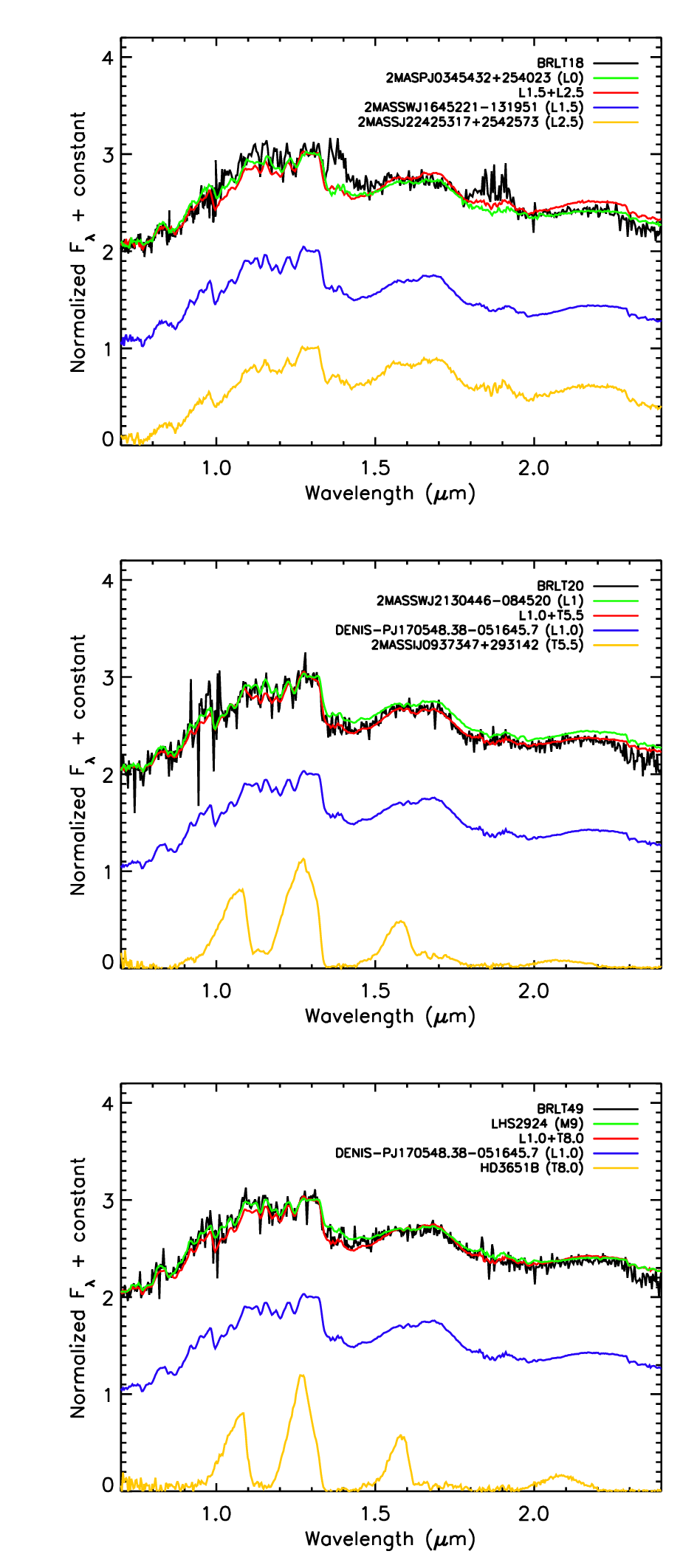}
\includegraphics[width=0.5\textwidth]{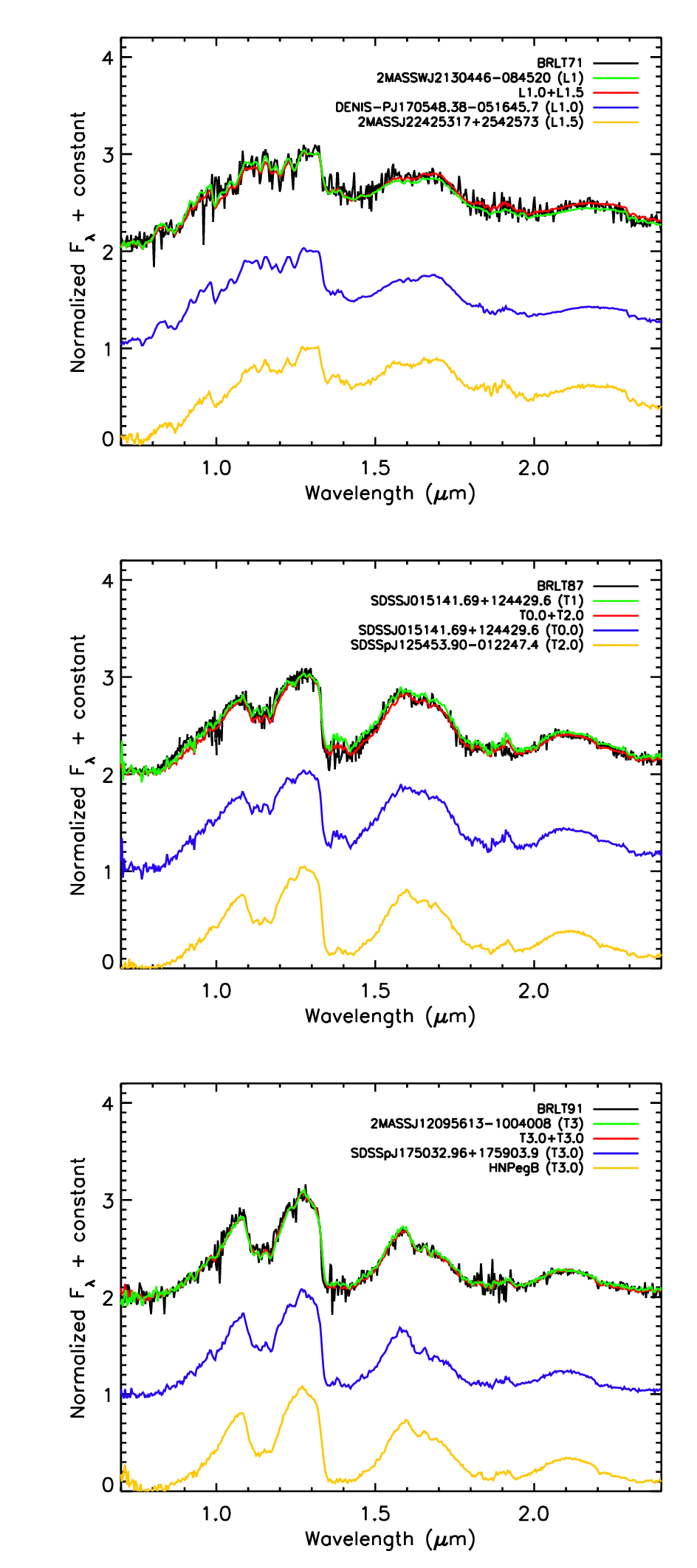}
}
\caption{The spectral deconvolution of the binary candidates. In each panel the target is plotted in black, the best-fit single template in green, the best-fit binary in red, and the two components of the best fit binary in blue and yellow, respectively.\label{bin_fit1}}
\end{figure*}

\begin{figure*}
\centerline{
\includegraphics[width=0.5\textwidth]{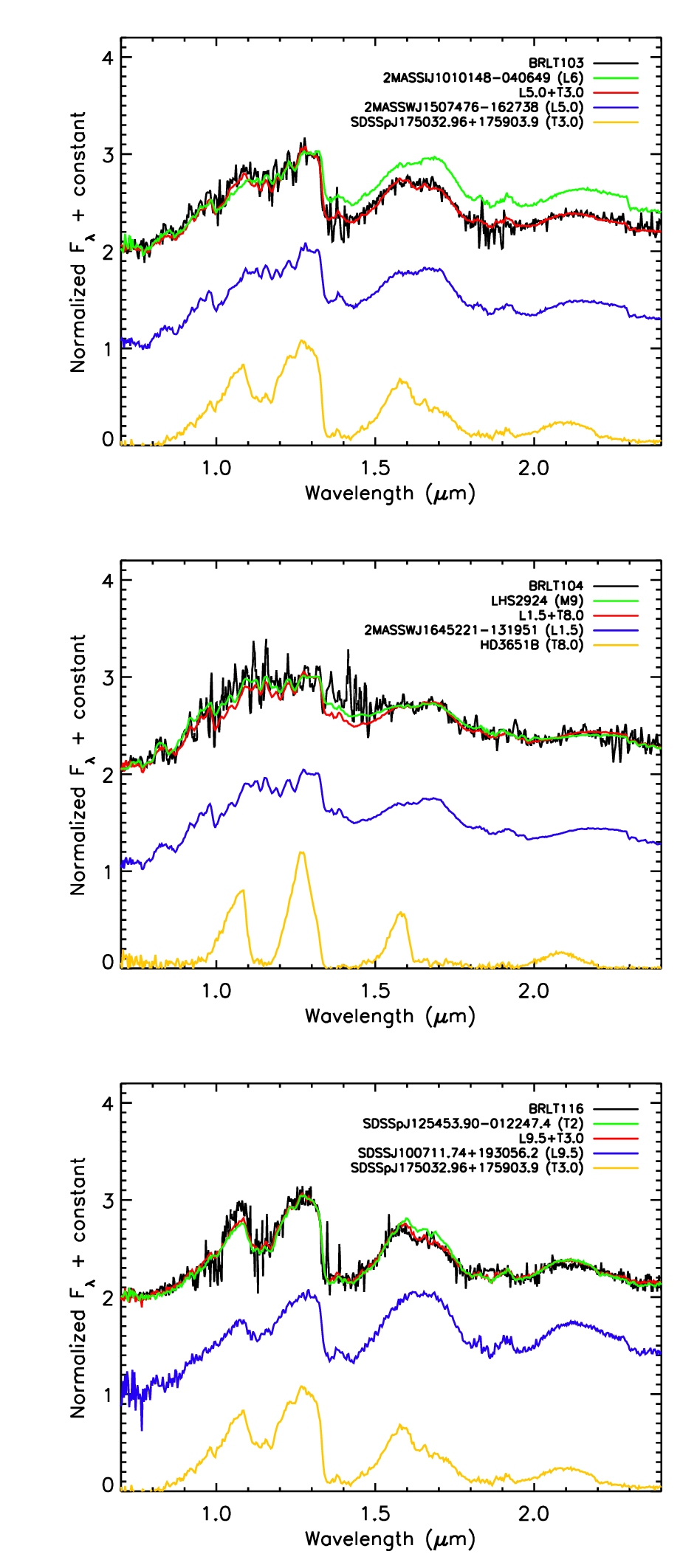}
\includegraphics[width=0.5\textwidth]{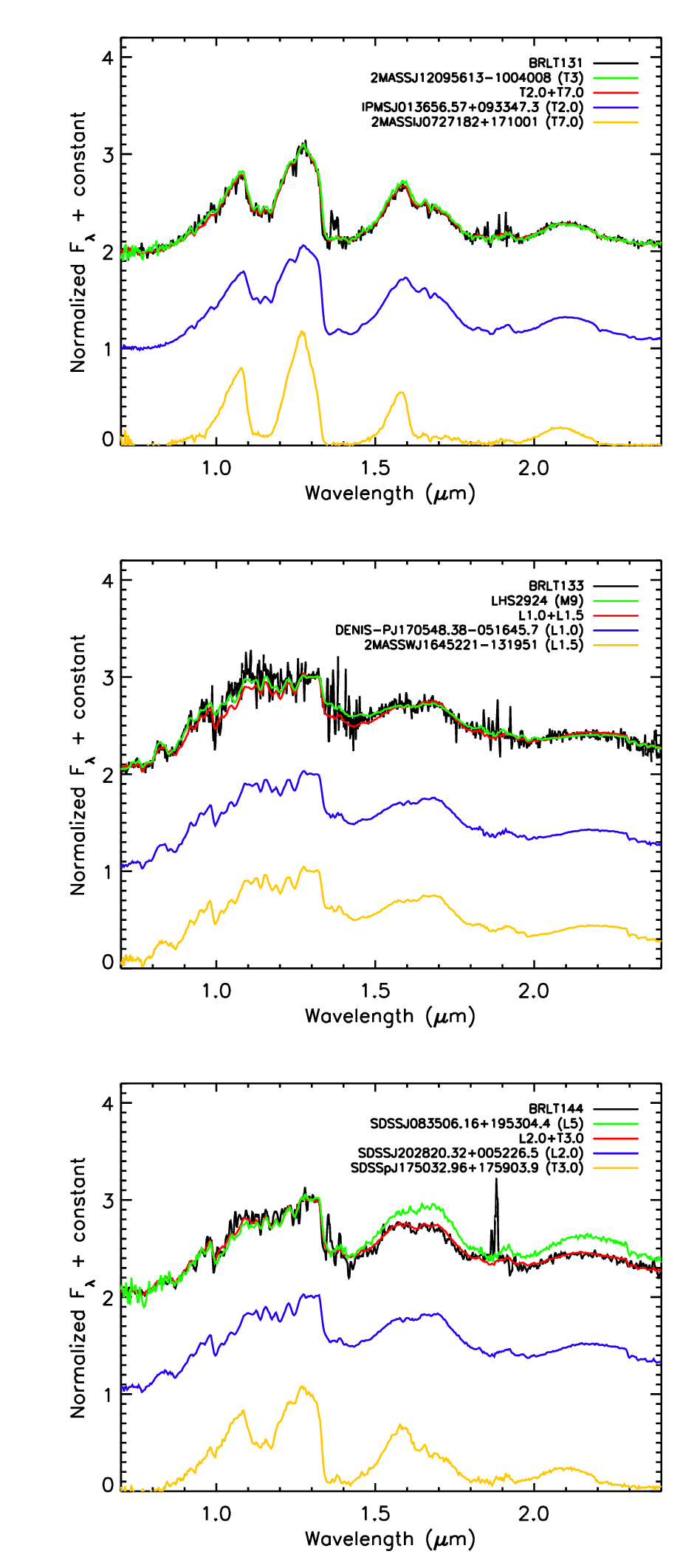}
}
\caption{The spectral deconvolution of the binary candidates. The colour coding is the same as in Figure \ref{bin_fit1}. \label{bin_fit2}}
\end{figure*}

\begin{figure*}
\centerline{
\includegraphics[width=0.5\textwidth]{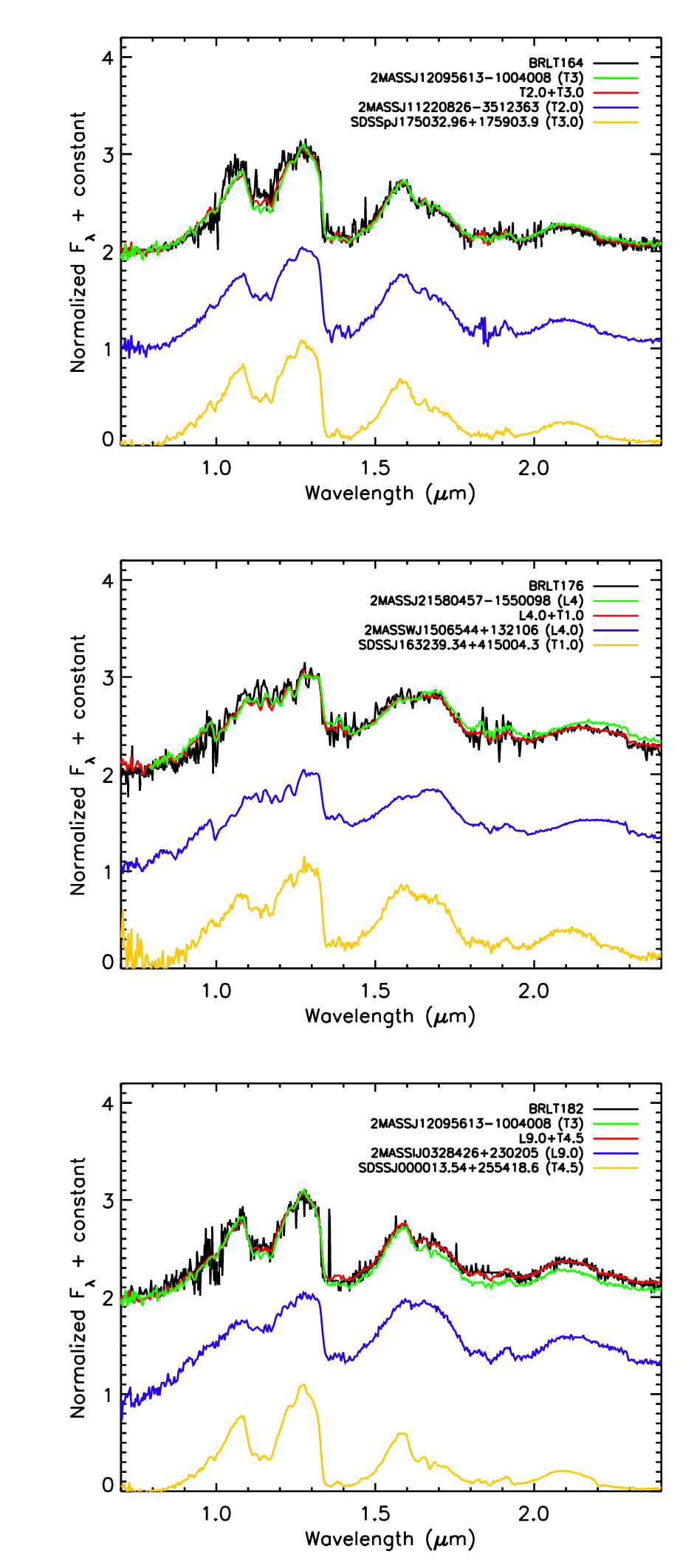}
\includegraphics[width=0.5\textwidth]{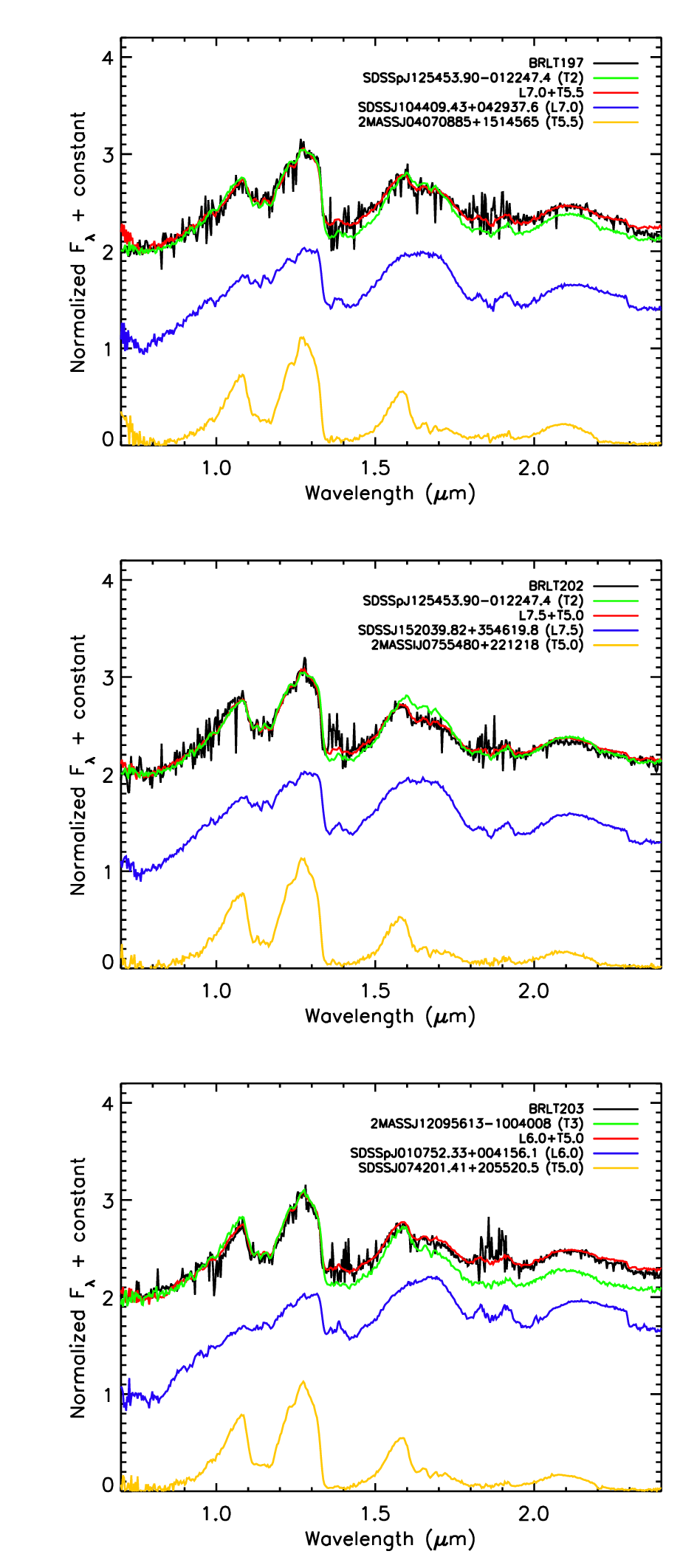}
}
\caption{The spectral deconvolution of the binary candidates. The colour coding is the same as in Figure \ref{bin_fit1}. \label{bin_fit3}}
\end{figure*}

\begin{figure*}
\centerline{
\includegraphics[width=0.5\textwidth]{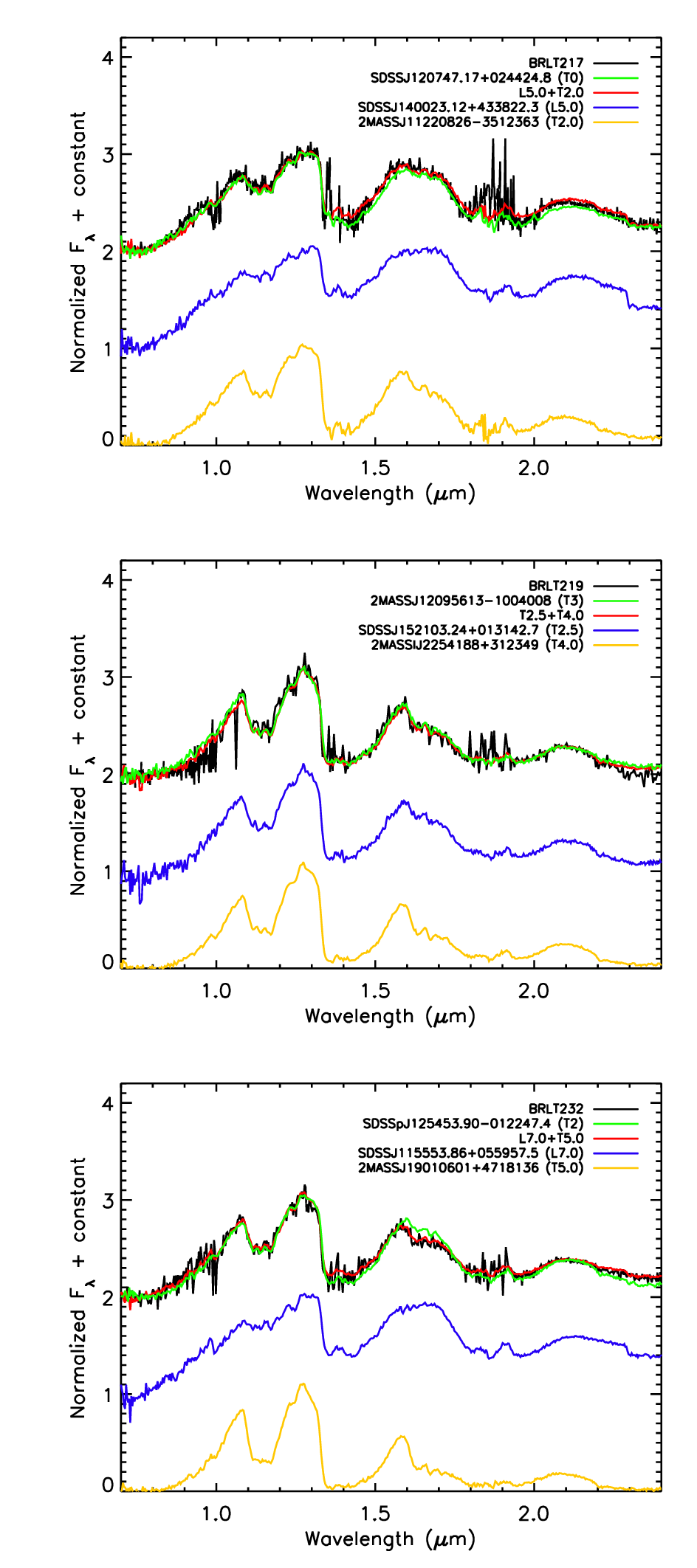}
\includegraphics[width=0.5\textwidth]{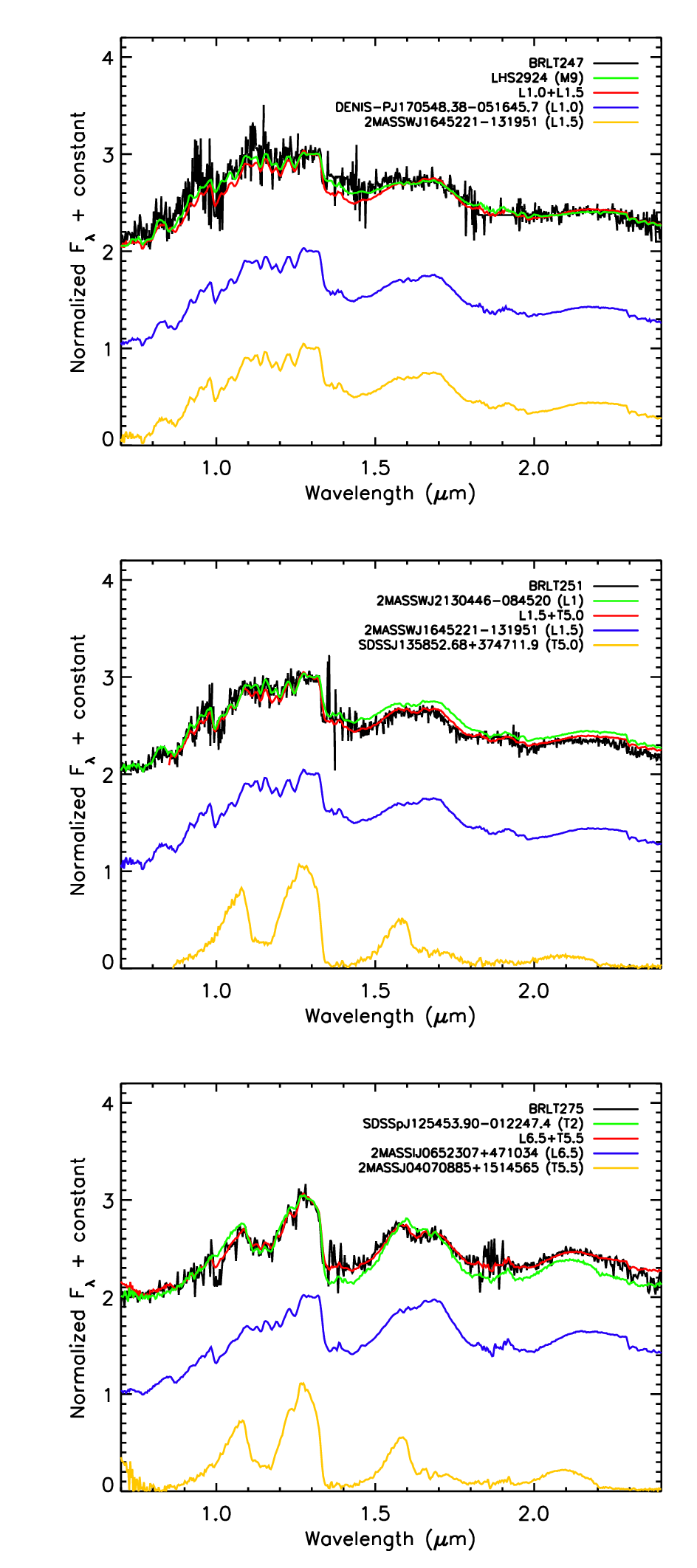}
}
\caption{The spectral deconvolution of the binary candidates. The colour coding is the same as in Figure \ref{bin_fit1}. \label{bin_fit4}}
\end{figure*}

\begin{figure}
\includegraphics[width=0.5\textwidth]{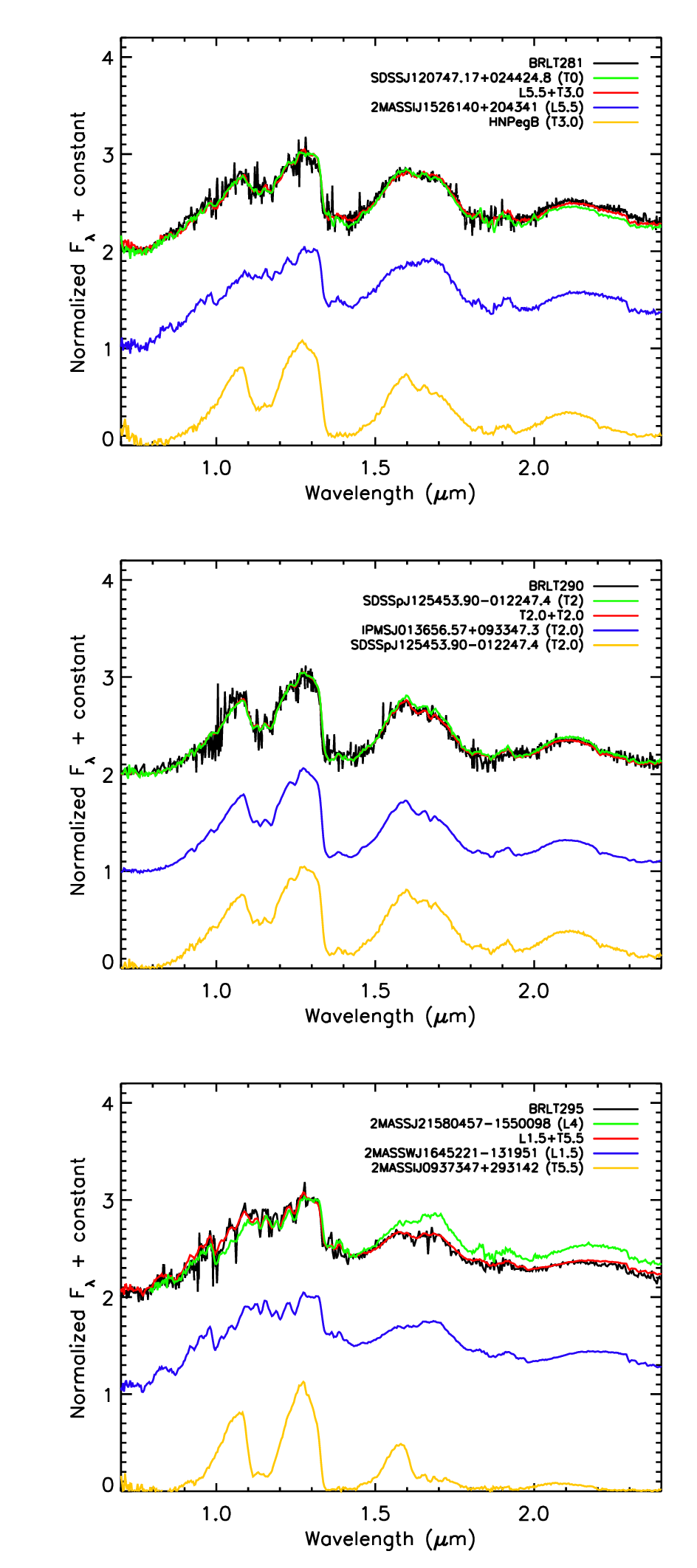}
\caption{The spectral deconvolution of the binary candidates. The colour coding is the same as in Figure \ref{bin_fit1}. \label{bin_fit5}}
\end{figure}

To deconvolve the spectra of the binary candidates and determine the types of the potential components we used the technique described in ADJ13. We created a library of synthetic unresolved binaries combining the spectral templates taken from the already mentioned SpeX-Prism library. All the templates were scaled to a common flux level using the $M_J$-spectral type relation defined in \citet[excluding both known and possible binaries]{2010A&A...524A..38M} and combined. Each candidate was then fitted with this new set of templates using a $\chi^2$ fitting technique, after normalizing both the candidate and the template at 1.28 $\mu$m. The fit are presented in Figure \ref{bin_fit1} $-$ \ref{bin_fit5}. The results of this fit were compared to the results obtained using the standard templates with a one-sided F test, to assess the statistical significance of the deconvolution. If the ratio of the two chi-squared fits ($\eta$) is greater than the critical value ($\eta_{\rm crit} = 1.15$), this represents a $99\%$ significance that the combined template fit is better than the standard template alone. The results are shown in Table \ref{bin_fit} where for each target we present the best fit standard template (with the associated $\chi^2$), the best fit combined template (with $\chi^2$) and the $\eta$ value of the F test. As one can see, 13 out of 27 dwarfs give a statistically ``better fit'' using combined templates ($\eta >$ 1.15) and are therefore the strongest binary candidates. 

Three of these candidates have previously been identified as binaries or binary candidates. BRLT131 was resolved into its two component via HST imaging by \citet{2006ApJS..166..585B}, and their spectral types were estimated to be $<$T2 and T5 based on the resolved photometry. This is in good agreement with the results of our deconvolution, suggesting types T2.0 and T7.0. BRLT275 and BRLT281 were identified as strong binary candidates in \citet{2010ApJ...710.1142B} and the spectral types of their deconvolution were L5.5+T5.0 for BRLT275 and L7.5+T2.5 for BRLT281. Again these results are in good agreement with ours, with the best fit template for BRLT275 being an L6.5+T5.5 and the best fit for BRLT281 being an L5.5+T3.0. BRLT275 was found to be $\sim$ 1 mag over-luminous compared to objects of similar ``unresolved type'' by \citet{2012ApJ...752...56F}, reinforcing the possibility of this object being a real binary. 

For the other candidates, as clearly stated in \citet{2010ApJ...710.1142B}, the results of this fitting must be taken with caution and a definitive confirmation of the binarity of these objects must come from high resolution imaging, radial velocity monitoring or spectro-astrometry.

\begin{table}
\centering
\begin{tabular}{l c c c}
\hline
Target & Single template & Combined template & F-test \\
name & best fit ($\chi^2$) & best fit ($\chi^2$) & $\eta$ \\
\hline
\multicolumn{4}{c}{Strong candidates} \\
\hline
BRLT87  & T1.0 (4.96) & T0.0+T2.0 (3.91) & 1.27 \\
BRLT116 & T2.5 (7.58) & L9.5+T3.0 (6.78) & 1.12 \\
BRLT133 & M9.0 (8.51) & L1.0+L1.5 (10.99) & 0.77 \\
BRLT144 & L5.0 (12.27) & L2.0+T3.0 (11.80) & 1.04 \\
BRLT182 & T3.0 (6.59) & L9.0+T4.5 (5.74) & 1.15 \\
BRLT197 & T2.0 (10.88) & L7.0+T5.5 (6.33) & 1.72 \\
BRLT202 & T2.5 (7.62) & L7.5+T5.0 (5.82) & 1.31 \\
BRLT203 & T3.0 (15.90) & L6.0+T5.0 (5.67) & 2.80 \\
BRLT232 & T2.5 (6.52) & L7.0+T5.0 (4.02) & 1.62 \\
BRLT275 & T2.0 (12.38) & L6.5+T5.5 (6.02) & 2.05 \\
\hline
\multicolumn{4}{c}{Weak candidates} \\
\hline
BRLT18  & L0.0 (37.43) & L1.5+L2.5 (42.29) & 0.88 \\
BRLT20  & L1.0 (12.05) & L1.0+T5.5 (9.32) & 1.29 \\
BRLT49  & M9.0 (4.65) & L1.0+T8.0 (6.31) & 0.74 \\
BRLT71  & L1.5 (5.80) & L1.0+L1.5 (5.82) & 0.99 \\
BRLT91  & T3.0 (3.71) & T3.0+T4.0 (3.41) & 1.09 \\
BRLT103 & L5.5 (8.66) & L5.0+T3.0 (5.97) & 1.45 \\
BRLT104 & M9.0 (26.58) & L1.5+T8.0 (32.18) & 0.83 \\
BRLT131 & T3.0 (2.95) & T2.0+T7.0 (2.25) & 1.31 \\
BRLT164 & T3.0 (7.23) & T2.0+T3.0 (6.17) & 1.17 \\
BRLT176 & L4.0 (7.15) & L4.0+T1.0 (6.76) & 1.06 \\
BRLT217 & T0.0 (11.81) & L5.0+T2.0 (10.60) & 1.11 \\
BRLT219 & T3.0 (9.51) & T2.5+T4.0 (8.26) & 1.15 \\
BRLT247 & M9.0 (12.75) & L1.0+L1.5 (17.95) & 0.71 \\
BRLT251 & L1.0 (9.03) & L1.5+T5.0 (6.41) & 1.41 \\
BRLT281 & T0.0 (5.03) & L5.5+T3.0 (3.77) & 1.33 \\
BRLT290 & T2.0 (4.77) & T2.0+T3.0 (4.45) & 1.07 \\
BRLT295 & L4.0 (13.23) & L1.5+T5.5 (8.94) & 1.48 \\
\hline
\end{tabular}
\caption{The results of the spectral fitting of the binary candidates with combined templates. If $\eta$ (last column) is greater than 1.15 the deconvolution is significant and the object highly likely to be a unresolved binary. \label{bin_fit}}
\end{table}

\subsection{Identification of peculiar objects \label{pec}}
As discussed in the previous section, one of the most common origins of peculiarities in the spectra of brown dwarfs is unresolved binarity. The other common sources are unusual values of surface gravity and metallicity.

The first attempts to quantify the effect of surface gravity on the spectra of brown dwarfs were conducted by \citet{1999AJ....118.2466M}, \citet{2000AJ....120..447K} and \citet{2003ApJ...593.1074G}. They showed that the absorption lines of \ion{K}{i} at 1.25 $\mu$m and of \ion{Na}{i} at 1.21 $\mu$m are very sensitive to gravity, while the bands of H$_2$O and CO at 1.35 $\mu$m and 2.30 $\mu$m are almost insensitive. In the same years \citet{2001MNRAS.326..695L} found that young objects tend to have ``triangular-shaped'' H band peaks, as opposed to the ``trapezoidal-shaped'' peaks of field dwarfs.

A few years later \citet{2009AJ....137.3345C} defined a gravity based classification scheme for early L dwarfs. A detailed study of the optical spectra of 23 young L dwarfs showed that low-gravity L dwarfs display weak \ion{Na}{i}, \ion{Cs}{i}, \ion{Rb}{i} lines. The prominent \ion{K}{i} doublet at 7665,7699 \AA{} has both weak line cores and weak pressure-broadened wings. The molecular bands of FeH and TiO are also weaker than in field L dwarfs while, at early types, VO is stronger. Using a set of 12 indices measuring the strength of the features described above, \citet{2009AJ....137.3345C} defined three gravity classes, labeled using Greek suffix notations. An $\alpha$ suffix denotes normal-gravity objects, $\beta$ indicates moderately low gravity, while $\gamma$ is used for very low-gravity objects.

More recently \citet{2013ApJ...772...79A} proposed an alternative classification using near-infrared spectra. In this fundamental work the authors analysed a sample of 73 M and L dwarfs, comparing in particular ``old'' field dwarfs with members of young moving group of different ages. By measuring the strength of the prominent absorption features in the near-infrared, using both spectral indices and direct equivalent width measurements, the authors confirmed that the H$_2$O bands are gravity-insensitive, and therefore used the ``water-based'' indices to define the spectral typing scheme. The gravity classification scheme is instead based on the spectral indices and the equivalent widths of the gravity-sensitive features, specifically the \ion{K}{i} and \ion{Na}{i} lines (weaker in low-gravity objects), the FeH (weaker) and VO bands (stronger), and the ``peakiness'' of the H band (i.e. quantifying the effect first seen by \citealt{2001MNRAS.326..695L}). Based on the combination of these indicators, M and L dwarfs are divided in three categories: \textsf{FLD-G} indicates normal field dwarfs (corresponding to $\alpha$ from \citealt{2009AJ....137.3345C}), \textsf{INT-G} labels intermediate gravity (like $\beta$ in \citealt{2009AJ....137.3345C}), while \textsf{VL-G} stands for low gravity (analogue to $\gamma$ in \citealt{2009AJ....137.3345C}). \citet{2013ApJ...772...79A} attempted to establish a rough correspondence between their classification and the ages of the dwarfs studied, indicating that \textsf{INT-G} objects appear to be $\sim$50-200 Myr old, while \textsf{VL-G} objects should be $\sim$10-30 Myr old.

To determine how the metallicity affects the spectral characteristics, although the theory is of great help, it is necessary to observe reference objects. Since metal poor objects must have formed early in life in the galaxy, they are members of the halo or thick disk and, in general, have higher proper motions than solar metallicity objects. The most effective way to discover them is therefore the kinematic study of large portions of sky. In \citet{2013MNRAS.434.1005Z} the authors used the SDSS DR8, scanning 9274 deg$^2$ of sky. By studying the large sample of late-M and early-L sub-dwarfs found, they conclude that sub-stellar sub-dwarfs tend to be brighter than their solar-metallicity counterparts of similar spectral type, especially in the optical bands. 

\citet{2010ApJS..190..100K} used multi-epoch 2MASS data covering 4030 deg$^2$ to look for high proper motion candidates. Among the various findings, they identified 15 late-M and L sub-dwarfs. All of these ultra-cool sub-dwarfs show stronger hydride bands (CaH, FeH, and CrH) compared to solar-metallicity objects, a result of the reduced opacity from oxides (e.g. VO and TiO). Counter-intuitively, metal-poor dwarfs show stronger alkali (\ion{Na}{i}, \ion{K}{i}, \ion{Cs}{i}, and \ion{Rb}{i}) and metal lines (in particular \ion{Ti}{i} and \ion{Ca}{i}), a consequence of a reduced condensate formation in those metal-deficient atmospheres. Another clear distinction is in the strength of the CIA of H$_2$. This particular phenomenon is very sensitive to metallicity, and is particularly strong in metal-poor dwarfs, resulting in bluer J$-$H and J$-$K colours and spectra for the sub-dwarfs compared to normal dwarfs. However the CIA of H$_2$ is also very sensitive to surface gravity, and older objects are more compact than field objects.

One way to disentangle the effects of surface gravity and metallicity is by studying binaries \citep[e.g.][]{2008MNRAS.388..838D,2011MNRAS.410..705D,2009MNRAS.395.1237B,2010AJ....139..176F,2010MNRAS.404.1817Z}. When a brown dwarf is found in a binary system with a brighter star, the study of the primary can provide valuable information. Depending on the type of the primary, one can put precise limits on age and metallicity of the system, thus identify the spectral signatures of these quantities in the spectrum of the dwarf.

One of the most famous binaries is probably the T7.5 HD~3651B, companion of a K0 star, discovered by \citet{2007ApJ...660.1507L}. What is particularly interesting is the comparison between HD~3651B and Gl~570D, a T7.5 which is part of another binary system \citep{2000ApJ...531L..57B}. The two dwarfs have very similar temperatures ($\sim$ 800 K), but quite different ages: Gl~570D is relatively young ($\sim$ 2 Gyr) while HD~3651B is relatively old ($\sim$ 6 Gyr). In addition, an estimate of the mass of the two \citep[based on the theoretical models of][]{1997ApJ...491..856B} led to the conclusion that HD~3651B is more massive. From all these considerations it follows that the first has a surface gravity greater than the second (log~$g$ = 5.35 against 5.0). As mentioned earlier it was expected a lower strength of the peak at 2.18$\mu$m in HD~3651B. \citet{2007ApJ...660.1507L} observed instead the opposite effect. What acts against gravity is metallicity. HD~3651B has a higher metallicity ([Fe/H] = 0.13 against 0.06) and this causes a decrease in the photospheric pressure \citep{2006ApJ...640.1063B} and suppress the CIA.

These first observations were followed by others \citep{2008MNRAS.390..304P,2009ApJ...695.1517L,2012MNRAS.422.1922P} which essentially confirmed the strong dependence of the CIA of H$_2$ on metallicity, and indicate that also the absorption of CO at 4.5$\mu$m is influenced, but in an opposite way.

Metallicity and gravity, therefore, have a similar effect on the infrared spectra of brown dwarfs and thus tend to ``hide'' each other. This makes the study of these parameters in isolated objects extremely complex.

Assuming that all the unresolved binaries in the sample have been successfully identified in Section \ref{unres_bins}, we now analyse the SEDs of the remaining objects to identify peculiar dwarfs. 

\subsubsection{Unusually blue L dwarfs}
A number of objects in our sample show unusually blue infrared colours, but do not present any clear sign of metal depletion. Hence they cannot be classified as sub-dwarfs. In particular, they do not present significant enhancement of the alkali absorption lines, while they still show significant suppression of the H and K band flux, and in some cases strong FeH and CrH absorption bands. 

Previous studies of the kinematics of such peculiar objects \citep[e.g. ][]{2009AJ....137....1F, 2010ApJS..190..100K} have pointed out that blue L dwarfs could be part of an older population compared to ``normal'' L dwarfs, but not as old as the halo population. The metal abundances of these peculiar objects would then be reduced, but not enough to be labelled as sub-dwarfs.

Another possible origin for the peculiarity of these brown dwarfs is a variation in the size and location of the dust grains in their atmosphere. Peculiarities in the dust content and the dust property can influence heavily the near-infrared spectra and photometry of L dwarfs \citep[e.g.][]{2014MNRAS.439..372M}.

A problem that arises immediately is how to classify these targets, as their spectra diverge significantly from those of standard objects. We adopted an hybrid way of classifying the blue L dwarfs in the sample. We fit the spectra of the targets with the standard templates, but instead of normalizing both the target and the template at a chosen point, we cut the spectra in three parts, roughly corresponding to the optical + J band, H band, and K band, and then separately normalize and fit these three parts. The final spectral type is given by the template that fits best the three separate portions.

\begin{figure}
\includegraphics[trim=1.7cm 0cm 0cm 0cm, clip=true, width=0.5\textwidth]{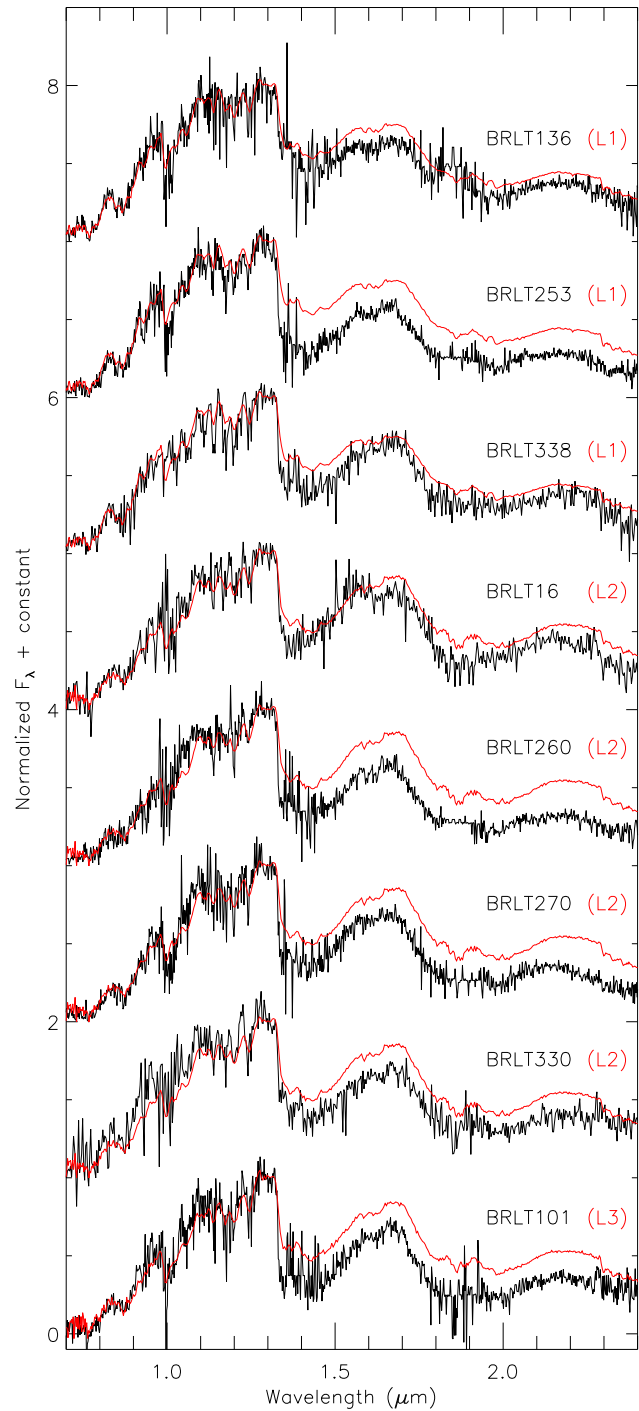}
\caption{The spectra of the peculiar blue L dwarfs. Overplotted in red we show the best fit template for each target. \label{blueL_1}}
\end{figure}

\begin{figure}
\includegraphics[trim=1.7cm 0cm 0cm 0cm, clip=true, width=0.5\textwidth]{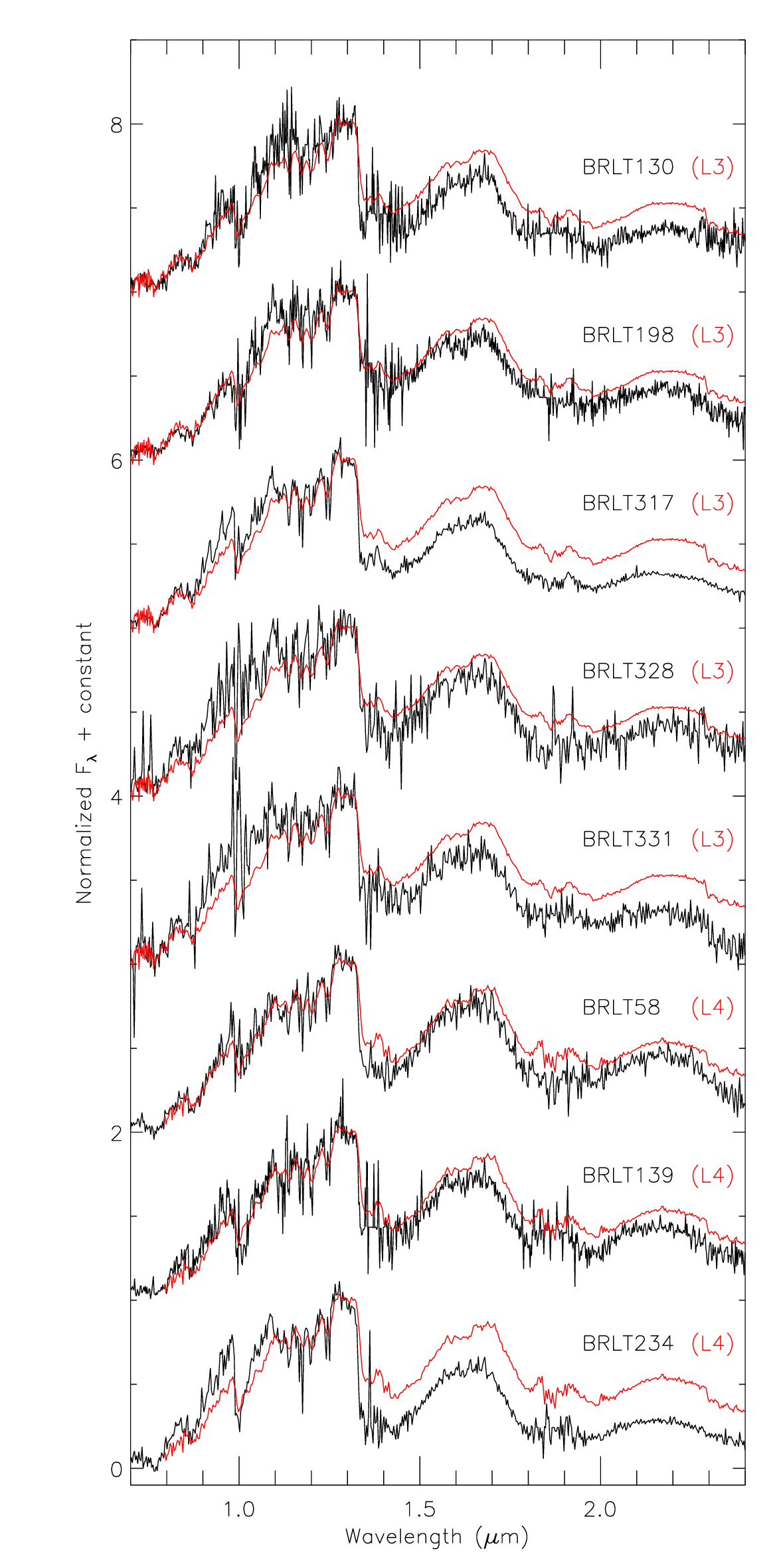}
\caption{The spectra of the peculiar blue L dwarfs. Overplotted in red we show the best fit template for each target. \label{blueL_2}}
\end{figure}

\begin{figure}
\includegraphics[trim=1.7cm 0cm 0cm 0cm, clip=true, width=0.5\textwidth]{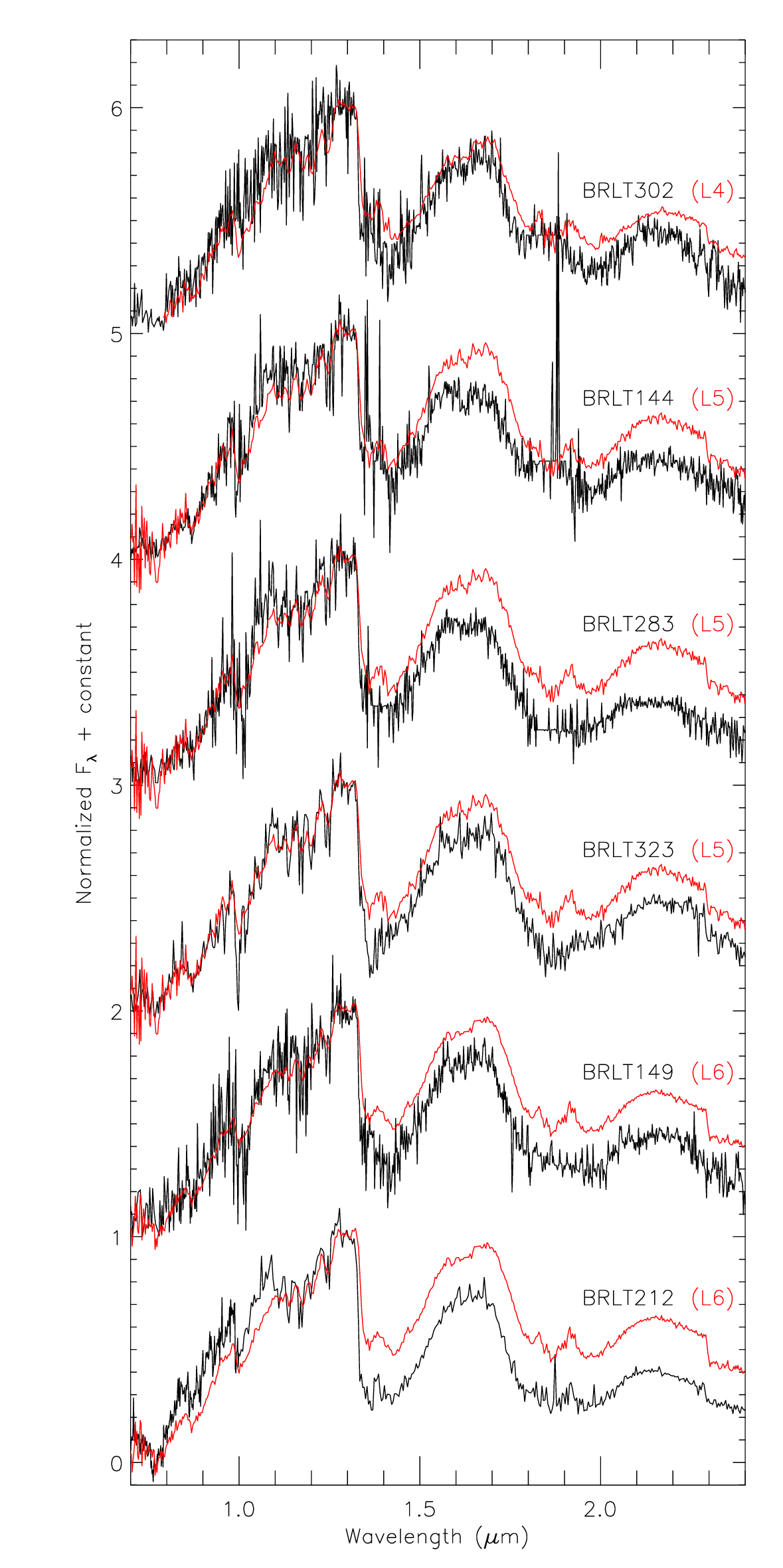}
\caption{The spectra of the peculiar blue L dwarfs. Overplotted in red we show the best fit template for each target. \label{blueL_3}}
\end{figure}

The spectra of the blue L dwarfs identified here are presented in Figures \ref{blueL_1} - \ref{blueL_3}. For each object we overplot in red the best fit standard template. The targets generally present suppressed H and K band fluxes, and enhanced J bands. The H and K band suppression can be an indication of an enhancement of the CIA of H$_2$, which is the proxy of metal depletion or high surface gravity, and this would be in agreement with the hypothesis of \citet{2009AJ....137....1F}, suggesting the membership of peculiar L dwarfs to a slightly older population.

Another common feature in all the blue L dwarfs is the presence of very strong H$_2$O absorption bands. When looking at Figure \ref{indices_plot_1}, it is evident how blue L dwarfs tend to lie below the ``main sequence'' in two of the three plots on the left, with the H$_2$O-H and the H$_2$O-K indices typical of objects of later spectral type.  This could be the effect of a reduced dust content in these metal poor atmospheres, that makes water the main source of opacity.

It must be noted at this point that an alternative explanation for unusually blue L dwarfs is unresolved binarity. The presence of a close T type companion would produce a similar effect. However, only one of the new blue L dwarfs matches the selection criteria for binaries (BRLT16), and its fit with unresolved binary templates is not significantly better than the one with a single template (see Section \ref{unres_bins}). We therefore conclude that our sample of blue L dwarfs is entirely made of intrinsically blue objects.

\subsubsection{Blue T dwarfs}
In the same way as for the blue L dwarfs, we identified 2 peculiar T dwarfs which show H and K band suppression. 

\begin{figure}
\includegraphics[width=0.5\textwidth]{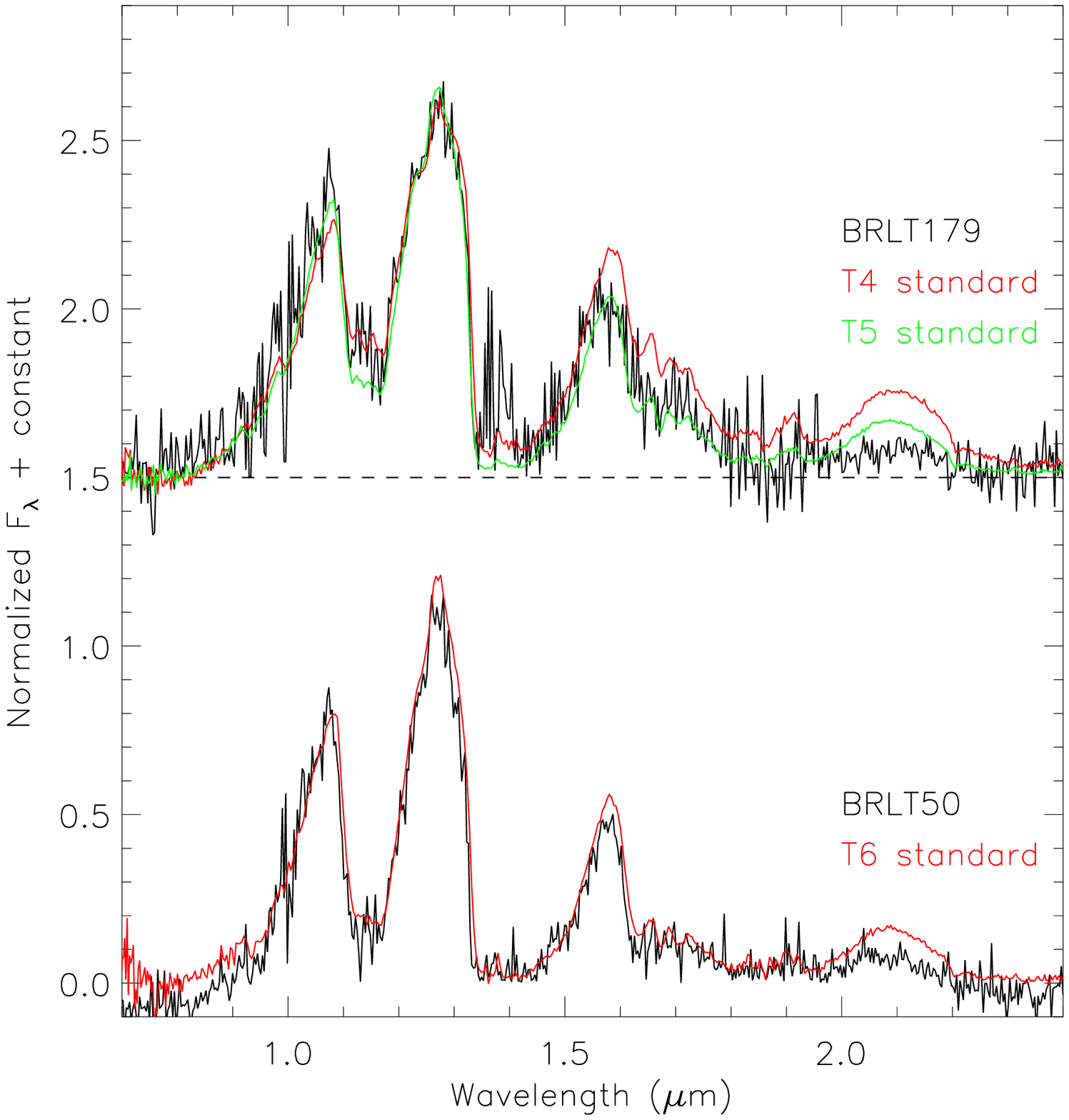}
\caption{The spectra of the peculiar blue T dwarfs. Overplotted in red and green we show the best fit template for each target. \label{blueT2}}
\end{figure}

A number of unusually blue T dwarfs have been presented in \citet{2011MNRAS.414..575M}, who selected the peculiar objects based on their MKO photometry. One of the two objects identified here, BRLT179, was indeed part of that sample. The spectra of the two blue T dwarfs in the sample are presented in Figure \ref{blueT2}. Both of them show a very suppressed K band flux, which is indicative of an enhanced CIA. Whether this enhancement is due to low metallicity or to a higher surface gravity is still a matter of debate \citep[see for instance][]{2011MNRAS.414..575M}. A way to distinguish between the two cases is the analysis of the kinematics of the brown dwarfs, as thick disk or halo-like space velocities would be suggestive of a metal-poor nature, while in the case of a thin disk-like space motion high gravity would be the preferred explanation. 

\textit{BRLT50}: the general shape of the spectrum of this object is well fitted by the T6 standard SDSSp~J162414.37+002915.6. However, the peak of the J and H band are slightly lower in the target, and the K band is clearly suppressed, all hints to metal depletion. The kinematics can generally offer insights into the interpretation of the nature of peculiar objects like this one, but with no measured proper motion, we cannot address the possibility of this object belonging to a older disk population.

\textit{BRLT179}: we assigned a spectral type of T4.5 to this object as the T5 standard reproduces quite well the general shape of the SED in the 0.7$-$1.8 $\mu$m range, except for the depth of the H$_2$O absorption at 1.15 and 1.35 $\mu$m. These features are much better fitted by the T4 standard. The flux level in the K band is extremely suppressed, with almost no flux left. The assigned spectral type is 1 subtype later than the one given in \citet{2010MNRAS.406.1885B}, but that is based on a 1.05$-$1.35 $\mu$m spectrum only. The kinematics analysis of BRLT179 performed by \citet{2011MNRAS.414..575M} suggests a young disk nature for this object, which is somewhat surprising as BRLT179 is the second bluest T dwarf known (J$-$K = -1.2 $\pm$ 0.1), and its K band spectrum is strongly suppressed. This apparent inconsistency is in common with the bluest T dwarf known, SDSS J1416+1348B \citep{2010MNRAS.404.1952B, 2010A&A...510L...8S} which has young disk kinematics as well. 

\subsubsection{Low gravity objects}
While unusually blue infrared colours are generally tracers of reduced metallicity or high surface gravity, unusually red spectra are the product of an increased metal content or a low surface gravity (which is typical of young objects). We refer the reader to \citet{2013ApJ...772...79A} and references therein for a more detailed description of the spectral signatures associated (or believed to be associated) with these two atmospheric parameters, and the classification scheme developed for this type of objects. 

\begin{figure}
\includegraphics[width=0.5\textwidth]{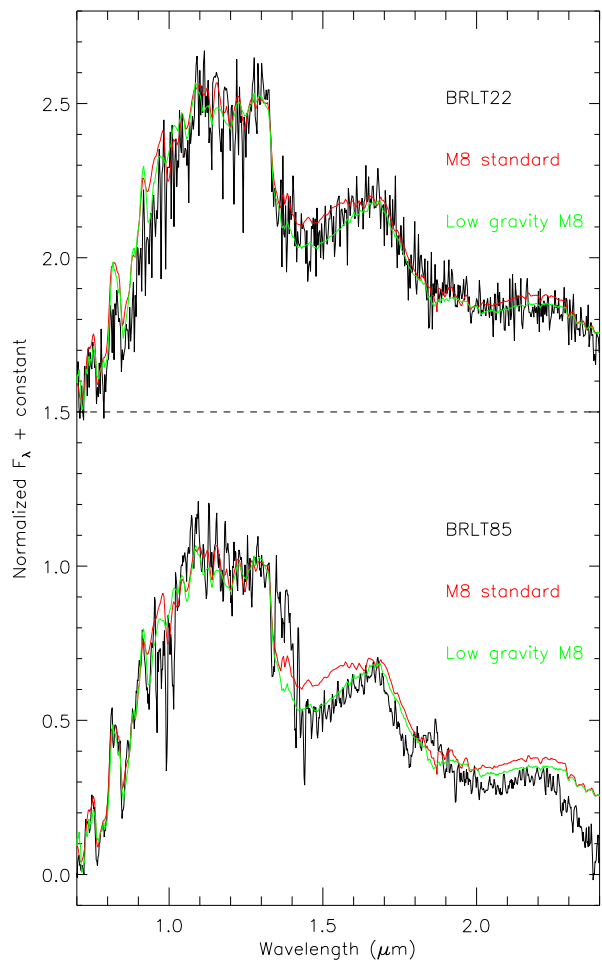}
\caption{The spectra of the low gravity M dwarfs. Overplotted in red and green we show the best fit field standard and the best fit low gravity standard \label{redL2}}
\end{figure}

We identified 2 peculiar low gravity objects within the sample, BRLT22 and BRLT85, and their spectra can be found in Figure \ref{redL2}.

These two late M dwarfs show the peculiar signs of low gravity objects. Specifically they have a somewhat triangular shaped H band, and shallower alkali lines in the J band (in particular in BRLT85). Both objects also show stronger water absorption when compared to the standard template (overplotted in red in Figure \ref{redL2}). In both cases the low gravity M8 template matches better the SED of the target. The gravity classification scheme defined in \citet{2013ApJ...772...79A} gives a classification of \textsf{INT-G} for BRLT22 and \textsf{LOW-G} for BRLT85, further highlighting the peculiar nature of these two targets. A definitive confirmation has to come from the kinematics, possibly associating the targets to known young moving groups in the solar neighbourhood. 

\section{Spectral indices and equivalent widths}
\label{ind_eqw}
\begin{figure*}
\includegraphics[width=\textwidth]{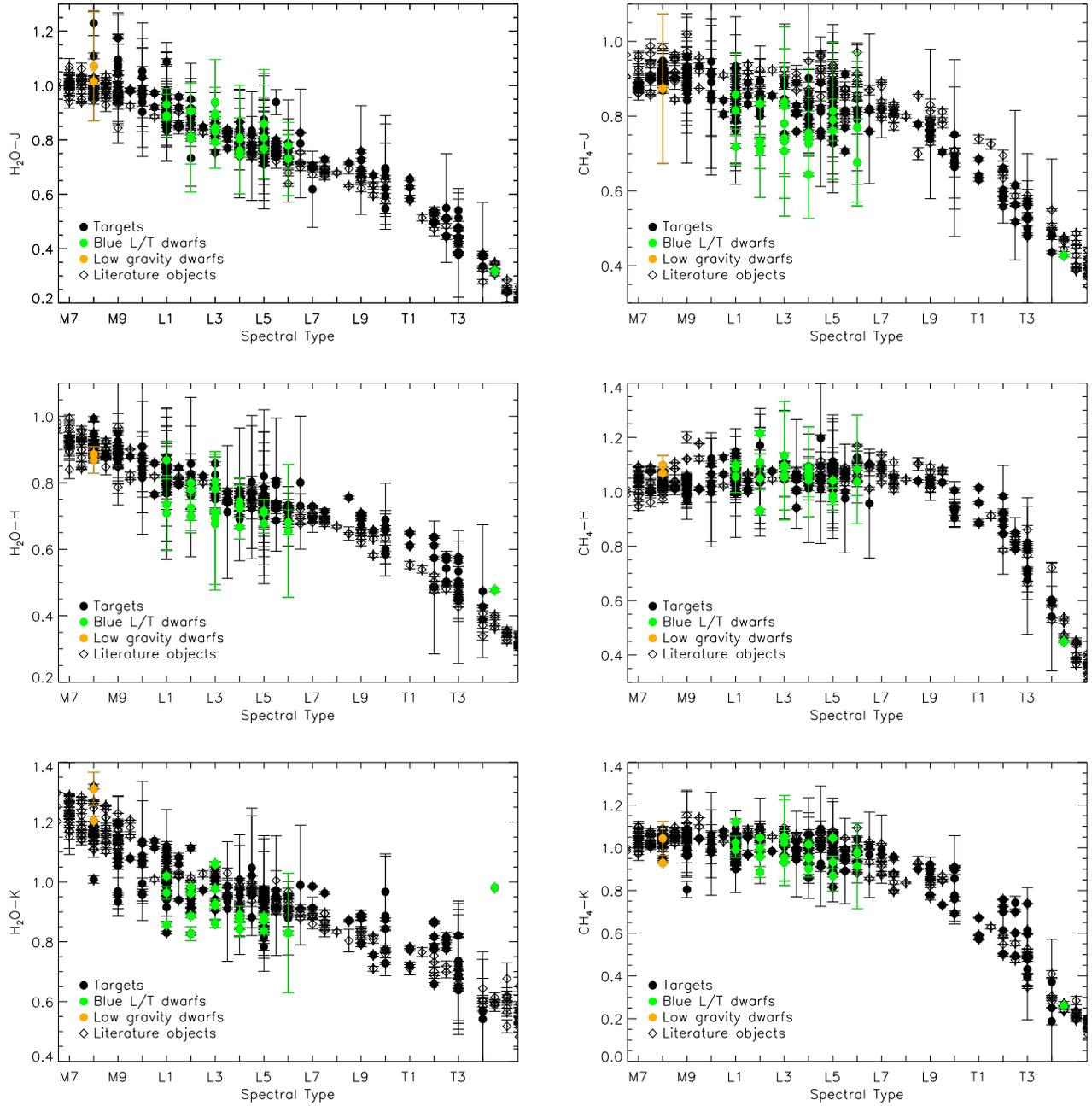}
\caption{The spectral indices as a function of spectral type. Peculiar objects are plotted in colours. The spectral indices calculated for a series of known L and T dwarfs from the literature are overplotted for reference. \label{indices_plot_1}}
\end{figure*}

\begin{figure*}
\includegraphics[width=\textwidth]{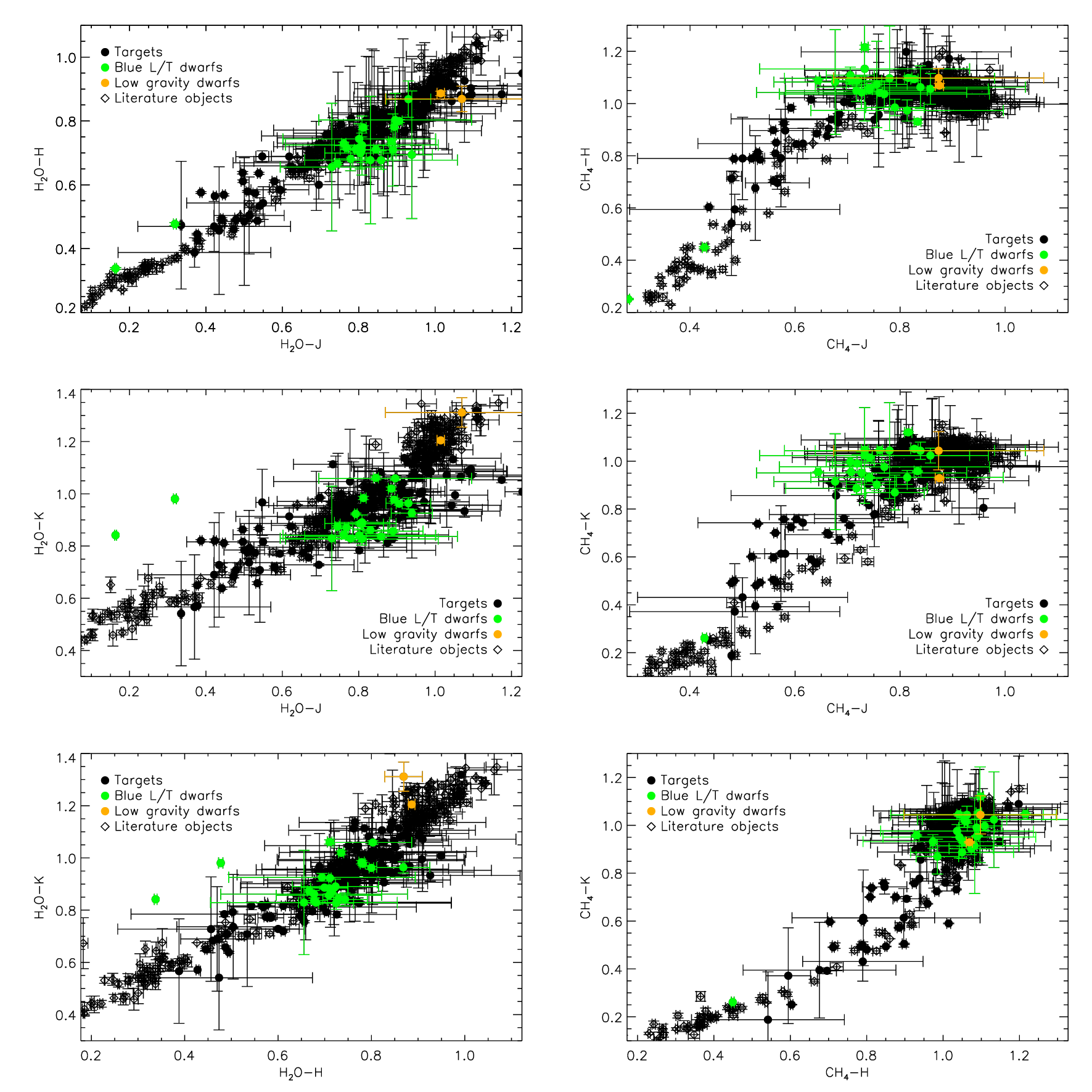}
\caption{Index-index plots. Peculiar objects are plotted in colours. The spectral indices calculated for a series of known L and T dwarfs from the literature are overplottedfor reference. The ``main sequence'' is clearly visible from the top-right to the bottom-left corner of each plot. \label{indices_plot_2}}
\end{figure*}

A way to quantify the evolution of spectral features across the spectral sequence is to use spectral indices to measure their strength. The spectral indices calculated for the targets are presented in Table \ref{indices}, and plotted in Figure \ref{indices_plot_1} and \ref{indices_plot_2}. The peculiar objects identified in the previous section are plotted in colour. 

In Figure \ref{indices_plot_1} one can see how the indices measuring the \emph{relative} strength of the water absorption bands (the three plots on the left hand side) correlate very well with spectral types. Blue L dwarfs tend to have stronger water absorption bands and their indices therefore are typical of later type objects (as late as T0$-$T1 in some cases), lying below the ``main sequence''. A purely index-based classification for these objects could therefore lead to systematically later types. 

The right hand side of Figure \ref{indices_plot_1} shows the indices measuring the \emph{relative} depth of the methane absorption bands. Not surprisingly, the correlation between those indices and spectral type is valid only in the T dwarf range, as there is very little methane absorption in L dwarfs, except at mid-infrared wavelengths. 

In Figure \ref{indices_plot_2} we present a series of index-index plots. It is easy to spot the ``main sequence'', from the late-Ms and early-Ls on the top-right to the mid-Ts in the bottom-left corner of each plot. Once again, the methane indices do not correlate in the L dwarfs regime, with all of the L dwarfs clustered in the 0.8$-$1.0 range for each methane index. When looking at the left hand side of the figure, blue L dwarfs tend to be clustered below the sequence in two of the three plots, further stressing the unusual strength of the $\sim 1.4 \mu$m and the $\sim 1.9 \mu$m water absorption bands, while blue T dwarfs sit above it. In particular the two blue T dwarfs have very high values of the H$_2$O-K index, which is the effect of their extreme flux suppression in the K band. With very little flux left, their K band spectra are almost flat, and their corresponding indices tend to one. 

While these indices give an indication of the evolution of broad molecular absorption bands, to measure the strength of narrow atomic lines we calculated their equivalent width. The main atomic lines in the spectra of brown dwarfs are due to \ion{Na}{i} and \ion{K}{i}. We calculated the equivalent width of the \ion{Na}{i} doublet at 1.139 $\mu$m, and the \ion{K}{i} lines at 1.169, 1.177, 1.244, and 1.253 $\mu$m, as these are the strongest and best detected lines. 

To measure the equivalent width, we fit each doublet and the region of the spectrum around it using a double Gaussian profile. We decided to fit the doublets together since the lines are too close to allow for a separate fit, as one would have to restrict the region to fit too much, leading to a more uncertain determination of the continuum. The continuum is a parameter of the fit, and is assumed to be changing linearly as a function of wavelength. This is to take into account that, especially in late type objects, some of the lines considered do not fall in regions of flat continuum. The centre of the lines is also a parameter of the fit, but the separation between them is fixed and assumed to be equal to the tabulated separation. The equation describing each doublet is therefore:

\begin{equation}
F(\lambda) = F_0 + \lambda \times C + a_1 e^{(\lambda - \lambda_1)^2/2\sigma_1^2} + a_2 e^{(\lambda - (\lambda_1 + \Delta\lambda))^2/2\sigma_2^2}
\end{equation}

where $F_0$ and $C$ are the two parameters describing the continuum, $\lambda_1$ is the centre of the first line in the doublet, $\Delta\lambda$ is the separation between the two lines, $\sigma_1$ and $\sigma_2$ are the width of the two lines, and $a_1$ and $a_2$ are the depth of the two lines, i.e. the minimum flux at the centre of the lines. $F_0, C, \lambda_1, a_1, a_2, \sigma_1$ and $\sigma_2$ are all parameters of the fit. 

\begin{figure*}
\includegraphics[width=\textwidth]{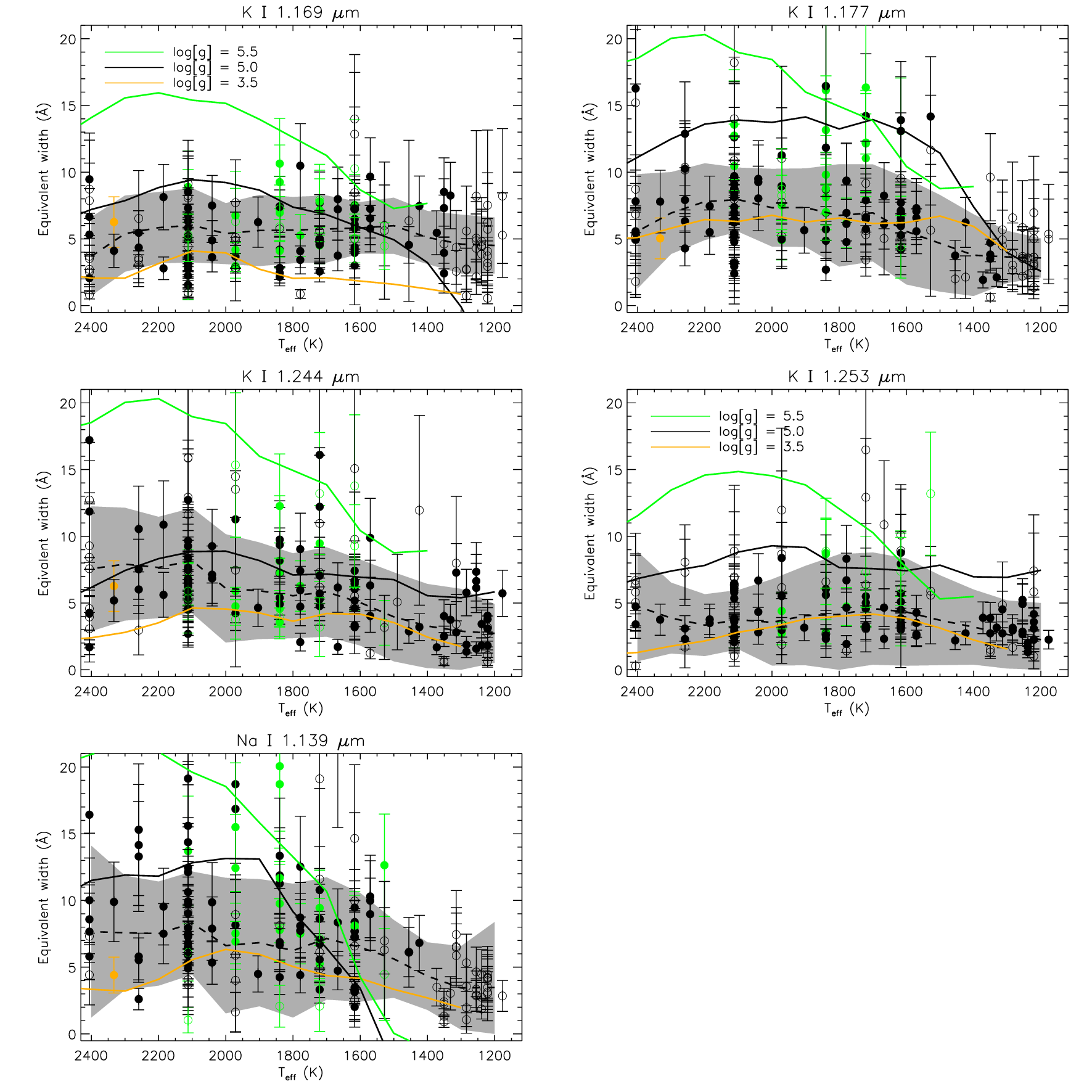}
\caption{The equivalent width of \ion{Na}{i} and \ion{K}{i} lines as a function of spectral type. Measurements with relative errors larger than 0.33 are plotted as open circles. Peculiar objects are labelled following the same colour scheme of Figure \ref{indices_plot_1} and \ref{indices_plot_2}. The median equivalent width as a function of $T_{\rm eff}$ is plotted as a dashed line, while the grey shaded area indicates one standard deviation from the median. Overplotted for comparison are the equivalent width measured from the BT-Settl atmospheric models \citep{2011ASPC..448...91A} for solar metallicity. The yellow line corresponds to a surface gravity log~$g$ = 3.5, the black line to log~$g$ = 5.0 and the green line to log~$g$ = 5.5. \label{eqw_plot}}
\end{figure*}

The equivalent width measured for the targets are presented in Table \ref{eqw_table} and plotted as a function of effective temperature in Figure \ref{eqw_plot}. Since the \ion{Na}{i} doublet at 1.139 $\mu$m is partly blended, the values presented are the total equivalent width of the doublet. The effective temperature of an object was determined from its spectral type using the type-to-temperature conversion presented in \citet{2013AJ....146..161M}. Objects with very low signal to noise, or with dubious detection of the lines have been omitted. Measurements with relative errors larger than 0.33 are plotted as open circles, while those with relative errors better than 0.33 are plotted as filled circles. Overplotted for reference are the equivalent width calculated for the BT-Settl atmospheric models \citep{2011ASPC..448...91A} for solar metallicity, and three different values of surface gravity. The median equivalent width as a function of effective temperature is plotted as a dashed black line, while one standard deviation around it is shown as a grey shaded area. The median is calculated by binning up our targets in 100 K wide bins. The equivalent widths show a large scatter, and there is no clear separation between blue/red L and T dwarfs and the rest of the sample. However, when looking at the median values, our sample appears to be mostly clustered between the log~$g$ = 5.0 and log~$g$ = 3.5 lines. These values are lower than what one would expect for thin disk objects, i.e. with intermediate ages ($\approx 1 - 3$ Gyr). Model isochrones predict values of log~$g$ typically around 5 or slightly above \citep[e.g.][]{2011ASPC..448...91A} for L/T transition dwarfs. The discrepancy could be due to a systematic overestimate of the lines equivalent width in the atmospheric models, possibly due to uncertainties in the measured oscillator strengths in the near-infrared regime \citep[e.g. Table 2,][]{1996MNRAS.280...77J}.

The models suggest that the lines should reach their maximum strength at $T_{\rm eff} \sim 2000$ K, and then slowly get weaker towards lower temperature. Looking at the values from the sample, only the \ion{K}{i} lines at 1.244 and 1.253 $\mu$m follow the expected trend, while the \ion{Na}{i} doublet and the \ion{K}{i} lines at 1.169 and 1.177 $\mu$m remain strong even at temperatures as low as $\sim$1200 K. However the discrepant measurements tend to have very large associated errors. This is because the mentioned lines fall in regions of growing H$_2$O and CH$_4$ absorption, so in late type (i.e. low $T_{\rm eff}$) objects the signal-to-noise ratio in those areas decreases sharply, and the fit to the doublet gets less reliable. This would not be a problem in the atmospheric models, nor for the \ion{K}{i} lines at 1.244 and 1.253 $\mu$m since they fall in a region where water and methane absorption is less prominent, and therefore follow the expected trend.

\onecolumn
\begin{deluxetable}{l c c c c c c c c c}
\tablewidth{0pt}
\tabletypesize{\normalsize}
\tablehead{
\colhead{Name} & \colhead{Spectral type} & \colhead{H$_2$O-J} & \colhead{H$_2$O-H} & \colhead{H$_2$O-K} & \colhead{CH$_4$-J} & \colhead{CH$_4$-H} & \colhead{CH$_4$-K} & \colhead{K/J} & \colhead{H-dip}
}
\tablecaption{Spectral indices for the objects in the sample. For the indices definition, see Table \ref{bin_indices}. \label{indices}}
\startdata
BRLT28  & L6.0 $\pm$ 0.5 & 0.72 & 0.69 & 0.88 & 0.79 & 1.04 & 0.91 & 0.62 & 0.48 \\
BRLT49  & M9.0 $\pm$ 0.5 & 0.99 & 0.87 & 1.09 & 0.86 & 1.00 & 1.06 & 0.36 & 0.49 \\
BRLT63  & L1.0 $\pm$ 0.5 & 0.97 & 0.82 & 1.11 & 0.87 & 1.01 & 0.98 & 0.41 & 0.49 \\
BRLT65  & M9.0 $\pm$ 0.5 & 0.99 & 0.90 & 1.09 & 0.87 & 1.00 & 1.07 & 0.37 & 0.49 \\
BRLT67  & L1.0 $\pm$ 0.5 & 0.97 & 0.81 & 1.01 & 0.88 & 1.06 & 1.05 & 0.39 & 0.48 \\
BRLT68  & L5.0 $\pm$ 0.5 & 0.82 & 0.70 & 0.88 & 0.82 & 1.04 & 0.92 & 0.64 & 0.51 \\
BRLT69  & L1.0 $\pm$ 0.5 & 0.95 & 0.86 & 1.08 & 0.92 & 1.04 & 1.07 & 0.41 & 0.48 \\
BRLT71  & L1.5 $\pm$ 0.5 & 0.96 & 0.80 & 1.06 & 0.89 & 1.00 & 0.97 & 0.43 & 0.49 \\
BRLT72  & M9.0 $\pm$ 0.5 & 1.05 & 0.91 & 1.13 & 0.90 & 1.04 & 1.07 & 0.39 & 0.49 \\
BRLT73  & L1.0 $\pm$ 0.5 & 0.93 & 0.85 & 0.99 & 0.84 & 1.12 & 0.92 & 0.44 & 0.50 \\
BRLT74  & L9.5 $\pm$ 1.0 & 0.67 & 0.66 & 0.76 & 0.70 & 1.03 & 0.73 & 0.50 & 0.53 \\
BRLT75  & M9.0 $\pm$ 1.0 & 1.00 & 0.89 & 1.14 & 0.91 & 1.04 & 1.04 & 0.34 & 0.49 \\
BRLT76  & L5.5 $\pm$ 0.5 & 0.78 & 0.81 & 0.97 & 0.82 & 1.08 & 0.96 & 0.57 & 0.49 \\
BRLT78  & L1.0 $\pm$ 0.5 & 0.98 & 0.84 & 1.11 & 0.80 & 1.07 & 1.04 & 0.35 & 0.51 \\
BRLT81  & M9.0 $\pm$ 0.5 & 1.07 & 0.88 & 1.07 & 0.89 & 1.01 & 0.99 & 0.41 & 0.49 \\
BRLT82  & L1.0 $\pm$ 0.5 & 0.94 & 0.85 & 1.07 & 0.88 & 1.05 & 1.02 & 0.43 & 0.49 \\
BRLT83  & M8.0 $\pm$ 1.0 & 1.11 & 0.99 & 1.32 & 0.94 & 1.02 & 0.94 & 0.37 & 0.49 \\
BRLT84  & L3.5 $\pm$ 0.5 & 0.77 & 0.77 & 0.95 & 0.81 & 1.06 & 1.03 & 0.49 & 0.49 \\
BRLT85  & M8.0 $\pm$ 0.5 & 1.07 & 0.87 & 1.31 & 0.87 & 1.10 & 1.04 & 0.28 & 0.50 \\
BRLT87  & T0.0 $\pm$ 0.5 & 0.59 & 0.58 & 0.77 & 0.66 & 0.90 & 0.69 & 0.41 & 0.48 \\
BRLT88  & L4.0 $\pm$ 1.0 & 0.83 & 0.78 & 1.01 & 0.83 & 1.06 & 1.06 & 0.50 & 0.48 \\
BRLT91  & T3.0 $\pm$ 0.5 & 0.47 & 0.46 & 0.68 & 0.53 & 0.79 & 0.49 & 0.25 & 0.47 \\
BRLT92  & L1.0 $\pm$ 0.5 & 0.86 & 0.83 & 1.00 & 0.82 & 1.08 & 1.02 & 0.39 & 0.48 \\
BRLT97  & L0.0 $\pm$ 1.0 & 0.99 & 0.84 & 1.13 & 0.92 & 1.03 & 0.98 & 0.39 & 0.49 \\
BRLT99  & L5.0 $\pm$ 0.5 & 0.81 & 0.71 & 0.85 & 0.84 & 1.08 & 0.98 & 0.55 & 0.48 \\
BRLT101 & L3.0 $\pm$ 1.0 & 0.83 & 0.68 & 0.86 & 0.73 & 1.13 & 1.02 & 0.34 & 0.49 \\
BRLT102 & L0.0 $\pm$ 0.5 & 0.94 & 0.88 & 1.06 & 0.91 & 1.03 & 1.05 & 0.42 & 0.49 \\
BRLT103 & L5.5 $\pm$ 0.5 & 0.74 & 0.74 & 0.92 & 0.71 & 1.00 & 0.89 & 0.36 & 0.50 \\
BRLT104 & M9.0 $\pm$ 0.5 & 1.08 & 0.93 & 0.93 & 0.96 & 0.98 & 0.80 & 0.40 & 0.46 \\
BRLT105 & L5.0 $\pm$ 0.5 & 0.82 & 0.78 & 0.96 & 0.84 & 1.07 & 0.99 & 0.53 & 0.48 \\
BRLT106 & M9.0 $\pm$ 0.5 & 0.97 & 0.88 & 0.97 & 0.84 & 1.03 & 1.00 & 0.37 & 0.48 \\
BRLT108 & L6.5 $\pm$ 0.5 & 0.79 & 0.73 & 0.91 & 0.82 & 1.10 & 1.01 & 0.68 & 0.47 \\
BRLT111 & L2.0 $\pm$ 0.5 & 0.86 & 0.77 & 0.90 & 0.81 & 1.05 & 1.06 & 0.53 & 0.48 \\
BRLT112 & L1.0 $\pm$ 0.5 & 0.86 & 0.79 & 0.98 & 0.85 & 1.04 & 0.95 & 0.43 & 0.48 \\
BRLT113 & M9.0 $\pm$ 0.5 & 1.17 & 0.88 & 1.05 & 0.93 & 1.02 & 1.05 & 0.36 & 0.47 \\
BRLT114 & L6.0 $\pm$ 0.5 & 0.76 & 0.73 & 0.91 & 0.82 & 1.08 & 0.93 & 0.68 & 0.52 \\
BRLT116 & T2.5 $\pm$ 0.5 & 0.55 & 0.54 & 0.82 & 0.62 & 0.85 & 0.74 & 0.33 & 0.46 \\
BRLT117 & L5.0 $\pm$ 1.0 & 0.88 & 0.70 & 0.81 & 0.78 & 1.03 & 0.89 & 0.56 & 0.48 \\
BRLT119 & L4.0 $\pm$ 0.5 & 0.87 & 0.76 & 0.90 & 0.86 & 1.03 & 1.01 & 0.51 & 0.46 \\
BRLT121 & L1.0 $\pm$ 0.5 & 0.91 & 0.80 & 0.99 & 0.81 & 1.01 & 1.03 & 0.36 & 0.47 \\
BRLT122 & L1.0 $\pm$ 0.5 & 0.91 & 0.84 & 0.95 & 0.82 & 1.04 & 0.90 & 0.40 & 0.50 \\
BRLT123 & L2.0 $\pm$ 0.5 & 0.95 & 0.80 & 0.94 & 0.85 & 1.03 & 1.05 & 0.47 & 0.46 \\
BRLT129 & L5.0 $\pm$ 1.0 & 0.80 & 0.70 & 0.95 & 0.79 & 1.04 & 0.95 & 0.58 & 0.49 \\
BRLT130 & L3.0 $\pm$ 1.0 & 0.94 & 0.69 & 0.93 & 0.78 & 1.10 & 1.04 & 0.34 & 0.49 \\
BRLT133 & M9.0 $\pm$ 0.5 & 1.04 & 0.86 & 0.96 & 0.91 & 0.99 & 0.98 & 0.42 & 0.46 \\
BRLT136 & L1.0 $\pm$ 1.0 & 0.93 & 0.87 & 0.96 & 0.86 & 1.06 & 1.02 & 0.37 & 0.48 \\
BRLT137 & L4.5 $\pm$ 0.5 & 0.76 & 0.72 & 0.88 & 0.75 & 1.04 & 0.91 & 0.52 & 0.50 \\
BRLT138 & L2.0 $\pm$ 1.0 & 0.88 & 0.82 & 0.97 & 0.86 & 1.06 & 0.98 & 0.50 & 0.49 \\
BRLT139 & L4.0 $\pm$ 1.0 & 0.80 & 0.74 & 0.84 & 0.73 & 1.04 & 0.95 & 0.41 & 0.49 \\
BRLT140 & L0.0 $\pm$ 0.5 & 0.94 & 0.86 & 0.96 & 0.84 & 1.07 & 1.00 & 0.41 & 0.48 \\
BRLT142 & L2.5 $\pm$ 0.5 & 0.85 & 0.82 & 0.97 & 0.85 & 1.08 & 0.98 & 0.54 & 0.49 \\
BRLT144 & L5.0 $\pm$ 1.0 & 0.86 & 0.68 & 0.84 & 0.81 & 0.97 & 0.93 & 0.46 & 0.47 \\
BRLT145 & L1.0 $\pm$ 0.5 & 0.93 & 0.83 & 1.01 & 0.84 & 1.02 & 0.99 & 0.42 & 0.49 \\
BRLT149 & L6.0 $\pm$ 1.0 & 0.78 & 0.68 & 0.83 & 0.77 & 1.04 & 0.98 & 0.43 & 0.49 \\
BRLT152 & L0.0 $\pm$ 0.5 & 1.05 & 0.91 & 0.99 & 0.91 & 1.03 & 1.03 & 0.41 & 0.49 \\
BRLT153 & L1.0 $\pm$ 0.5 & 0.91 & 0.84 & 1.02 & 0.82 & 1.06 & 1.02 & 0.34 & 0.48 \\
BRLT155 & L3.0 $\pm$ 1.0 & 0.86 & 0.75 & 1.05 & 0.88 & 1.09 & 1.00 & 0.50 & 0.49 \\
BRLT159 & L9.0 $\pm$ 0.5 & 0.73 & 0.67 & 0.80 & 0.78 & 1.02 & 0.89 & 0.62 & 0.50 \\
BRLT162 & L0.5 $\pm$ 0.5 & 0.97 & 0.86 & 1.11 & 0.84 & 1.06 & 1.06 & 0.35 & 0.49 \\
BRLT163 & L1.0 $\pm$ 0.5 & 0.83 & 0.83 & 1.02 & 0.90 & 1.05 & 1.09 & 0.44 & 0.48 \\
BRLT164 & T3.0 $\pm$ 0.5 & 0.54 & 0.53 & 0.71 & 0.57 & 0.70 & 0.39 & 0.24 & 0.46 \\
BRLT165 & L2.0 $\pm$ 0.5 & 0.92 & 0.86 & 0.95 & 0.90 & 1.04 & 1.03 & 0.46 & 0.50 \\
BRLT168 & L4.0 $\pm$ 0.5 & 0.79 & 0.75 & 0.92 & 0.80 & 1.09 & 1.03 & 0.56 & 0.46 \\
BRLT171 & L5.0 $\pm$ 0.5 & 0.79 & 0.74 & 0.94 & 0.82 & 1.08 & 1.00 & 0.55 & 0.49 \\
BRLT176 & L4.0 $\pm$ 1.0 & 0.87 & 0.77 & 0.96 & 0.79 & 1.01 & 0.96 & 0.44 & 0.49 \\
BRLT181 & L1.0 $\pm$ 1.0 & 1.09 & 0.88 & 1.09 & 0.89 & 1.03 & 1.01 & 0.42 & 0.49 \\
BRLT182 & T3.0 $\pm$ 0.5 & 0.51 & 0.50 & 0.74 & 0.57 & 0.79 & 0.61 & 0.37 & 0.44 \\
BRLT186 & L1.0 $\pm$ 1.0 & 0.92 & 0.85 & 1.04 & 0.85 & 1.04 & 1.03 & 0.39 & 0.49 \\
BRLT190 & T4.0 $\pm$ 0.5 & 0.34 & 0.47 & 0.54 & 0.48 & 0.54 & 0.19 & 0.22 & 0.36 \\
BRLT197 & T2.0 $\pm$ 1.0 & 0.54 & 0.64 & 0.87 & 0.66 & 0.88 & 0.70 & 0.44 & 0.47 \\
BRLT198 & L3.0 $\pm$ 1.0 & 0.90 & 0.80 & 1.06 & 0.84 & 1.06 & 1.04 & 0.39 & 0.47 \\
BRLT202 & T2.5 $\pm$ 0.5 & 0.45 & 0.57 & 0.81 & 0.52 & 0.79 & 0.60 & 0.32 & 0.44 \\
BRLT203 & T3.0 $\pm$ 1.0 & 0.39 & 0.58 & 0.82 & 0.53 & 0.81 & 0.74 & 0.46 & 0.44 \\
BRLT206 & L2.0 $\pm$ 0.5 & 0.90 & 0.83 & 0.91 & 0.82 & 1.06 & 1.10 & 0.45 & 0.48 \\
BRLT210 & L4.5 $\pm$ 0.5 & 0.77 & 0.75 & 0.96 & 0.82 & 1.09 & 1.01 & 0.54 & 0.49 \\
BRLT216 & M9.0 $\pm$ 0.5 & 1.10 & 0.88 & 1.07 & 0.90 & 1.00 & 0.99 & 0.39 & 0.49 \\
BRLT217 & T0.0 $\pm$ 0.5 & 0.69 & 0.66 & 0.89 & 0.75 & 0.94 & 0.78 & 0.50 & 0.52 \\
BRLT218 & L6.0 $\pm$ 0.5 & 0.79 & 0.71 & 0.95 & 0.80 & 1.13 & 1.04 & 0.62 & 0.52 \\
BRLT219 & T3.0 $\pm$ 0.5 & 0.43 & 0.46 & 0.73 & 0.52 & 0.68 & 0.40 & 0.26 & 0.44 \\
BRLT220 & L2.0 $\pm$ 0.5 & 0.89 & 0.77 & 0.83 & 0.85 & 1.09 & 1.07 & 0.51 & 0.47 \\
BRLT227 & L3.0 $\pm$ 0.5 & 0.83 & 0.76 & 0.91 & 0.85 & 1.10 & 1.07 & 0.55 & 0.50 \\
BRLT229 & M8.0 $\pm$ 0.5 & 1.23 & 0.95 & 1.01 & 0.95 & 1.05 & 1.04 & 0.37 & 0.50 \\
BRLT231 & L5.0 $\pm$ 0.5 & 0.87 & 0.77 & 0.91 & 0.85 & 1.07 & 1.00 & 0.59 & 0.49 \\
BRLT234 & L4.0 $\pm$ 1.0 & 0.74 & 0.67 & 0.87 & 0.64 & 1.09 & 0.95 & 0.26 & 0.50 \\
BRLT236 & L3.5 $\pm$ 0.5 & 0.82 & 0.71 & 0.93 & 0.76 & 1.07 & 0.97 & 0.43 & 0.49 \\
BRLT237 & L4.0 $\pm$ 0.5 & 0.84 & 0.69 & 0.88 & 0.84 & 1.09 & 0.99 & 0.49 & 0.51 \\
BRLT240 & L3.0 $\pm$ 0.5 & 0.94 & 0.82 & 0.93 & 0.90 & 1.05 & 1.05 & 0.55 & 0.47 \\
BRLT243 & T0.0 $\pm$ 0.5 & 0.70 & 0.60 & 0.73 & 0.69 & 1.01 & 0.76 & 0.50 & 0.51 \\
BRLT247 & M9.0 $\pm$ 0.5 & 1.09 & 0.90 & 1.09 & 0.93 & 0.97 & 1.00 & 0.41 & 0.48 \\
BRLT249 & L5.0 $\pm$ 0.5 & 0.84 & 0.71 & 0.84 & 0.84 & 1.05 & 0.92 & 0.57 & 0.50 \\
BRLT250 & L1.0 $\pm$ 0.5 & 0.85 & 0.77 & 0.83 & 0.79 & 1.05 & 0.94 & 0.46 & 0.49 \\
BRLT251 & L1.0 $\pm$ 0.5 & 0.92 & 0.80 & 0.94 & 0.80 & 1.00 & 0.94 & 0.34 & 0.48 \\
BRLT253 & L1.0 $\pm$ 1.0 & 0.89 & 0.71 & 0.86 & 0.72 & 1.09 & 0.99 & 0.27 & 0.48 \\
BRLT254 & L5.0 $\pm$ 0.5 & 0.84 & 0.82 & 0.95 & 0.89 & 1.06 & 1.04 & 0.54 & 0.49 \\
BRLT260 & L2.0 $\pm$ 1.0 & 0.80 & 0.70 & 0.89 & 0.71 & 1.11 & 1.00 & 0.29 & 0.49 \\
BRLT262 & L0.0 $\pm$ 0.5 & 0.90 & 0.82 & 0.96 & 0.89 & 1.12 & 1.02 & 0.47 & 0.49 \\
BRLT265 & L2.0 $\pm$ 0.5 & 0.89 & 0.80 & 0.92 & 0.80 & 1.04 & 1.07 & 0.47 & 0.48 \\
BRLT269 & L7.0 $\pm$ 0.5 & 0.62 & 0.69 & 0.91 & 0.82 & 1.10 & 0.99 & 0.76 & 0.50 \\
BRLT270 & L2.0 $\pm$ 1.0 & 0.81 & 0.72 & 0.83 & 0.72 & 1.05 & 0.89 & 0.33 & 0.50 \\
BRLT274 & L2.0 $\pm$ 0.5 & 0.73 & 0.76 & 1.11 & 0.89 & 1.17 & 1.07 & 0.63 & 0.48 \\
BRLT276 & L0.0 $\pm$ 0.5 & 0.96 & 0.83 & 1.06 & 0.85 & 1.02 & 1.03 & 0.38 & 0.49 \\
BRLT279 & L1.0 $\pm$ 0.5 & 0.91 & 0.83 & 1.01 & 0.82 & 1.05 & 1.03 & 0.39 & 0.48 \\
BRLT283 & L5.0 $\pm$ 1.0 & 0.81 & 0.71 & 0.88 & 0.79 & 0.99 & 0.87 & 0.37 & 0.49 \\
BRLT285 & L5.0 $\pm$ 0.5 & 0.80 & 0.73 & 0.78 & 0.82 & 1.05 & 0.96 & 0.62 & 0.52 \\
BRLT290 & T2.0 $\pm$ 0.5 & 0.50 & 0.49 & 0.78 & 0.58 & 0.90 & 0.61 & 0.35 & 0.50 \\
BRLT295 & L4.0 $\pm$ 2.0 & 0.83 & 0.80 & 0.99 & 0.76 & 0.96 & 0.97 & 0.33 & 0.47 \\
BRLT296 & L4.0 $\pm$ 0.5 & 0.84 & 0.75 & 0.89 & 0.82 & 1.05 & 0.99 & 0.46 & 0.49 \\
BRLT297 & L4.5 $\pm$ 0.5 & 0.83 & 0.76 & 0.95 & 0.89 & 1.06 & 1.00 & 0.52 & 0.50 \\
BRLT299 & L4.0 $\pm$ 1.0 & 0.77 & 0.74 & 0.90 & 0.76 & 1.03 & 0.92 & 0.46 & 0.49 \\
BRLT301 & L1.0 $\pm$ 0.5 & 0.92 & 0.77 & 0.92 & 0.82 & 1.03 & 1.03 & 0.42 & 0.47 \\
BRLT302 & L4.0 $\pm$ 1.0 & 0.81 & 0.72 & 0.89 & 0.76 & 1.07 & 0.90 & 0.39 & 0.49 \\
BRLT308 & L5.0 $\pm$ 0.5 & 0.74 & 0.72 & 0.94 & 0.82 & 1.09 & 1.03 & 0.57 & 0.50 \\
BRLT319 & T3.0 $\pm$ 0.5 & 0.42 & 0.47 & 0.69 & 0.50 & 0.79 & 0.43 & 0.28 & 0.47 \\
BRLT340 & L4.0 $\pm$ 0.5 & 0.79 & 0.75 & 0.97 & 0.90 & 1.08 & 1.06 & 0.49 & 0.49 \\
\enddata
\end{deluxetable}
\twocolumn

\onecolumn
\begin{deluxetable}{c c c c c c c}
\tabletypesize{\normalsize}
\tablecolumns{7}
\tablewidth{0pt}
\tablehead{ & & \multicolumn{5}{c}{Equivalent width (\AA)} \\ \colhead{Name} & \colhead{Spectral type} & \colhead{\ion{Na}{i}} & \colhead{\ion{K}{i}} & \colhead{\ion{K}{i}} & \colhead{\ion{K}{i}} & \colhead{\ion{K}{i}} \\ & & \colhead{1.139$\mu$m} & \colhead{1.169$\mu$m} & \colhead{1.177$\mu$m} & \colhead{1.244$\mu$m} & \colhead{1.253$\mu$m}}
\tablecaption{The equivalent width obtained from the spectra. Missing entries indicate the non detection of the line, due either to the line being too weak or the spectrum being too noisy. Numbers in \emph{italics} indicate measurements with relative errors larger than 0.33. For the details on how these values were calculated, see Section \ref{ind_eqw}. \label{eqw_table}}
\startdata
BRLT1 & L9.0 $\pm$ 0.5 &  \emph{0.90} &   8.52 &   3.83 &   2.50 &   2.73 \\
BRLT2 & L1.0 $\pm$ 1.0 &   6.79 &   5.23 &   4.77 &   9.27 &   2.38 \\
BRLT3 & L9.0 $\pm$ 1.0 &  \emph{1.33} &   3.95 &   4.72 &  \emph{0.57} &   3.87 \\
BRLT6 & L3.0 $\pm$ 1.0 &   6.88 &   2.15 &   2.70 &   9.36 &   2.58 \\
BRLT7 & M8.0 $\pm$ 1.0 &   6.41 &  \emph{1.26} &   5.69 &   3.04 &   4.95 \\
BRLT8 & L8.5 $\pm$ 0.5 &  \emph{3.50} &   5.46 &   1.92 &   1.68 &   3.89 \\
BRLT9 & L1.0 $\pm$ 1.0 &   7.38 &   4.97 &   8.39 &   5.20 &   6.09 \\
BRLT10 & L9.0 $\pm$ 0.5 &  \emph{2.41} &   2.42 &   3.52 &  \emph{0.67} &  \emph{1.32} \\
BRLT12 & L3.0 $\pm$ 1.0 &   4.24 &   2.82 &   8.70 &   5.39 &   5.61 \\
BRLT14 & L0.0 $\pm$ 0.5 &    $\ldots$ &    $\ldots$ &    $\ldots$ &    $\ldots$ &    $\ldots$ \\
BRLT15 & T2.0 $\pm$ 2.0 &    $\ldots$ &  \emph{4.29} &  \emph{2.20} &   1.57 &   2.66 \\
BRLT16 & L2.0 $\pm$ 1.0 &   6.92 &   6.76 &   4.91 &   5.89 &   2.72 \\
BRLT18 & L0.0 $\pm$ 1.0 &    $\ldots$ &    $\ldots$ &    $\ldots$ &    $\ldots$ &    $\ldots$ \\
BRLT20 & L1.0 $\pm$ 1.0 &   5.90 &   5.08 &   8.69 &   5.77 &   2.06 \\
BRLT21 & L3.5 $\pm$ 0.5 &   8.09 &   4.08 &   9.37 &   7.41 &   8.32 \\
BRLT22 & M8.0 $\pm$ 0.5 &   4.41 &   6.26 &   5.04 &   6.27 &    $\ldots$ \\
BRLT24 & L3.5 $\pm$ 0.5 &  12.53 &   3.42 &   6.94 &   5.96 &   6.70 \\
BRLT26 & L5.5 $\pm$ 0.5 &   8.97 &   5.77 &   5.57 &   4.03 &   2.66 \\
BRLT27 & T0.0 $\pm$ 0.5 &  \emph{3.55} &  \emph{5.59} &  \emph{3.87} &   5.78 &  \emph{1.67} \\
BRLT28 & L6.0 $\pm$ 0.5 &    $\ldots$ &  \emph{5.62} &  \emph{5.65} &    $\ldots$ &    $\ldots$ \\
BRLT30 & L5.0 $\pm$ 0.5 &    $\ldots$ &    $\ldots$ &    $\ldots$ &  \emph{4.72} &    $\ldots$ \\
BRLT31 & L4.0 $\pm$ 1.0 &   3.32 &   4.61 &   6.50 &   5.72 &   4.59 \\
BRLT32 & L1.5 $\pm$ 0.5 &   7.89 &   3.63 &   9.26 &   9.26 &   6.69 \\
BRLT33 & L3.5 $\pm$ 0.5 &   4.41 &  \emph{0.87} &   9.33 &   9.04 &   4.65 \\
BRLT35 & M9.5 $\pm$ 0.5 &   9.89 &   4.11 &   7.79 &   5.19 &   3.75 \\
BRLT37 & L5.0 $\pm$ 0.5 &   8.38 &   3.00 &   6.83 &   6.46 &   5.72 \\
BRLT38 & T0.0 $\pm$ 0.5 &  \emph{5.33} &  \emph{2.69} &  \emph{3.04} &   1.36 &   2.95 \\
BRLT39 & L5.0 $\pm$ 1.0 &   3.29 &   6.08 &   7.12 &   5.18 &   7.94 \\
BRLT42 & M9.0 $\pm$ 0.5 &   8.58 &  \emph{0.81} &   5.57 &   1.68 &   3.40 \\
BRLT44 & L5.0 $\pm$ 1.0 &   2.02 &   7.24 &   7.71 &   3.34 &   3.00 \\
BRLT45 & T1.0 $\pm$ 0.5 &  \emph{1.07} &  \emph{0.73} &  \emph{2.63} &   1.88 &   3.14 \\
BRLT46 & L0.5 $\pm$ 0.5 &   9.54 &   8.13 &   7.45 &  10.87 &   3.50 \\
BRLT48 & L4.5 $\pm$ 0.5 &    $\ldots$ &    $\ldots$ &    $\ldots$ &    $\ldots$ &    $\ldots$ \\
BRLT49 & M9.0 $\pm$ 0.5 &    $\ldots$ &  \emph{2.72} &  \emph{5.23} &  \emph{3.95} &  \emph{5.04} \\
BRLT50 & T6.0 $\pm$ 0.5 &    $\ldots$ & $\ldots$ & $\ldots$ & $\ldots$ & $\ldots$ \\
BRLT51 & L3.0 $\pm$ 1.0 &  11.88 &   4.20 &   4.86 &   9.75 &   2.31 \\
BRLT52 & L5.5 $\pm$ 0.5 &    $\ldots$ &   6.53 &   7.08 &  \emph{1.24} &   3.51 \\
BRLT56 & L1.5 $\pm$ 1.0 &   5.34 &   4.89 &   9.50 &   7.13 &   4.33 \\
BRLT57 & L0.0 $\pm$ 1.0 &   5.53 &   4.35 &   4.29 &   7.53 &   3.09 \\
BRLT58 & L4.0 $\pm$ 1.0 &   5.26 &   4.51 &  11.05 &   9.47 &   4.71 \\
BRLT60 & L1.0 $\pm$ 1.0 &  12.12 &   5.83 &   2.94 &   9.67 &   2.43 \\
BRLT62 & L5.0 $\pm$ 1.0 &   9.46 &   4.39 &   6.54 &   7.42 &   3.37 \\
BRLT63 & L1.0 $\pm$ 0.5 &  \emph{6.03} &  \emph{3.11} &  \emph{5.03} &  \emph{6.67} &  \emph{6.96} \\
BRLT64 & L4.0 $\pm$ 0.5 &  10.78 &   2.56 &   6.51 &    $\ldots$ &   5.02 \\
BRLT65 & M9.0 $\pm$ 0.5 &    $\ldots$ &    $\ldots$ &   5.44 &  \emph{4.58} &  \emph{5.82} \\
BRLT66 & L5.0 $\pm$ 0.5 &   7.67 &   3.95 &   5.28 &   6.04 &   4.49 \\
BRLT67 & L1.0 $\pm$ 0.5 &  \emph{8.17} &  \emph{4.98} &  \emph{3.20} &  \emph{7.99} &  \emph{8.89} \\
BRLT68 & L5.0 $\pm$ 0.5 & \emph{14.65} &  \emph{4.99} &    $\ldots$ &    $\ldots$ &  \emph{3.22} \\
BRLT69 & L1.0 $\pm$ 0.5 &   9.03 &   2.97 &   5.19 &   5.65 &   5.35 \\
BRLT71 & L1.5 $\pm$ 0.5 &   9.86 &   7.50 &   8.03 &   6.84 &   2.80 \\
BRLT72 & M9.0 $\pm$ 0.5 &   7.66 &   2.06 &   7.81 &   4.20 &   4.75 \\
BRLT73 & L1.0 $\pm$ 0.5 &   9.70 &   6.78 &   9.71 &  12.73 &   6.47 \\
BRLT74 & L9.5 $\pm$ 1.0 &  \emph{3.04} &   8.25 &   2.12 &   3.74 &   3.16 \\
BRLT75 & M9.0 $\pm$ 1.0 &   5.80 &   5.32 &   4.95 &   4.26 &  \emph{0.29} \\
BRLT76 & L5.5 $\pm$ 0.5 &   9.98 &   7.26 &   6.58 &   9.89 &   2.55 \\
BRLT78 & L1.0 $\pm$ 0.5 &  15.59 &   8.53 & \emph{14.00} &  \emph{8.82} &    $\ldots$ \\
BRLT81 & M9.0 $\pm$ 0.5 &  \emph{4.43} &  \emph{2.52} &  \emph{4.94} &  \emph{7.57} &    $\ldots$ \\
BRLT82 & L1.0 $\pm$ 0.5 &   7.93 &   3.65 &   7.38 &   7.66 &   3.86 \\
BRLT83 & M8.0 $\pm$ 1.0 &   7.97 &   6.65 &   5.46 &   5.51 &   5.07 \\
BRLT84 & L3.5 $\pm$ 0.5 &   7.73 &    $\ldots$ &   4.77 &   2.07 &   1.93 \\
BRLT85 & M8.0 $\pm$ 0.5 &  \emph{5.84} &  \emph{5.77} &  \emph{4.03} &  \emph{5.36} &    $\ldots$ \\
BRLT87 & T0.0 $\pm$ 0.5 &  \emph{1.96} &  \emph{2.74} &  \emph{6.58} &    $\ldots$ &    $\ldots$ \\
BRLT88 & L4.0 $\pm$ 1.0 &   5.57 &   7.69 &   7.89 &   4.77 &   3.30 \\
BRLT91 & T3.0 $\pm$ 0.5 &  \emph{4.00} &  \emph{6.22} &  \emph{2.96} &  \emph{0.62} &   4.17 \\
BRLT92 & L1.0 $\pm$ 0.5 &   6.48 &   3.16 &   6.49 &   7.47 &   3.77 \\
BRLT97 & L0.0 $\pm$ 1.0 &   2.60 &   5.46 &   6.33 &   6.02 &   2.30 \\
BRLT98 & T4.0 $\pm$ 0.5 &    $\ldots$ &  \emph{8.21} &  \emph{5.37} &    $\ldots$ &    $\ldots$ \\
BRLT99 & L5.0 $\pm$ 0.5 &    $\ldots$ &  \emph{8.47} &  \emph{6.21} &  \emph{5.50} &    $\ldots$ \\
BRLT101 & L3.0 $\pm$ 0.5 &  18.71 &  \emph{7.25} &  13.16 &  12.27 &    $\ldots$ \\
BRLT102 & L0.0 $\pm$ 0.5 &  14.15 &  \emph{6.53} &  12.87 &    $\ldots$ &    $\ldots$ \\
BRLT103 & L5.5 $\pm$ 0.5 &  10.30 &   9.66 &   7.30 &   6.31 &   4.27 \\
BRLT104 & M9.0 $\pm$ 0.5 &  21.60 &   9.47 &  \emph{1.81} & \emph{12.72} &  \emph{6.38} \\
BRLT105 & L5.0 $\pm$ 0.5 &   7.70 &   6.49 &   6.97 &   6.64 &   3.53 \\
BRLT106 & M9.0 $\pm$ 0.5 &  \emph{7.32} &  \emph{8.73} &  \emph{8.72} &  17.21 &    $\ldots$ \\
BRLT108 & L6.5 $\pm$ 0.5 &    $\ldots$ &    $\ldots$ &  \emph{2.14} & \emph{11.95} &    $\ldots$ \\
BRLT111 & L2.0 $\pm$ 0.5 &  \emph{5.89} &  \emph{5.12} &  11.27 &    $\ldots$ & \emph{11.96} \\
BRLT112 & L1.0 $\pm$ 0.5 &  \emph{6.48} &    $\ldots$ &   9.63 & \emph{15.86} &  \emph{4.92} \\
BRLT113 & M9.0 $\pm$ 0.5 &    $\ldots$ &  \emph{7.86} & \emph{15.20} &  11.86 &  \emph{6.69} \\
BRLT114 & L6.0 $\pm$ 0.5 &  \emph{6.29} &    $\ldots$ &  14.17 &    $\ldots$ &  \emph{5.59} \\
BRLT116 & T2.5 $\pm$ 0.5 &    $\ldots$ &  \emph{4.81} &    $\ldots$ &    $\ldots$ &    $\ldots$ \\
BRLT117 & L5.0 $\pm$ 0.5 &  \emph{2.56} & \emph{13.99} &  13.90 &  \emph{9.77} &  \emph{8.93} \\
BRLT119 & L4.0 $\pm$ 0.5 & \emph{19.11} &  \emph{5.58} &  \emph{5.31} & \emph{10.97} & \emph{16.48} \\
BRLT121 & L1.0 $\pm$ 0.5 &  29.10 &  \emph{8.06} &  12.60 &  \emph{8.68} &    $\ldots$ \\
BRLT122 & L1.0 $\pm$ 0.5 &  10.63 &   4.97 &   9.10 &   6.53 &  \emph{1.04} \\
BRLT123 & L2.0 $\pm$ 0.5 &  18.71 &    $\ldots$ & \emph{11.03} &    $\ldots$ &  \emph{1.89} \\
BRLT129 & L5.0 $\pm$ 1.0 &   3.13 &   7.27 &   6.11 &   3.19 &   3.86 \\
BRLT130 & L3.0 $\pm$ 0.5 &  20.06 &  10.64 &   8.77 &  \emph{3.99} &    $\ldots$ \\
BRLT131 & T3.0 $\pm$ 0.5 &  \emph{4.47} &  \emph{4.07} &  \emph{3.07} &   3.79 &   3.09 \\
BRLT133 & M9.0 $\pm$ 0.5 &  10.02 &  \emph{3.52} &  \emph{5.27} &  \emph{8.43} &    $\ldots$ \\
BRLT135 & T2.5 $\pm$ 0.5 &  \emph{1.87} &  \emph{3.86} &  \emph{3.57} &    $\ldots$ &   1.32 \\
BRLT136 & L1.0 $\pm$ 1.0 &  \emph{6.10} &  \emph{5.32} &  12.74 &  \emph{5.37} &    $\ldots$ \\
BRLT137 & L4.5 $\pm$ 0.5 &   8.36 &   7.98 &   6.09 &   1.71 &   4.66 \\
BRLT138 & L2.0 $\pm$ 1.0 &   8.16 &   3.67 &   4.97 &   4.23 &   2.98 \\
BRLT139 & L5.0 $\pm$ 0.5 &  \emph{7.14} &  \emph{7.92} &  12.15 & \emph{13.20} &  \emph{7.34} \\
BRLT140 & L0.0 $\pm$ 0.5 &  15.30 &  \emph{3.56} & \emph{10.37} &    $\ldots$ &  \emph{1.69} \\
BRLT142 & L2.5 $\pm$ 0.5 &   4.49 &   6.25 &   5.65 &   4.64 &   3.15 \\
BRLT144 & L5.0 $\pm$ 0.5 &    $\ldots$ & \emph{10.26} &  \emph{4.40} & \emph{13.78} &  \emph{6.10} \\
BRLT145 & L1.0 $\pm$ 0.5 &  \emph{7.44} &  \emph{5.15} &  \emph{9.39} & \emph{15.89} &    $\ldots$ \\
BRLT147 & T3.0 $\pm$ 0.5 &    $\ldots$ &  \emph{5.72} &  \emph{5.02} &   1.82 &   2.28 \\
BRLT149 & L6.0 $\pm$ 0.5 &  12.64 &  \emph{4.46} &    $\ldots$ &    $\ldots$ & \emph{13.19} \\
BRLT152 & L0.0 $\pm$ 0.5 &  13.29 &  \emph{3.42} & \emph{10.16} &    $\ldots$ &  \emph{8.06} \\
BRLT153 & L1.0 $\pm$ 0.5 &  \emph{3.69} &  \emph{3.85} & \emph{10.86} & \emph{12.93} &    $\ldots$ \\
BRLT155 & L3.0 $\pm$ 1.0 &  11.26 &   2.45 &   5.72 &   8.17 &    $\ldots$ \\
BRLT159 & L9.0 $\pm$ 0.5 &    $\ldots$ &  \emph{4.61} &  \emph{9.62} &    $\ldots$ &    $\ldots$ \\
BRLT162 & L0.5 $\pm$ 0.5 &   7.51 &   2.79 &   5.51 &   5.62 &   3.77 \\
BRLT163 & L1.0 $\pm$ 0.5 &  19.13 &  \emph{5.04} &  \emph{5.45} &  \emph{7.36} &  \emph{6.29} \\
BRLT164 & T3.0 $\pm$ 0.5 &  \emph{3.14} &    $\ldots$ &  \emph{6.96} &  \emph{2.72} &    $\ldots$ \\
BRLT165 & L2.0 $\pm$ 0.5 &  \emph{1.63} &    $\ldots$ &    $\ldots$ & \emph{14.47} &   5.52 \\
BRLT168 & L4.0 $\pm$ 0.5 &   6.73 &  \emph{4.78} &  \emph{9.16} &  12.21 & \emph{12.93} \\
BRLT171 & L5.0 $\pm$ 0.5 &   7.22 &   5.79 &   5.95 &   4.61 &   3.26 \\
BRLT176 & L4.0 $\pm$ 1.0 &    $\ldots$ &    $\ldots$ &    $\ldots$ &    $\ldots$ &    $\ldots$ \\
BRLT179 & T4.5 $\pm$ 0.5 & $\ldots$ & $\ldots$ & $\ldots$ &    $\ldots$ &    $\ldots$ \\
BRLT181 & L1.0 $\pm$ 1.0 &   7.28 &   7.31 &   6.98 &   4.58 &   3.60 \\
BRLT182 & T3.0 $\pm$ 0.5 &  \emph{3.02} &  \emph{2.93} &  \emph{5.61} &  \emph{3.18} &  \emph{7.41} \\
BRLT186 & L1.0 $\pm$ 1.0 &   7.48 &   2.24 &   5.39 &   4.51 &   2.92 \\
BRLT190 & T4.0 $\pm$ 0.5 &    $\ldots$ &    $\ldots$ &    $\ldots$ &    $\ldots$ &    $\ldots$ \\
BRLT197 & T2.0 $\pm$ 1.0 &  \emph{2.84} &  \emph{4.87} &  \emph{3.53} &   6.65 &   2.85 \\
BRLT198 & L3.0 $\pm$ 0.5 &  \emph{2.07} &  \emph{5.90} &  16.15 &  \emph{8.15} &  \emph{8.87} \\
BRLT202 & T2.5 $\pm$ 0.5 &  \emph{4.49} &  \emph{4.41} &  \emph{3.57} &   1.82 &   2.02 \\
BRLT203 & T3.0 $\pm$ 1.0 &  \emph{4.15} &  \emph{3.20} &  \emph{3.63} &   4.05 &  \emph{1.01} \\
BRLT206 & L2.0 $\pm$ 0.5 &  16.86 &  \emph{4.89} &   8.94 &  \emph{7.47} &  \emph{3.43} \\
BRLT207 & L7.0 $\pm$ 0.5 &   6.14 &   4.55 &   3.75 &   2.88 &   2.19 \\
BRLT210 & L4.5 $\pm$ 0.5 &   4.74 &   3.75 &   8.24 &   6.16 &   2.32 \\
BRLT212 & L6.0 $\pm$ 2.0 &  \emph{4.46} &    $\ldots$ &    $\ldots$ &  \emph{3.52} &    $\ldots$ \\
BRLT216 & M9.0 $\pm$ 0.5 &    $\ldots$ &   6.66 &  \emph{6.84} &  \emph{3.07} &    $\ldots$ \\
BRLT217 & T0.0 $\pm$ 0.5 &    $\ldots$ &    $\ldots$ &  \emph{4.93} &  \emph{3.43} &    $\ldots$ \\
BRLT218 & L6.0 $\pm$ 0.5 &  \emph{6.97} &  \emph{6.01} & \emph{11.66} &  \emph{3.17} &  \emph{5.89} \\
BRLT219 & T3.0 $\pm$ 0.5 &    $\ldots$ &    $\ldots$ &    $\ldots$ &    $\ldots$ &    $\ldots$ \\
BRLT220 & L2.0 $\pm$ 0.5 &  \emph{4.01} &  \emph{7.71} &  \emph{7.76} &    $\ldots$ &  \emph{8.67} \\
BRLT227 & L3.0 $\pm$ 0.5 &  \emph{8.12} &  \emph{5.75} &  16.46 &  \emph{7.24} &  \emph{6.59} \\
BRLT229 & M8.0 $\pm$ 0.5 &  \emph{5.02} &    $\ldots$ &  11.39 &  \emph{4.85} &  \emph{2.87} \\
BRLT231 & L5.0 $\pm$ 0.5 &  \emph{6.58} & \emph{12.85} &  \emph{9.54} & \emph{15.08} & \emph{10.09} \\
BRLT232 & T2.5 $\pm$ 0.5 &  \emph{2.74} &  \emph{4.51} &  \emph{2.32} &   3.42 &   1.71 \\
BRLT234 & L4.0 $\pm$ 1.0 &   9.44 &   7.80 &  12.22 &   7.10 &   4.57 \\
BRLT236 & L3.5 $\pm$ 0.5 &    $\ldots$ &    $\ldots$ &    $\ldots$ &    $\ldots$ &    $\ldots$ \\
BRLT237 & L4.0 $\pm$ 0.5 &  \emph{6.12} &  \emph{4.88} &  14.21 &  \emph{6.18} &    $\ldots$ \\
BRLT240 & L3.0 $\pm$ 0.5 &  13.35 &   7.40 &  11.84 &    $\ldots$ &  \emph{2.69} \\
BRLT243 & T0.0 $\pm$ 0.5 &  \emph{6.55} &    $\ldots$ &    $\ldots$ &  \emph{7.99} &  \emph{4.86} \\
BRLT247 & M9.0 $\pm$ 0.5 &  16.42 &  \emph{3.70} &  16.27 &  \emph{9.29} &    $\ldots$ \\
BRLT249 & L5.0 $\pm$ 0.5 &  \emph{4.02} &  \emph{7.47} &    $\ldots$ &    $\ldots$ &    $\ldots$ \\
BRLT250 & L1.0 $\pm$ 0.5 &  \emph{5.99} &  \emph{9.14} &  \emph{8.69} &  \emph{6.73} &  \emph{6.40} \\
BRLT251 & L1.0 $\pm$ 0.5 &  \emph{4.05} &    $\ldots$ &  \emph{3.27} &    $\ldots$ &  \emph{6.88} \\
BRLT253 & L1.0 $\pm$ 0.5 &  \emph{1.03} &  \emph{5.08} &  13.56 &  \emph{5.89} &    $\ldots$ \\
BRLT254 & L5.0 $\pm$ 0.5 &  \emph{9.96} &    $\ldots$ &  13.09 &    $\ldots$ &    $\ldots$ \\
BRLT258 & L5.0 $\pm$ 1.0 &    $\ldots$ &    $\ldots$ &    $\ldots$ &    $\ldots$ &    $\ldots$ \\
BRLT260 & L2.0 $\pm$ 0.5 &  15.49 &   4.18 &  \emph{8.72} &  \emph{3.77} &    $\ldots$ \\
BRLT262 & L0.0 $\pm$ 0.5 &   5.80 &  \emph{3.06} &  10.26 &  10.55 &  \emph{7.30} \\
BRLT265 & L2.0 $\pm$ 0.5 &  \emph{8.94} &  \emph{2.86} &   8.95 &  11.27 &   8.38 \\
BRLT269 & L7.0 $\pm$ 0.5 & \emph{14.63} &    $\ldots$ &  \emph{5.99} &  \emph{7.25} &    $\ldots$ \\
BRLT270 & L2.0 $\pm$ 0.5 &  12.43 &  \emph{6.30} &  \emph{8.25} & \emph{15.34} &    $\ldots$ \\
BRLT274 & L2.0 $\pm$ 0.5 &    $\ldots$ &  \emph{2.74} &  \emph{5.22} & \emph{13.52} &  \emph{6.16} \\
BRLT275 & T2.0 $\pm$ 2.0 &  \emph{4.64} &  \emph{2.16} &  \emph{3.69} &   6.13 &   4.94 \\
BRLT276 & L0.0 $\pm$ 0.5 &    $\ldots$ &    $\ldots$ &   7.91 &  \emph{2.96} &    $\ldots$ \\
BRLT279 & L1.0 $\pm$ 0.5 &  \emph{8.03} &  \emph{4.44} &  \emph{7.97} &  \emph{5.10} &    $\ldots$ \\
BRLT281 & T0.0 $\pm$ 1.0 &  \emph{7.44} &    $\ldots$ &  \emph{5.68} &    $\ldots$ &    $\ldots$ \\
BRLT283 & L5.0 $\pm$ 0.5 &   8.11 &  \emph{5.03} &  13.08 &  \emph{9.28} &  \emph{7.37} \\
BRLT285 & L5.0 $\pm$ 0.5 &  \emph{3.25} &  \emph{7.24} &    $\ldots$ &    $\ldots$ &   8.77 \\
BRLT287 & T3.0 $\pm$ 0.5 &  \emph{3.48} &  \emph{3.73} &  \emph{3.63} &   3.09 &   3.58 \\
BRLT290 & T2.0 $\pm$ 0.5 & \emph{10.65} &  \emph{3.05} &  \emph{5.31} &    $\ldots$ &    $\ldots$ \\
BRLT295 & L4.0 $\pm$ 2.0 &    $\ldots$ &    $\ldots$ &    $\ldots$ &  \emph{5.07} &    $\ldots$ \\
BRLT296 & L4.0 $\pm$ 0.5 &  \emph{4.88} &  \emph{4.97} &  \emph{3.42} &  16.10 &  \emph{3.07} \\
BRLT297 & L4.5 $\pm$ 0.5 &  22.31 &   7.22 &  \emph{5.03} &    $\ldots$ & \emph{10.87} \\
BRLT299 & L4.0 $\pm$ 1.0 &   7.06 &   5.02 &   5.73 &   5.28 &   3.10 \\
BRLT301 & L1.0 $\pm$ 0.5 &  \emph{9.14} &  \emph{8.24} & \emph{18.21} & \emph{11.62} &    $\ldots$ \\
BRLT302 & L4.0 $\pm$ 0.5 &  \emph{2.08} &   7.14 &  16.35 &  \emph{3.20} &    $\ldots$ \\
BRLT305 & L5.5 $\pm$ 1.0 &    $\ldots$ &    $\ldots$ &    $\ldots$ &    $\ldots$ &    $\ldots$ \\
BRLT306 & L4.5 $\pm$ 1.0 &    $\ldots$ &    $\ldots$ &    $\ldots$ &    $\ldots$ &    $\ldots$ \\
BRLT307 & L1.0 $\pm$ 0.5 &   9.93 &  \emph{0.99} &   2.40 &   8.40 &   3.16 \\
BRLT308 & L5.0 $\pm$ 0.5 &  \emph{7.27} &    $\ldots$ &  \emph{5.15} &    $\ldots$ &    $\ldots$ \\
BRLT309 & L7.0 $\pm$ 0.5 &  \emph{6.11} &  \emph{6.35} &  \emph{2.00} &    $\ldots$ &    $\ldots$ \\
BRLT311 & T3.0 $\pm$ 0.5 &  \emph{4.25} &  \emph{0.55} &  \emph{1.36} &  \emph{1.09} &   2.25 \\
BRLT312 & T0.0 $\pm$ 0.5 &  \emph{6.46} &  \emph{2.57} &  \emph{4.47} &   7.28 &   4.52 \\
BRLT313 & L3.5 $\pm$ 0.5 &   8.72 &  10.48 &   6.12 &   4.73 &  \emph{1.41} \\
BRLT314 & L7.5 $\pm$ 0.5 &   6.82 &   7.46 &   6.24 &   3.20 &   2.73 \\
BRLT315 & L1.0 $\pm$ 1.0 &   5.33 &   1.56 &   8.78 &   6.81 &   5.95 \\
BRLT316 & L1.0 $\pm$ 0.5 &  12.46 &   2.74 &   3.19 &   2.69 &   5.79 \\
BRLT317 & L3.0 $\pm$ 1.0 &   9.77 &   6.95 &   9.81 &   4.57 &   8.67 \\
BRLT318 & L1.0 $\pm$ 0.5 &  14.36 &   6.33 &   6.22 &   6.20 &   6.10 \\
BRLT319 & T3.0 $\pm$ 0.5 &  \emph{4.19} &  \emph{7.52} &  \emph{3.21} &    $\ldots$ &  \emph{5.84} \\
BRLT320 & L1.0 $\pm$ 0.5 &   4.90 &  \emph{0.90} &   8.12 &   6.34 &  \emph{1.88} \\
BRLT321 & T4.0 $\pm$ 0.5 &  \emph{2.85} &  \emph{5.28} &  \emph{4.99} &   5.73 &   2.26 \\
BRLT322 & L5.0 $\pm$ 0.5 &   3.48 &   4.19 &   4.20 &   6.66 &   5.10 \\
BRLT323 & L5.0 $\pm$ 1.0 &   7.97 &   7.54 &   6.30 &   6.35 &   7.85 \\
BRLT325 & T2.0 $\pm$ 1.0 &  \emph{3.35} &  \emph{8.20} &  \emph{5.85} &   7.34 &   4.93 \\
BRLT328 & L3.0 $\pm$ 1.0 &   7.78 &   9.25 &   8.12 &   7.30 &   2.85 \\
BRLT330 & L2.0 $\pm$ 1.0 &   7.53 &   2.97 &   7.51 &   4.76 &   4.40 \\
BRLT331 & L3.0 $\pm$ 1.0 &  11.68 &   4.10 &   7.02 &   3.56 &   8.74 \\
BRLT332 & L2.0 $\pm$ 1.0 &   6.67 &   7.44 &   6.96 &   4.92 &   6.05 \\
BRLT333 & T2.0 $\pm$ 0.5 &  \emph{2.92} &  \emph{5.06} &  \emph{3.55} &    $\ldots$ &   5.24 \\
BRLT334 & L3.5 $\pm$ 0.5 &   7.51 &   5.26 &   6.19 &   6.27 &   4.74 \\
BRLT335 & L4.0 $\pm$ 1.0 &   8.62 &   4.56 &   6.62 &   7.05 &   5.50 \\
BRLT338 & L1.0 $\pm$ 1.0 &  13.70 &   8.90 &  10.46 &   6.94 &   2.44 \\
BRLT340 & L4.0 $\pm$ 0.5 & \emph{11.60} &    $\ldots$ &  \emph{6.78} &  \emph{7.30} &  \emph{6.37} \\
BRLT343 & L9.0 $\pm$ 1.0 &  \emph{2.24} &   7.31 &  \emph{0.62} &   4.02 &  \emph{1.53} \\
BRLT344 & T0.0 $\pm$ 1.0 &  \emph{5.87} &  \emph{4.36} &    $\ldots$ &   2.80 &   2.73 \\
\enddata                      
\end{deluxetable}
\twocolumn

\section{Radial velocities}
\begin{figure}
\includegraphics[width=0.5\textwidth]{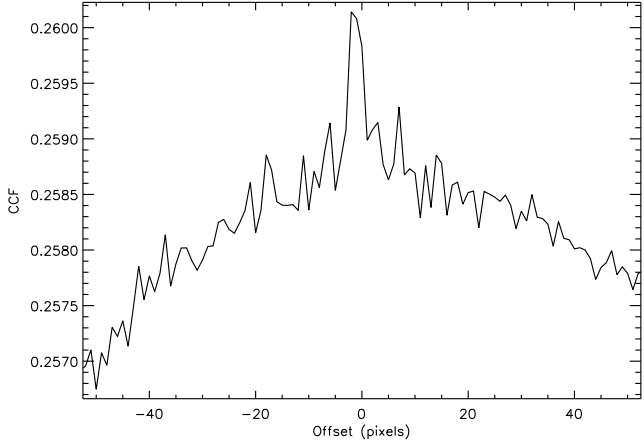}
\caption{An example of a cross-correlation function (CCF) obtained for one of our targets. The offset is measured in pixels and then converted into a radial velocity using the wavelength dispersion of the instrument. \label{ccf} }
\end{figure}

Using a cross-correlation technique we calculated the radial velocities for the objects in the sample. An example of a cross-correlation function (hereafter CCF) obtained is shown in Figure \ref{ccf}. The CCF shows a clear sharp peak around -1 pix, highlighting the precision of the radial velocity obtained. The exact position of the CCF peak was determined using the procedure described in \citet{1992RSPTA.341..117T} and \citet{1986nras.book.....P}. The two main telluric bands at 1.35$-$1.45 $\mu$m and 1.80$-$1.95 $\mu$m are not considered when evaluating the CCF, to avoid a possible systematic bias towards smaller velocities.

The radial velocities have been measured relative to Kelu-1, for which we obtained a very high signal-to-noise ratio spectrum on the night of 07-04-2013. The radial velocity of Kelu-1 is given in \citet{2010ApJ...723..684B} and is 6.37$\pm$0.35 km s$^{-1}$. Kelu-1 is a known binary with an estimated radial velocity semi-amplitude of 3$-$4 km s$^{-1}$ over a period of 38 years \citep{2006PASP..118..611G,2009AIPC.1094..561S}. Since this semi-amplitude is similar to the precision of our observations which are made over a much shorter time span of around 3 years, we neglect this systematic error.

\begin{figure}
\includegraphics[width=0.5\textwidth]{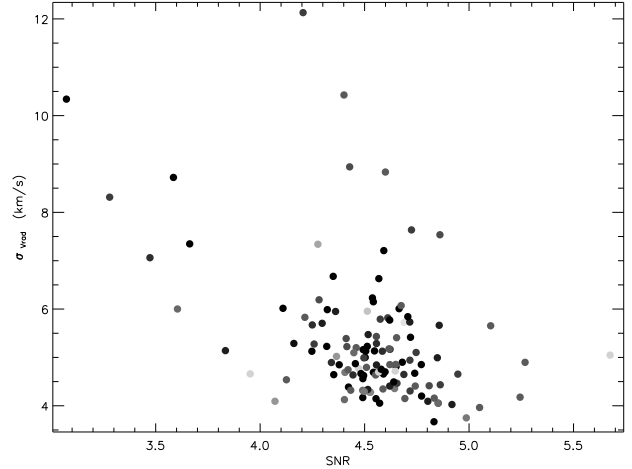}
\caption{The distribution of the radial velocity estimated precision given by the CCF algorithm as a function of the signal-to-noise ratio (SNR) of the spectra. The difference in spectral types between the target and the standard is indicated by the colour of the point, with black points indicating a difference of zero and light grey points indicating a difference of 12 spectral subtypes. With a SNR as little as $\sim$5 we can achieve radial velocity estimated precisions of 4$-$6 km s$^{-1}$. \label{sigma_rv} }
\end{figure}

\begin{figure}
\includegraphics[width=0.5\textwidth]{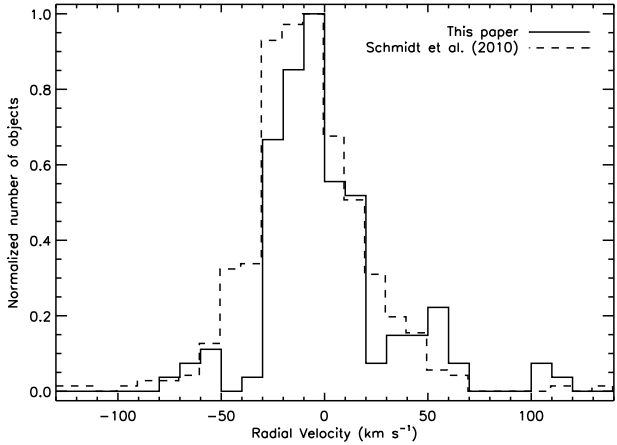}
\caption{The radial velocity distribution of our targets. Overplotted as a dashed line is the distribution obtained by \citet{2010AJ....139.1808S} from a sample of 484 L dwarfs from SDSS. Both distributions are normalized to their peak value to allow for direct comparison. The two samples have very similar dispersions of 31.5 and 34.3 km s$^{-1}$ respectively. The slightly lower dispersion of our sample could be a geometric effect (see text for further details). \label{histo_rv} }
\end{figure}

The results obtained can be seen in Figure \ref{sigma_rv}, where we plot the estimated precision given by the CCF algorithm as a function of the SNR of the spectra, and in Figure \ref{histo_rv} where we plot the radial velocity distribution of our targets. In Figure \ref{sigma_rv} the difference in spectral types between the target and the standard is indicated by the colour of the point, with black points indicating a difference of zero and light grey points indicating a difference of 12 spectral subtypes. Even with an SNR as low as $\sim$5, we can obtain an average estimated precision of $\sim$5 km s$^{-1}$. It is clear however that the scatter is very high, and that is probably due to the difference between the late-type targets and the ``standard'' adopted. In late type objects (i.e. the light grey points in Figure \ref{sigma_rv}) the \ion{Na}{i} and \ion{K}{i} lines become shallower and the correlation between the standard and the target is therefore weaker. 

The radial velocity distribution of our sample, plotted in Figure \ref{histo_rv}, peaks at -1.7$\pm$1.2 km s$^{-1}$ with a dispersion of 31.5 km s$^{-1}$. The dispersion in our distribution is slightly narrower than that obtained by \citet[overplotted as a dashed line for comparison]{2010AJ....139.1808S}, who derived a dispersion of 34.3 km s$^{-1}$ from a sample of 484 L dwarfs from SDSS. This discrepancy could be due to a geometric effect. Our sample is drawn from a smaller area of sky covering predominantly the northern galactic cap. The radial velocity of our sample is therefore dominated by the $W$ component of the galactic velocity, which is known to have a narrower dispersion than the $U$ and $V$ components \citep[see e.g.][]{1998MNRAS.298..387D}. We tested this hypothesis using the Besan\c{c}on Model of stellar population synthesis of the Galaxy \citep{2003A&A...409..523R}. For O to M type dwarfs with J $<$ 18.1 we obtain a dispersion of 34.8 km s$^{-1}$ when considering a sample spread over the SDSS footprint (i.e. the area covered by the \citealt{2010AJ....139.1808S} sample) and a dispersion of 31.8 km s$^{-1}$ when considering a sample spread over the right ascension and declination limits of our sample (see Section \ref{sp_dens}). Both numbers are in good agreement with the observed ones, and the measured difference between our sample and Schmidt's one seems therefore to be due to a geometric effect.

Another possible explanation is the fact that our sample is focused on field (i.e. thin disk) objects, and is therefore biased towards slightly younger dwarfs, i.e. towards a narrower $V_{\rm rad}$ distribution. Finally, the discrepancy could be due to errors in the determination of our radial velocities, introduced by uncorrected instability of X-shooter \citep[which is however believed to be stable down to 0.5 km s$^{-1}$,][]{2011A&A...536A.105V}.

The results presented in this section represent a feasibility study for a larger project to determine accurate and precise radial velocities for the brown dwarfs that will be observed by the European Space Agency mission Gaia. Further details on the project can be found in Marocco et al. (MmSAIt, in press).

\section{Constraining the sub-stellar IMF and formation history}
The Initial Mass Function (IMF) of stellar objects more massive than the Sun was first derived by \citet{1955ApJ...121..161S}, who parameterized it as $\Psi(m)\Delta m \propto (m/M_{\rm Sun})^{-\alpha}$ with $\alpha = 2.35$. The Salpeter IMF was later extended to less massive stars by \citet{1979ApJS...41..513M} who suggested that the IMF flattened below one solar mass. Currently the most widely accepted parameterizations are the log-normal IMF derived by \citet{2003PASP..115..763C,2005ASSL..327...41C} and the broken power law introduced by \citet{2001MNRAS.322..231K}.

When trying to extend the IMF further into the sub-stellar regime, one is faced by one fundamental challenge. Sub-stellar objects do not form a main sequence, and keep on cooling down while evolving through the spectral sequence. As a result, there is no unique mass-luminosity relationship that one can use to convert the observed luminosity function (LF) into the IMF (this issue is sometimes referred to as age-mass-luminosity degeneracy). The observed luminosity function is therefore influenced by the formation history (or Birth Rate, hereafter BR) of this kind of objects, and one needs to take this into account while trying to constrain the IMF. While the BR is often assumed to be constant for stars \citep[e.g.][]{1979ApJS...41..513M}, it is unconstrained in the sub-stellar regime. Moreover, the formation mechanism of brown dwarfs and giant planets is not well understood. The different formation scenarios proposed would leave their imprints in the IMF and BR, so to distinguish between them it is fundamental to constrain these two quantities in the sub-stellar regime. 

One way to break the age-mass-luminosity degeneracy is to look at young clusters and associations, since their known ages and metallicities allow the use of a mass-luminosity relation based on the cluster age, removing the dependency of the LF on the BR. Therefore they have been the target of many observational campaigns,  e.g. \citet{2011A&A...527A..24L,2009A&A...505.1115L,2007MNRAS.380..712L,2009AIPC.1094..912C,2009ApJ...703..399L,2013A&A...559A.126A,2012A&A...539A.151A}. These clusters allow a relatively direct measurement of the sub-stellar IMF, but the initial conditions and accretion histories of individual objects introduce uncertainties regarding the ages, and hence masses, of such young objects \citep[e.g.][]{2010HiA....15..755B}. Moreover, very high and variable extinction increases contamination by reddened field stars. Evolutionary models are also very uncertain at young ages, and the effect of magnetic activity or episodic accretion on the determination of luminosity are not yet fully understood. Finally, some of these regions are still forming stars, introducing further uncertainties and possible biases \citep[see e.g.][]{2013MmSAI..84..905A}.

Studying the IMF of the field populations has significant advantages, since there is a larger number of benchmark systems, and therefore the evolutionary and atmospheric models are more mature. Reddening is not an issue, given that even the deepest surveys can only probe the solar neighbourhood. On the other hand, the LF of field brown dwarfs depends on their BR, because only few field sub-stellar objects have age constraints, either as binaries \citep[e.g.][]{2011MNRAS.414.3590B,2010MNRAS.404.1952B,2009MNRAS.395.1237B,2010MNRAS.404.1817Z,2011MNRAS.410..705D,2003A&A...398L..29S,2013MNRAS.431.2745G,2013A&A...553L...5D} or as members of moving groups \citep[e.g.][]{2014ApJ...783..121G,2014arXiv1402.6053M,2013ApJ...762...88M,2010MNRAS.402..575C,2010MNRAS.409..552G},. The assessment of completeness, contamination, and other observational biases can introduce further uncertainties.

Since the first attempt by \citet{1999ApJ...521..613R}, several groups have made measurements of the sub-stellar mass function in the field. Due to the limited samples available, these measurements were either covering only L dwarfs \citep[e.g][]{2007AJ....133..439C} or only T dwarfs \citep[e.g.][]{2008ApJ...676.1281M,2013MNRAS.433..457B,2012ApJ...753..156K,2010A&A...522A.112R}. Considering the full spectral sequence is in fact a challenge, and those studies that attempted this \citep[e.g.][]{2010A&A...522A.112R} have been battling with high associated uncertainties and had to compromise with large bin sizes in order to get statistically significant sampling of the spectral sequence.

With modern large-scale near- and mid-infrared surveys, such as DENIS \citep{1999A&A...349..236E}, SDSS \citep{2000AJ....120.1579Y}, 2MASS \citep{2006AJ....131.1163S}, UKIDSS \citep{2007MNRAS.379.1599L}, VHS \citep{2013Msngr.154...35M}, and WISE \citep{2010AJ....140.1868W}, which have identified large numbers of brown dwarfs it is now possible to provide the necessary sample of such objects. In particular, while surveys like 2MASS and SDSS were more sensitive to the detection of L dwarfs than T dwarfs, the ULAS probes to greater depth and can provide a more even and statistically robust sampling of the full early-L to late-T range,. The L/T transition region, which is well populated by the sample presented here, is most sensitive to the BR. 

This section outlines the efforts to use the sample of mid-L$-$mid-T dwarfs presented here to empirically constrain the Galactic brown dwarf formation history. This sample is an obvious choice because it covers a large spectral type range (crucially focused on the L/T transition) with a good sampling of each spectral type bin, and it is complete (see Section \ref{completeness}), unbiased (see Section \ref{cand_selection}) and uncontaminated, since its members have been followed up with spectroscopy. 

\subsection{Determining the space density of L/T transition dwarfs \label{sp_dens}}
The spectroscopic follow-up of the full sample is incomplete. However, there are areas of sky where the follow-up is complete. So in order to determine the space density of brown dwarfs we divided the full sample in three sub-samples: between RA = 15h50m to 9h20m the follow-up is complete down to the limit of J = 18.1; between RA = 9h20m to 12h20m the follow-up is complete down to J = 17.87; finally between RA = 12h20m to 15h50m the follow-up is complete down to J = 17.7. These RA ranges correspond to an area of $\sim$620, 375 and 712 deg$^2$ in ULAS DR7, and account for 88, 29, and 50 objects respectively. 

To determine the volume sampled we calculated the maximum distance at which an object of a given spectral type could have been detected (assuming the given magnitude limit), using the $M_{\rm J}$-NIR spectral type relation from \citet{2010A&A...524A..38M}. With this distance limit we then calculated the volume sampled by each spectral type bin, and the corresponding space density of objects.

The derived space densities were then corrected for the Malmquist and Eddington biases following the approach described in \citet{2008MNRAS.390..304P}. The Eddington bias is caused by the photometric uncertainties on the magnitudes of objects near our cut (i.e. J $<$ 18.1). However, since the magnitude cut imposed is bright (it corresponds to a $\sim$12$\sigma$ detection in the ULAS), the uncertainties at the J = 18.1 limit are typically less than $\sigma$ = 0.05 and therefore the Eddington bias correction is less than 1 per cent. This is negligible compared to the other sources of uncertainty. We estimated the Malmquist bias correction considering the mean scatter of the sample of known L and T dwarfs around the adopted $M_{\rm J}$-NIR spectral type relation. This represents an increase in the volume sampled of 22 per cent.

To increase the number of objects per bin, and therefore reduce the Poisson errors, we binned up the sample in four spectral type bins: L0-L3, L4-L6, L7-T0, and T1-T4. These bins correspond roughly to effective temperature ranges of $\sim$150 K.

\subsection{Completeness correction \label{completeness}}
\begin{figure}
\includegraphics[width=0.5\textwidth]{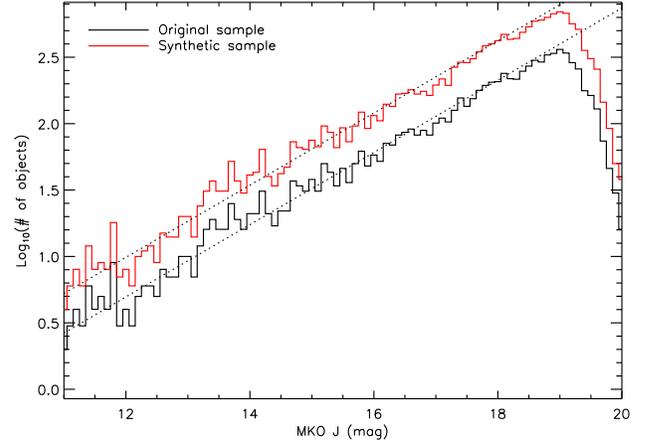}
\caption{The number of objects detected as a function of the J magnitude in the images used for the sample selection. The black histogram shows the results for the original images, while the red histogram shows the results in the synthetic images created by duplicating the number of objects. The dotted lines represent a fit to the bright tail of the distribution, i.e. for $14 <$ J $<  17$. \label{completeness_fig}}
\end{figure}

In order to check the completeness of the sample, first we need to estimate the number of objects lost due to missed detections. As stated above, the imposed magnitude limit (J $<$ 18.1) is bright compared to the limit of the ULAS, and therefore we do not expect to lose any object because of missed detections. This is well demonstrated by Figure \ref{completeness_fig} where we show the number of objects detected in the original ULAS images as a function of MKO J magnitude. The number of faint sources increases as a power of ten (note that the y axis is in logarithmic scale) up to J $\sim$ 19 \citep[as a consequence of the larger volume probed at fainter magnitudes,][]{1981gask.book.....M}, where it sharply drops. The dotted line is the fit to the bright tail of the distribution, i.e. for $14 <$ J $<  17$. Extrapolating the fit up to J=18.1 and comparing the ``expected'' number of objects with the measured one gives a completeness of $> 99\%$. The number of objects lost due to incomplete detection in therefore negligible.

Another possible issue, especially when searching for faint objects, is the possible blending with bright sources. However the typical object density in the fields considered is very low, because we are probing regions outside the galactic plane, therefore blending should not be an issue. To quantitatively assess its impact we adopted the following approach. We used the ULAS J band images containing the selected objects. We run the Cambridge Astronomy Survey Unit (CASU) pipeline on the images to detect and extract all the sources in the field. We then doubled the number of objects in every image by taking a 20$\times$20 pixels cut out around every object and copying it into a random position in the image, re-scaling it appropriately to blend the background level and avoid artefacts. We then re-run the CASU pipeline on the images and compared the number of sources identified (as a function of their J magnitude) with the number of sources in the original images. One would obviously expect to detect twice as many objects in the new synthetic images, with no dependence on the objects magnitude. This is indeed the case, as can be seen in Figure \ref{completeness_fig}, where the number of detected objects in the synthetic images is plotted in red. With an average number of sources detected in the synthetic images of $\sim$1.987 time the number of sources detected in the original images (recovered sources after doubling / recovered sources prior to doubling = 15427/7764), and no clear dependence on the J magnitude, the incompleteness due to objects blending is 0.3$\%$, which is again negligible compared to the other causes of incompleteness considered below.

To assess the completeness of the photometric selection criteria, the sample was compared to a control sample of known L and T dwarfs taken from www.DwarfArchives.org, for a magnitude limit of J $\leq$ 16, removing any objects that are known to be members of unresolved binary systems. The control sample was cross-matched with the ULAS and SDSS in order to obtain photometry on the same colour system as the selection criteria used. The same set of colour cuts described in Section \ref{cand_selection} was imposed to reveal the level of completeness of the sample selection. We retain all of the L4 dwarfs from the control sample, but only some of the L0-L3 dwarfs, indicating that the sample selection is complete for L4 spectral types and later. Similarly, the selection is largely incomplete beyond spectral types of T5. We therefore only consider the three spectral type bins covering the L4$-$T4 range.

The loss of objects due to photometric scattering of colours was also considered. For L4-L6 types one would expect to lose 3.7 dwarfs, this corresponds to a completeness level of 88$\%$. The L7-T0 range would lose 0.55 dwarfs, corresponding to a 94$\%$ completeness; for T1-T4 the expected loss is 0.05 dwarfs, corresponding to a completeness of 99$\%$. 

Pixel-noise correlation is not an issue, as demonstrated by \citet{2014PASA...31....4A}, who estimated the randomness of background noise in the ULAS images by visually selecting 11 empty 7$\times$7 pixel regions from the mosaics. They computed the standard deviation of the mean pixel value of each region (calling it $q$) and compared it against a similar calculation after randomly swapping pixels between regions. A $q/q_{\rm swapped}$ of 1 indicates perfectly uncorrelated noise while $q/q_{\rm swapped} \gg 1$ is due to non-pixel scale systematic variations. For ULAS images they found $q/q_{\rm swapped} \sim 1$.

\subsection{Correction for unresolved binarity}
We also corrected the results for the presence of binaries by first considering objects identified as possible binaries (Section \ref{unres_bins}) for which the spectral deconvolution gives a statistically ``better fit''. We derived the J magnitude of the two components given the unresolved photometry and the two spectral types determined with the deconvolution, and removed from the sample all companions and those primaries that would fall beyond the magnitude limit.

To assess the completeness of this correction we performed numerical simulations, using the spectral templates taken from the SpeX-Prism library. The spectra were combined to create a sample of synthetic unresolved binaries, following the procedure described in ADJ13 and in Section \ref{unres_bins} of this work. The synthetic templates were ``degraded'' to the typical SNR of the observed spectra by adding gaussian noise. We then run the binary identification process on each of the synthetic binaries to calculate the rate of successful detections. To avoid false positive detections in low mass ratio binaries, when fitting a given synthetic binary we removed from the template list all the synthetic binaries that had the same primary as the ``target'' one. For example, when fitting the synthetic binary SDSS~J165329.69+623136.5 + 2MASSI~J0415195$-$093506 (L1.0 + T8.0) we removed from the set of templates all the synthetic binaries that had SDSS~J165329.69+623136.5 as a primary. This is because one can expect that the synthetic L1.0 + T8.0 SDSS~J165329.69+623136.5 + HD~3651B would fit better the target than an L1.0 template alone, not because the synthetic binary genuinely fits better, but because the contribution from the T8.0 component is negligible and we would essentially be fitting the L1.0 component with itself.

\begin{figure}
\includegraphics[width=0.5\textwidth]{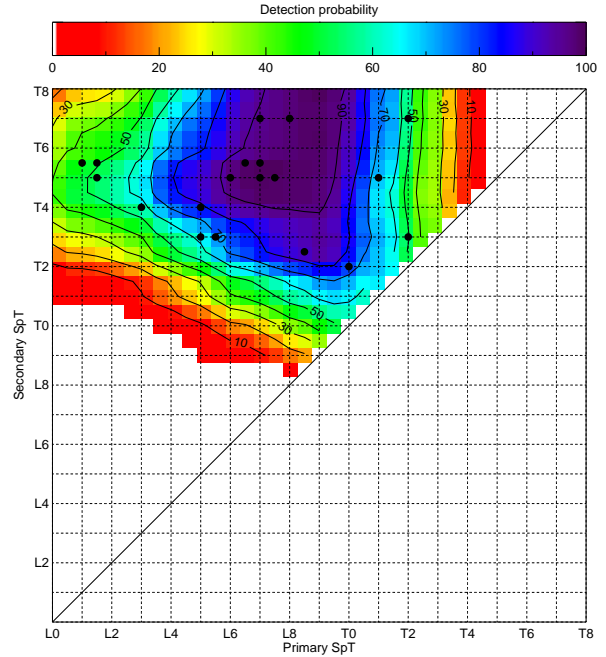}
\caption{The detection probability for unresolved binaries as a function of the spectral types of the two components, using the detection and deconvolution technique described in Section \ref{unres_bins}. Interpolated contour level are overplotted to ease the reading of the figure. Overplotted as black circles are the binary candidates identified in Section \ref{unres_bins}. \label{bin_simulations}}
\end{figure}

The results are shown in Figure \ref{bin_simulations}, where we plot the fraction of synthetic binaries retrieved as a function of the spectral type of the two components. Interpolated contour level are overplotted to ease the reading of the Figure. As expected, the technique is most efficient at the L/T transition, and the fraction of detected binaries steeply declines when moving towards very low mass ratios and early L type binaries. Equal spectral type binaries are also not detectable with this method. Overplotted as black circles are the binary candidates identified in Section \ref{unres_bins}. Not surprisingly the candidates are concentrated mostly in the high detection fraction area.

The sample of binary candidates is probably contaminated by peculiar objects, and therefore the derived binary fraction is somewhat higher than the ``true'' one. To assess the level of contamination we run the binary identification method on a sample of L and T dwarfs that have been previously targeted by high-resolution imaging campaigns, and have not showed evidences of binarity. The control sample consists of 40 objects covering the L0.0 to T7.5 spectral range, and includes objects taken from \citet{2003AJ....126.1526B}, \citet{2003AJ....125.3302G}, and \citet{2006ApJS..166..585B}. Two out of 40 objects are flagged as binaries by the detection method, implying a level of contamination of $5\%$. 

We can now use the detected binaries to constrain the binary fraction. To do that we combine the detection probability from Figure \ref{bin_simulations} with the mass-ratio distribution of sub-stellar binaries from Figure 3 of \citet{2007prpl.conf..427B}. First of all we correct the observed number of binaries for contamination using the fraction derived above, and then for completeness using the detection probability. All these binaries have mass ratio $q <$1, if not they would have equal spectral type. Using the distribution from \citet{2007prpl.conf..427B} we could estimate the number of undetected equal mass/equal spectral type binaries and therefore derive the binary fraction. \emph{However} the exact mass ratio of a system depends on its age, which is unconstrained. So we can only correct the number of observed binaries using the ratio between the number of $q$=1  binaries over the number of $q <$1 binaries, which is $\sim$1.2.

The numbers derived are presented in Table \ref{bin_fraction}. We calculated the binary fraction in the three spectral type ranges considered above (i.e. L4-L6, L7-T0, and T1-T4). The fraction is 24$\%$ in the L4-L6 range but rises to $\sim$70$\%$ in the L7-T0 range, before dropping down to $\sim$40$\%$ in the early T regime. This variations could be partly due to an underestimate of the number of equal spectral type binaries in the early L regime, due to the fact that the detected binary candidates lie in the $q \ll$1 range, and the ratio of $(q = 1)/(q \ll 1)$ binaries is higher than the assumed value of $\sim$1.2. 

High-resolution imaging and radial velocity surveys typically detect a binary fraction of $\sim$10-20$\%$ for brown dwarfs \citep[e.g.][]{2008A&A...492..545J,2006MNRAS.372.1879K,2006AJ....132..663B}. Many studies however report an higher observed binary fraction in the L/T transition regime \citep[e.g.][]{2006ApJ...647.1393L,2007ApJ...659..655B,2006ApJS..166..585B}, with values around $\approx 40 \%$. The binary fraction obtained here is even higher, peaking at $\sim$70$\%$ in the L7-T0 range. The reason for this discrepancy could be in an higher false positive ratio than estimated here. The control sample used to determine the contamination is in fact limited (only 40 objects) and the L/T transition in particular is poorly sampled. When correcting the derived space densities for unresolved binarity we will therefore use both the binary fraction we measured, and the values published in the literature.

\begin{table*}
\centering
\begin{tabular}{c c c c c c c}
\hline
Spectral   & Total number & \multicolumn{2}{c}{$q <$1} & \multicolumn{2}{c}{$q =$1} \\
type range & of objects & Binary fraction & Binaries & Binary fraction & Binaries \\
\hline
L4$-$L6 & 54 & 11$\%$ & 5.8 & 13$\%$ & 7.0 \\
L7$-$T0 & 19 & 34$\%$ & 7.1 & 40$\%$ & 8.5 \\
T1$-$T4 & 16 & 19$\%$ & 4.7 & 23$\%$ & 5.6 \\
\hline
\end{tabular}
\caption{The derived binary fraction. For each spectral type range we indicate the binary fraction and the expected number of binaries in the sample. \label{bin_fraction}}
\end{table*}

To take into account the presence of the undetected equal spectral type binaries, which would fall beyond the J limit if they were single objects, we used the definition of ``observed binary fraction'' given by \citet{2003ApJ...586..512B},

\begin{equation} 
\frac{N_B}{N_m} = \frac{\gamma}{\gamma + (1/{\rm BF}) -1}
\end{equation}

\noindent where $N_B$ and $N_m$ are the observed binaries and the total number of systems, respectively, BF is the ``true'' binary fraction, and $\gamma$ is the fractional increase in volume due to inclusion of binaries in the sample. The number of binaries that fall within the magnitude limit because of their increased brightness ($N_D$) is

\begin{equation}
\frac{N_D }{N_B} = \frac{\gamma - 1}{\gamma}
\end{equation}

Therefore, the fraction of objects to be excluded from the sample ($f_{\rm excl}$) is 

\begin{equation}
f_{\rm excl} = \frac{N_B}{N_m} \frac{N_D}{N_B} = \frac{\gamma - 1}{\gamma + (1/{\rm BF}) -1}
\end{equation}

For equal spectral type binaries $\gamma = 2\sqrt{2}$. 

As stated above, the final correction applied was derived assuming BF$ = 26 \pm 13 \%$, i.e. the mid point between the upper and lower limit to the $q = 1$ binary fraction derived in this work, and BF$ = 14 \pm 10 \%$, i.e. the weighted average of the values presented in the literature. The corrections applied are therefore $f_{\rm excl} = 0.30 \pm 0.10$ and $f_{\rm excl} = 0.18 \pm 0.12$.

\subsection{Comparison with numerical simulations}
We compared the space densities obtained above with the results of numerical simulations computed assuming different IMFs and birth rates from \citet{2006MNRAS.371.1722D}. Details of the simulations are briefly summarized here.

\begin{figure*}
\includegraphics[width=\textwidth]{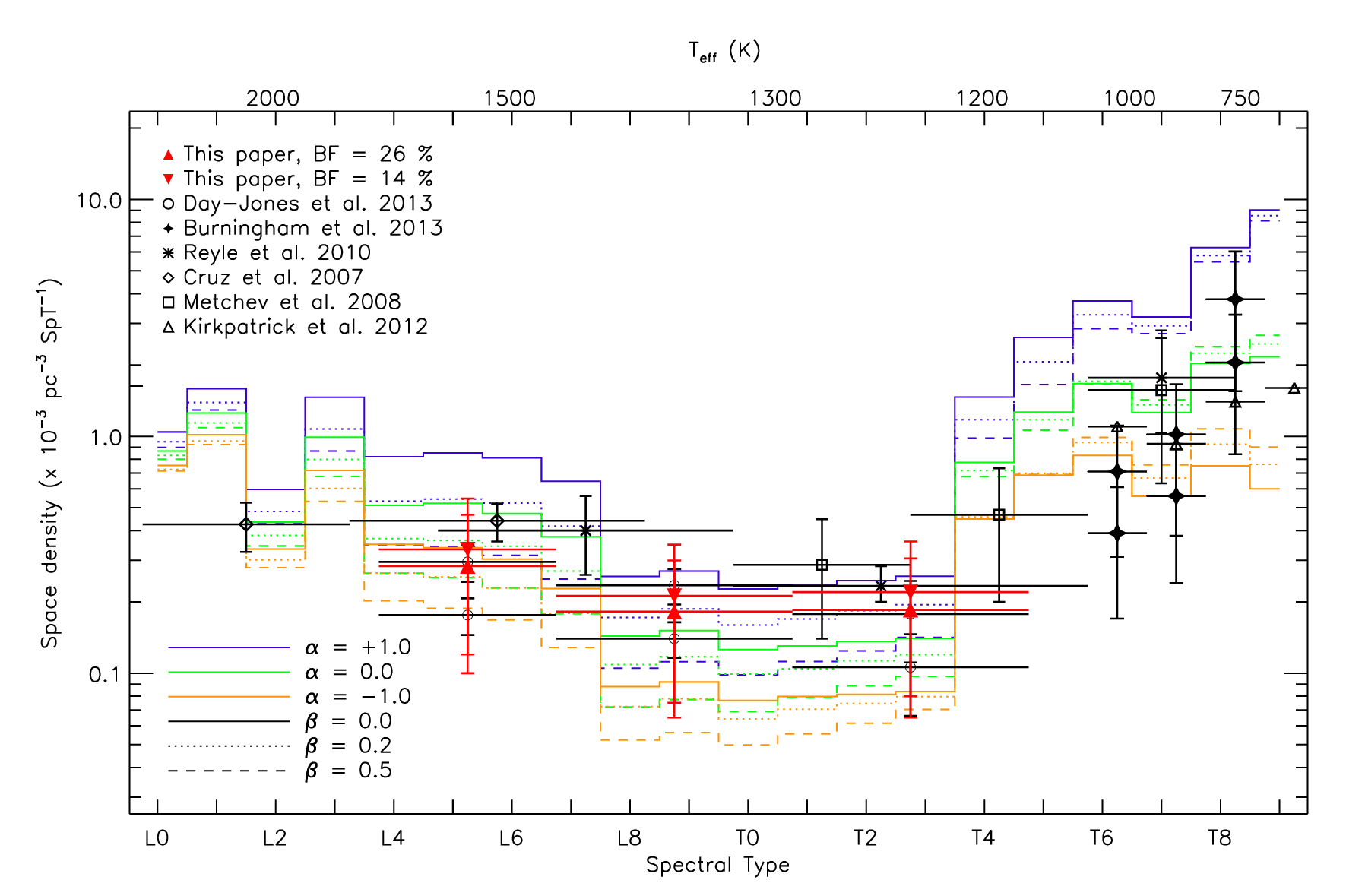}
\caption{A comparison between measured space densities of L and T dwarfs with simulations from \citet{2006MNRAS.371.1722D} with $\alpha$ = +1.0, 0.0, 1.0 and $\beta$ = 0.0, 0.2, 0.5. On the top axis we show an indicative temperature scale. \label{spt_histo}}
\end{figure*}

They assume a power-law IMF in the form 

\begin{equation}
\Psi(M) \propto M^{-\alpha} \left(pc^{-3} M^{-1}_\odot \right)
\end{equation}

where $\Psi(M)$ is the number of objects per unit volume in a given mass interval. They also assumed an exponential birth rate of the form

\begin{equation}
b(t) \propto e^{-\beta t}
\end{equation}

where $t$ is in Gyr and $\beta$ is the inverse of the birth rate scale time $\tau$ (in Gyr, since the galaxy was formed). Each simulated object was assigned an age based on the birth rate and a mass based on the IMF, giving a final creation function $C$ given by the equation 

\begin{equation}
C(M, t) = \Psi(M) \frac{b(t)}{T_G}
\end{equation}

where $T_G$ is the age of the Galaxy. $C$ is therefore the number of objects created per unit time per unit mass. The evolution of each object and its parameters (i.e. $T_{\rm eff}$ and absolute magnitudes) were calculated using the evolutionary models from \citet{1998A&A...337..403B}. Any model-dependent systematics would be introduced, but these should not affect the overall trend. The $T_{\rm eff}$ of an object was then converted into a spectral type using the $T_{\rm eff}$-NIR spectral type relation presented in \citet[equation 3]{2009ApJ...702..154S}. The number densities obtained for each bin were finally normalized to 0.0024 pc$^{-3}$ in the 0.1-0.09 $M_\odot$ mass range, according to \citet{2008A&A...486..283D}. We consider the simulations for three different values of $\beta$ (0.0, 0.2, and 0.5 corresponding to $\tau$ = $\infty$, 5, and 2 Gyr, respectively) and three values of $\alpha$ (+1.0, 0.0, -1.0). The results obtained are shown in Figure \ref{spt_histo}, where different colours represent different values of $\alpha$ and different line styles represent different values of $\beta$.

We compared the calculated space densities, taking into account the completeness and contribution from unresolved binaries, with those presented by various authors in the literature. We considered five different studies: \citet{2007AJ....133..439C}, \citet{2008ApJ...676.1281M}, \citet{2010A&A...522A.112R}, \citet{2012ApJ...753..156K}, \citet{2013MNRAS.433..457B}, and \citet{2013MNRAS.430.1171D}.

The \citet{2007AJ....133..439C} space densities probe down to the 2MASS limit (J = $\sim$16) and cover the M9-L8 dwarfs, likely suffering from incompleteness at the later types due to colour scattering. The binary correction uses the observed binary fraction of $\sim 17\%$  derived via high-resolution imaging of their sample. \citet{2008ApJ...676.1281M} cross-matched 2MASS with SDSS DR1 and used a series of colour selection criteria to select a sample of L and T dwarfs down to $z \leq$21. The correction for binarity assumes only the existence of equal mass/equal spectral type binaries for reason of the strong peak in the $q$ distribution \citep[Figure 3]{2007prpl.conf..427B}. The adopted binary fraction is assumed to decline from $50\%$ in the T0-T2.5 range, down to $21\%$ in the T3-T5.5 range, to $13\%$ in the T6-T8 range, and is therefore comparable to the numbers derived here. \citet{2010A&A...522A.112R} used CFBDS to select and classify a sample of $\sim$100 $>$L5 dwarfs down to $z' <$22.5, a comparable depth to this sample. They chose not make any correction for binarity, given the large uncertainty in the measured binary fraction. \citet{2012ApJ...753..156K} focused on the late T and Y dwarfs, using the WISE-selected sample of nearby objects. Assuming a binary fraction of $30\%$ and correcting for the incompleteness at the faint end of their sample, they derive the space density in the T6 to Y0.5 range. The \citet{2013MNRAS.433..457B} space densities use the same $M_{\rm J}$-spectral type relations we adopted. They correct for binarity assuming an upper limit on the binary fraction of 45$\%$ \citep{2005MNRAS.362L..45M} and a lower limit of 5$\%$ \citet{2003ApJ...586..512B}, hence deriving two values of the space density in each spectral type bin. They also probe down to a magnitude limit comparable to this sample. Finally, \citet{2013MNRAS.430.1171D} represent an early result from this sample, obtained from the sub-sample falling in the RA = 22h to 4h range. The only difference in the treatment of the data is in the binary correction, since in \citet{2013MNRAS.430.1171D} we followed the approach of \citet{2013MNRAS.433..457B} and derived two values for each spectral type range.

Our results and those listed above are summarized in Table \ref{sp_density}, and in Figure \ref{spt_histo}. It is important to notice that the numbers in Table \ref{sp_density} are \emph{integrated} over the spectral range quoted, while those plotted in Figure \ref{spt_histo} are \emph{per spectral type}, to allow a direct comparison with the simulations. A first look at the plot shows that our space densities do not differ drastically (within uncertainties) from those previously measured and discussed earlier. The differences between our derived densities and those previously published are mostly due to the use of different $M_{\rm J}$-SpT conversions and different binary fractions by the various groups. 

\begin{table*}
\centering
\begin{tabular}{c c c}
\hline
Reference & Spectral type range & Space density ($\times 10^{-3}$ pc$^{-3}$) \\
\hline
\multirow{2}{*}{\citet{2007AJ....133..439C}} & L0-L3   & 1.7 $\pm$ 0.4 \\
                                             & L3.5-L8 & 2.2 $\pm$ 0.4 \\
                                             \hline
\multirow{3}{*}{\citet{2008ApJ...676.1281M}} & T0-T2.5 & 0.86$^{+0.48}_{-0.44}$ \\
                                             & T3-T5.5 & 1.4$^{+0.8}_{-0.8}$ \\
                                             & T6-T8   & 4.7$^{+3.1}_{-2.8}$ \\
                                             \hline
\multirow{3}{*}{\citet{2010A&A...522A.112R}} & L5-T0   & 2.0$^{+0.8}_{-0.7}$ \\
                                             & T0.5-T5.5 & 1.4$^{+0.3}_{-0.2}$ \\
                                             & T6-T8   & 5.3$^{+3.1}_{-2.2}$ \\
                                             \hline
\multirow{4}{*}{\citet{2012ApJ...753..156K}} & T6-T6.5 & 1.1 \\
                                             & T7-T7.5 & 0.93 \\
                                             & T8-T8.5 & 1.4 \\
                                             & T9-T9.5 & 1.6 \\
                                             \hline
\multirow{3}{*}{\citet{2013MNRAS.433..457B}} & T6-T6.5 & 0.39 $\pm$ 0.22 $-$ 0.71 $\pm$ 0.40 \\
                                             & T7-T7.5 & 0.56 $\pm$ 0.32 $-$ 1.02 $\pm$ 0.64 \\
                                             & T8-T8.5 & 2.05 $\pm$ 1.21 $-$ 3.79 $\pm$ 2.24 \\
                                             \hline
\multirow{3}{*}{\citet{2013MNRAS.430.1171D}} & L4-L6.5 & 0.53 $\pm$ 0.10 $-$ 0.88 $\pm$ 0.16 \\
                                             & L7-T0.5 & 0.56 $\pm$ 0.10 $-$ 0.94 $\pm$ 0.16 \\
                                             & T1-T4.5 & 0.42 $\pm$ 0.16 $-$ 0.71 $\pm$ 0.27 \\       
                                             \hline
\multirow{3}{*}{This paper, BF = 26 $\pm$ 13} & L4-L6.5 & 0.85 $\pm$ 0.55 \\
                                             & L7-T0.5 & 0.73 $\pm$ 0.47 \\
                                             & T1-T4.5 & 0.74 $\pm$ 0.48 \\                                             
\hline
\multirow{3}{*}{This paper, BF = 14 $\pm$ 10} & L4-L6.5 & 1.00 $\pm$ 0.64 \\
                                             & L7-T0.5 & 0.85 $\pm$ 0.55 \\
                                             & T1-T4.5 & 0.88 $\pm$ 0.56 \\                                             
\hline
\end{tabular}
\caption{The space density derived here compared to values presented in the literature. The numbers are integrated over the spectral range quoted in the second column. \label{sp_density}}
\end{table*}

The most apparent feature is the absence of a significant drop in the number of objects between L7 and T4. The number of L/T transition dwarfs decreases, but not as much as expected. For the predicted theoretical deficit to be realised a higher binary fraction than assumed for the L/T transition would be necessary. That would lead to a larger correction and therefore to lower space densities. Conversely, a lower binary fraction at early types would bring up the density of objects in the L4-L6 range increasing the drop. 

However, this second scenario would lead to a preference for $\alpha >0$, which would be inconsistent with the results for late type objects, that consistently point towards $\alpha < 0$. On the other hand, $\alpha > 0$ is found also in nearby young clusters \citep[e.g.][]{2007MNRAS.378.1131C,2010ARA&A..48..339B} and by microlensing surveys \citep[$\alpha = 0.49^{+0.24}_{-0.27}$,][]{2011Natur.473..349S}. 

To reconcile the results in the two temperature regimes, one could assume that the binary fraction in the L/T transition is much higher than currently estimated. An alternative explanation is that objects in the high-mass end and low-mass end form in different environments, with the high-mass brown dwarfs forming predominantly in dense clusters (i.e. resulting in an $\alpha > 0$ IMF) and the low-mass brown dwarfs forming in low density environments, leading to a $\alpha < 0$ IMF \citep{2013pss5.book..115K}. Another possibility, as suggested by \citet{2013MNRAS.433..457B}, is that the cooling times assumed to transform the IMF into field luminosity function are affected by systematic errors.

As regards the formation history, it is not currently possible to place robust constraints on the birth rate with this sub-sample. One of the largest sources of uncertainties is the binary fraction. This could be resolved with the follow-up of the unresolved binary candidates identified here, by either AO imaging or radial velocity. 

The other main source of uncertainty is the absolute magnitude$-$spectral type calibration. Although based on an increasing number of objects with well measured parallaxes, the scatter around the current polynomial relation is still large, with typical rms of 0.4 magnitudes \citep{2012ApJS..201...19D}, and this propagates into a factor of $\sim$1.5 in the volume sampled. 

\section{Conclusions}
In this contribution we have presented the spectroscopic analysis of a sample of 196 late-M, L, and T dwarfs from the UKIDSS LAS DR7. One hundred and twenty two of these represent new discoveries. Among this large sample of objects we have identified 22 peculiar blue L dwarfs and 2 blue T dwarfs, that further increase the population of this class of objects. Suspected to pertain to a slightly older disk population (therefore slightly metal depleted) the kinematics are fundamental to fully characterize these new objects.

We have also identified 2 peculiar low-gravity late-M dwarfs, potentially young objects that can constitute useful benchmarks to study low-gravity atmospheres and constrain early evolution models. Once again the kinematics will be fundamental to confirm or refute their youth.

Using an index-based selection technique coupled with spectral deconvolution, we also identified 27 unresolved binary candidates among our targets. These objects are particularly important as their follow-up constraint on the population properties of multiple sub-stellar systems, and offer hints into the understanding of their formation mechanism. 

The sample presented here, being complete, unbiased, and uncontaminated, represents an opportunity to measure the luminosity function of field sub-stellar objects. Our attempt to use the measured space density has however been limited by two fundamental uncertainties. One is the lack of knowledge of the binary fraction. Following up the binary candidates identified here can represent a first step forward towards a more precise constraint of this important observable parameter. In the near future the ESA mission Gaia will provide a more accurate measurement, significantly reducing this source of uncertainty. The other is the use of photometric distances to compute the volume sampled, and the large associated uncertainty due to the scatter of objects around the ``main sequence''. Only measuring trigonometric parallaxes for a large sample of brown dwarfs would remove this uncertainty, and the astrometric programs focusing on sub-stellar objects represent an encouraging step forward in this direction.

Although limited by these uncertainties, the space densities derived here nevertheless point toward an higher than expected ratio of L/T transition dwarfs to late-Ts. If we assume a power-law IMF with a negative exponent (as suggested by the LF of late-T dwarfs), then the observed density of L/T transition objects is almost a factor of two higher than expected. This discrepancy can be suggestive of a higher than expected binary fraction in the L/T transition range, or that the cooling times assumed to transform the IMF into field LF are affected by systematic errors, or that low-mass and high-mass brown dwarfs are predominantly the product of different formation mechanisms and therefore derive from different underlying IMFs.

The full exploitation of present surveys is revealing larger and larger populations of L, T, and Y dwarfs, and new facilities like SPHERE and GPI will push the boundaries of our observations towards lower and lower masses. Moreover the already mentioned ESA/Gaia mission will provide a more accurate calibration of the absolute magnitude sequence and a more robust constraint on the binary fraction. Therefore it seems we are now approaching a more reliable determination of the sub-stellar IMF and BR, that will lead to a better understanding of their formation.

\section*{Acknowledgments}
We thank the referee, John Gizis, for comments that have significantly improved the quality of this paper.

FM would like to thank Antonio Chrysostomou and Philippe Delorme for the useful discussions and their valuable suggestions.

This research is based on observations collected at the European Organisation for Astronomical Research in the Southern Hemisphere, Chile programs 086.C-0450, 087.C-0639, 088.C-0048, and 091.C-0452. 

The authors would like to acknowledge the Marie Curie 7th European Community Framework Programme grant n.247593  Interpretation and Parameterization of Extremely Red COOL dwarfs (IPERCOOL) International Research Staff Exchange Scheme. FM would like to acknowledge the support received from the European Science Foundation (ESF) within the framework of the ESF activity entitled ``Gaia Research for European Astronomy Training'', Exchange Grant number 4641. 

This research has made use of: the SIMBAD database operated at CDS France; the SpeX Prism Spectral Libraries, maintained by Adam Burgasser at http://pono.ucsd.edu/$\sim$adam/browndwarfs/spexprism; and, the M, L, and T dwarf compendium housed at dwarfArchives.org and maintained by Chris Gelino, Davy Kirkpatrick, and Adam Burgasser.


\appendix
\section{Observations log}
\label{app_log}
We present in Table \ref{obs_log} the log of the observations for our targets. For each object we show the date of observation, the standard used for telluric correction with its spectral type indicated in brackets, and the spectrophotometric standard used for flux calibration. Objects are referred to using their short ID, for the full ID and coordinates, please see Table \ref{types}.

\onecolumn
\begin{longtable}{l c c c}
\hline
ID & Observation date & Telluric & Spectrophotometric \\
 & (YYYY-MM-DD) & standard (type) & standard \\
\hline
\endfirsthead
\multicolumn{4}{c}{Continued from the previous page.} \\
\hline
ID & Observation date & Telluric & Spectrophotometric \\
 & (YYYY-MM-DD) & standard (type) & standard \\
\hline
\endhead
\hline
\multicolumn{4}{c}{Continued on the next page.} \\
\endfoot
\hline
\caption{The log of the observations. For each target we specify the observation date, the telluric standard, and the spectrophotometric standard used. \label{obs_log}}
\endlastfoot
BRLT1   & 2011-09-19 & HIP~014898 (B3V) & LTT~7987 \\
BRLT2   & 2011-09-20 & HIP~105164 (B5V) & LTT~7987 \\
BRLT3   & 2010-11-28 & HIP~038896 (B3V) & GD~71 \\
BRLT5   & 2010-11-27 & HIP~018926 (B3V) & GD~71 \\
BRLT6   & 2011-09-18 & HIP~013327 (B7V) & LTT~7987 \\
BRLT7   & 2011-09-20 & HIP~105164 (B5V) & LTT~7987 \\
BRLT8   & 2011-09-21 & HD~1160 (A0V)    & LTT~7987 \\
BRLT9   & 2011-09-19 & HIP~014898 (B3V) & LTT~7987 \\
\end{longtable}
\twocolumn

\section{Additional photometry}
\label{app_phot}
We present in Table \ref{add_phot} additional photometric data for our targets. SDSS data were obtained from the DR10, while WISE data were obtained from the All-Sky Data Release \citep{2010AJ....140.1868W}. None of our targets is detected in the WISE W4 band. Objects are referred to using their short ID, for the full ID and coordinates, please see Table \ref{types}.

\onecolumn
\begin{longtable}{l c c c c c c c}
\hline
ID & SDSS $u$ & SDSS $g$ & SDSS $r$ & WISE W1 & WISE W2 & WISE W3 & Notes \\
\hline
\endfirsthead
\multicolumn{8}{c}{Continued from the previous page.}\\
\hline
ID & SDSS $u$ & SDSS $g$ & SDSS $r$ & WISE W1 & WISE W2 & WISE W3 & Notes \\
\hline
\endhead
\hline
\multicolumn{8}{c}{Continued on the next page.}\\
\endfoot
\hline
\caption{Additional photometry for our targets, obtained from the SDSS DR10 and the \textit{WISE} All-Sky Data Release. Notes: (a) WISE magnitudes likely affected by blending with nearby source or background galaxy; (b) WISE magnitudes likely affected by diffraction spikes of nearby source. \label{add_phot}}
\endlastfoot
BRLT1   & 23.838$\pm$0.626  &  24.458$\pm$0.483  &  24.103$\pm$0.500 & 15.326 $\pm$ 0.047 & 14.870 $\pm$ 0.093 & $>$ 12.144 &  \\
BRLT2   & 24.875$\pm$0.749  &  25.306$\pm$0.526  &  24.139$\pm$0.549 & 16.484 $\pm$ 0.111 & 16.172 $\pm$ 0.325 & 12.452 $\pm$ 0.495 &  \\
BRLT3   & 24.177$\pm$0.789  &  25.244$\pm$0.524  &  24.198$\pm$0.700 & 14.949 $\pm$ 0.039 & 14.340 $\pm$ 0.060 & $>$ 12.359 &  \\
BRLT6   & 22.220$\pm$0.575  &  25.576$\pm$0.535  &  23.935$\pm$0.480 & 16.054 $\pm$ 0.074 & 16.187 $\pm$ 0.364 & $>$ 12.400 &  \\
BRLT7   & 24.241$\pm$0.806  &  25.025$\pm$0.570  &  25.135$\pm$0.651 & 15.073 $\pm$ 0.044 & 14.845 $\pm$ 0.167 & $>$ 11.062 &  \\
BRLT8   & 25.520$\pm$0.593  &  25.621$\pm$0.527  &  25.402$\pm$0.556 & 15.035 $\pm$ 0.039 & 14.529 $\pm$ 0.155 & 11.380 $\pm$ 0.404 &  \\
BRLT9   & 24.663$\pm$0.718  &  25.126$\pm$0.594  &  24.757$\pm$0.627 & 15.818 $\pm$ 0.057 & 15.934 $\pm$ 0.307 & $>$ 12.472 &  \\
BRLT10  & 25.035$\pm$1.039  &  25.436$\pm$0.724  &  24.274$\pm$0.674 & 14.261 $\pm$ 0.032 & 13.669 $\pm$ 0.042 & $>$ 12.220 &  \\
\end{longtable}
\twocolumn

\label{lastpage}

\end{document}